\documentclass[preprint,authoryear,11pt]{elsarticle}
\pdfoutput = 1

\usepackage{geometry}
\usepackage{cmbright}
\usepackage[utf8]{inputenc}
\usepackage[T1]{fontenc}
\usepackage[english]{babel}
\usepackage{amsmath}
\usepackage{graphicx}
\usepackage{fancyhdr}
\usepackage{pdflscape}
\usepackage{natbib}
\usepackage{adjustbox}
\usepackage{setspace}
\usepackage{float}
\usepackage[]{caption}
\usepackage{dingbat}
\usepackage{tabularx}
\usepackage{booktabs}
\usepackage{tabto}
\usepackage{nameref}
\usepackage{enumitem}
\usepackage[title]{appendix}

\usepackage{amsmath}
\usepackage{amsfonts}
\usepackage{amssymb}
\usepackage{amsthm}
\usepackage{latexsym}

\usepackage{sectsty}
\sectionfont{\large}
\subsubsectionfont{\normalsize \bf}

\newtheoremstyle
{style}   					
{5pt}                     	
{5pt}                     	
{\normalfont}          		
{}                         	
{\sffamily}					
{.}  						
{ } 						
{}							
\theoremstyle{style}
\newtheorem{definition}{\textit{Definition}}

\usepackage{lineno}
\usepackage{soul}

\usepackage[usenames, dvipsnames]{xcolor}

\usepackage[colorlinks]{hyperref}

\addto\extrasenglish{												%
}

\addto\captionsenglish{}

\newcommand*{\fullnameref}[1]{\hyperref[{#1}]{\autoref*{#1}: \nameref*{#1}}}

\newcommand*{\fullref}[1]{\hyperref[{#1}]{\autoref*{#1}}}

\newcommand*{\defref}[1]{\hyperref[{#1}]{Definition \autoref*{#1}}}

\begin{document}

\hypersetup{colorlinks, linkcolor = blue, citecolor = Gray} 


\begin{frontmatter}
\title{Random field theory-based p-values: A review of the SPM implementation}

\author{Dirk Ostwald\corref{cor1}}\ead{dirk.ostwald@ovgu.de}
\author{Sebastian Schneider}
\author{Rasmus Bruckner}
\author{Lilla Horvath}

\address{
Institute of Psychology and Center for Behavioral Brain Sciences, \\
Otto-von-Guericke Universität Magdeburg, Germany}

\cortext[cor1]{Corresponding author}

\begin{abstract}
P-values and null-hypothesis significance testing are popular data-analytical tools in functional neuroimaging. Sparked by the analysis of resting-state fMRI data, there has been a resurgence of interest in the validity of some of the p-values evaluated with the widely used software SPM in recent years. The default parametric p-values reported in SPM are based on the application of results from random field theory to statistical parametric maps, a framework commonly referred to as \textit{RFT}. While RFT was established two decades ago and has since been applied in a plethora of fMRI studies, there does not exist a unified documentation of the mathematical and computational underpinnings of RFT as implemented in current versions of SPM. Here, we provide such a documentation with the aim of contributing to contemporary efforts towards higher levels of computational transparency in functional neuroimaging.
\end{abstract}

\begin{keyword}
fMRI \sep GLM \sep SPM \sep p-values \sep random field theory
\end{keyword}
\end{frontmatter}

\section{Introduction}
\thispagestyle{empty}
Despite their debatable value in a scientific context, p-values and statistical hypothesis testing remain popular data-analytical techniques \citep[][]{Fisher1925, Neyman1933, Benjamin2018}. Given recent discussions about the reproducibility of quantitative empirical findings \citep[e.g.][]{Ioannidis2005, Button2013}, there has  been a resurgence of interest in the validity of p-values and statistical hypothesis tests involved in the analysis of fMRI data with the popular software packages SPM and FSL \citep[][]{Eklund2016, Mumford2016, Brown2017, Eklund2017, Cox2017, Kessler2017, Flandin2017, Bowring2018, Geuter2018, Turner2018, Slotnick2017, Slotnick2017a, Mueller2017, Eklund2018, Gopinath2018a, Gopinath2018b, bansal_clusterlevel_2018, gong_statistical_2018, geerligs_improving_2021, davenport_expected_2021}. The default parametric p-values reported in SPM and FSL are based on the application of results from random field theory to statistical parametric maps and derive from the fundamental aim to control the multiple testing problem entailed by the mass-univariate analysis of fMRI data \citep{Friston1994a, Poline2012, Monti2011,  Ashburner2012, Jenkinson2012,  Nichols2012}. In brief, the p-values reported by SPM and FSL are exceedance probabilities of topological features of a data-adapted null hypothesis random field model. Historically, this framework was established in a series of landmark papers in the 1990's, with some additional contributions in the early 2000's \citep{Friston1991, Worsley1992, Friston1994, Worsley1996, Friston1996, Kiebel1999, Jenkinson2000, Hayasaka2003, Hayasaka2004}. In the following, we will collectively refer to this framework as ``random field theory-based p-value evaluation'', which is commonly abbreviated \textit{RFT} in the neuroimaging literature \citep[cf.][]{Nichols2012}.

While RFT was established almost two decades ago and has since been applied in a plethora of functional neuroimaging studies, there does not exist a unified documentation of the mathematical and computational underpinnings of RFT. More specifically, the early landmark papers on the approach are characterized by an evolution of ideas which were often superseded by later developments. In these later developments, however, the RFT framework was not constructed from first principles again, such that many important implementational details were rendered opaque.  As concrete examples, the conceptual papers by \citet{Friston1994} and \citet{Friston1996} mention the expected Euler characteristic, which is central to the current SPM implementation of RFT, only in passing. They are also concerned with Gaussian random fields only, rather than the \textit{T} and \textit{F} random fields that are routinely assessed using RFT. Similarly, the technical papers by \citet{Kiebel1999}, \citet{Hayasaka2003}, and \citet{Hayasaka2004} present variations on the estimation of RFT's null model parameters without a definite commitment to the approach that is currently implemented in SPM. Finally, reviews of the approach, such as provided by \citet{Nichols2003}, \citet{Friston2007b}, \citet{Worsley2007}, and \citet{Flandin2015} only cover certain aspects of RFT and usually omit many computational details that would be required for its de-novo implementation. Given this unsatisfying state of the RFT literature and the recent commotion with regards to the approach, we reasoned that a comprehensive documentation of the mathematical and computational underpinnings of RFT may help to alleviate uncertainties about the conceptual and implementational details of the approach for neuroimaging analysis novices, practitioners and theoreticians alike. Here, we thus aim to provide such a documentation.

\begin{figure}[htp!]
\centerline{\includegraphics[width= 15.5 cm]{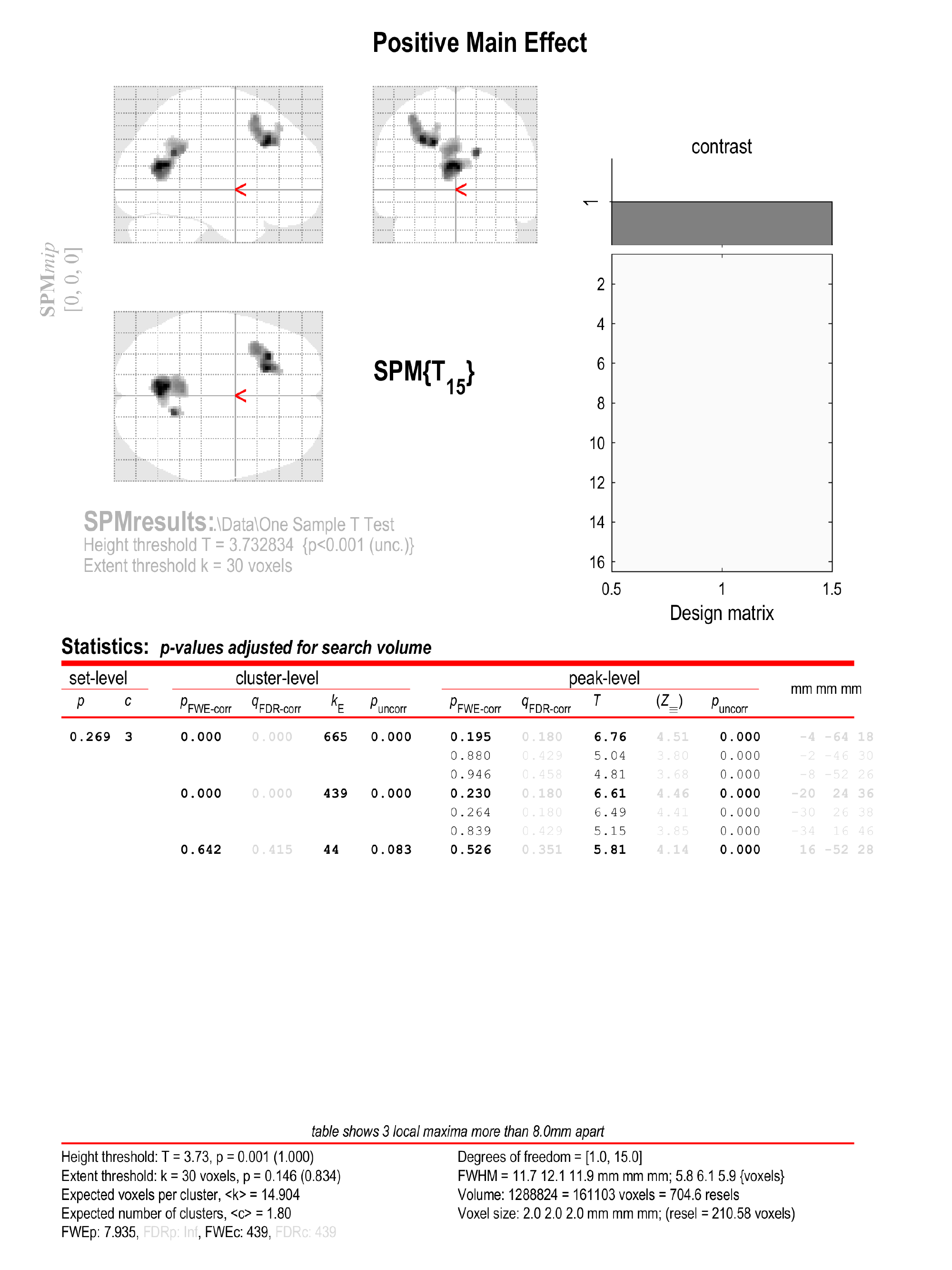}}
\caption{The SPM results table for a second-level one-sample $T$-test design relating to the positive main effect of visual stimulus coherence in a perceptual decision making fMRI study \citep{Georgie2018}. The central aim of this work is to document the mathematical and computational underpinnings of the p-values listed in the SPM results table. As indicated by the whitened parts of the table, we only consider corrected p-values related to FWER control. For implementational details of the evaluation of the SPM results table using the SPM12 Version 7771 distribution, please see \textit{rft\_1.m}.}
\label{fig:rft_spm_table}
\end{figure}

We constrain our treatment by what probably constitutes the most important practical aspect of the RFT approach: the SPM results table (\autoref{fig:rft_spm_table}). We focus on the RFT implementation in SPM rather than FSL, because SPM remains the most commonly used software tool in the neuroimaging community and because the origins of RFT are closely linked to the development of SPM \citep{Carp2012, Borghi2018, Ashburner2012}. More specifically, we focus on the mathematical theory and statistical evaluation of the p-values reported in the set-level, cluster-level, and peak-level columns, as well as the footnote of the results table of SPM12 Version 7771. Note that in the following, all references to SPM imply a reference to SPM12 Version 7771 run under Matlab R2021a (Version 9.10.0.1710957, Update 4). We particularly address the application of RFT in the context of GLM-based fMRI data analysis, and do not discuss other applications, such as in the analysis of M/EEG data \citep[e.g.,][]{Kilner2010}. Our guiding example is a second-level one-sample $T$-test design as shown in \autoref{fig:rft_spm_table}. With respect to multiple testing control, we focus on the family-wise error rate (FWER)-corrected p-values and do not cover the later addition of false discovery rate-corrected q-values \citep{Chumbley2009, Chumbley2010}. With respect to FWER control, we focus on the evaluation of p-values rather than statistical hypothesis tests. This is warranted by the fact that SPM de-facto only evaluates p-values, while statistical hypothesis tests are performed by the SPM user, who typically decides to report only activations at a given level of significance. Finally, from an implementational perspective, we primarily focus on the functionality of the \textit{spm\_list.m}, \textit{spm\_P\_RF.m}, \textit{spm\_est\_smoothness.m}, and \textit{spm\_resel\_vol.c} functions and cover computational aspects of more low-level SPM routines only in passing. With these constraints in place, we next review the outline of our treatment and preview the central points of each section.

\newpage
\subsubsection*{Overview}

Overall, we proceed in three main parts, \fullnameref{sec:background}, \fullnameref{sec:theory}, and \fullnameref{sec:application}. In \fullnameref{sec:background}, we establish a few key definitions from the two main pillars of RFT: random field theory and geometry \citep{Adler1981, Adler2007}. Specifically, in \fullnameref{sec:random_fields}, we define the concept of a real-valued random field and discuss some subtle but important aspects of this definition. Our definition rests on mathematical probability theory, in which random variables are understood as mappings between probability spaces. Readers unfamiliar with these measure-theoretic underpinnings of probability theory are recommended to consult appropriate references from the general literature, such as 
\citet{Billingsley1978, Rosenthal2006, Fristedt2013, kallenberg_foundations_2021}. After discussing a few analytical features of random fields (expectation and covariance functions, stationarity), we introduce the central classes of random fields for RFT in \fullnameref{sec:gaussian_random_fields}. As a toy example that we will return to repeatedly, we introduce two-dimensional Gaussian random fields with Gaussian covariance functions. Such Gaussian random fields have the feature that their \textit{smoothness} in the RFT-sense can be captured by a single parameter of their covariance function. We also introduce the central notions of $Z\text{-}, T\text{-}$ and $F\text{-}$ fields, which we will often refer to as \textit{Gaussian-related random fields} \citep[cf.][]{Adler2017} and consider as the theoretical analogues to statistical parametric maps that derive from the practical analysis of fMRI data. In \fullnameref{sec:volumetrics} we then introduce two geometric foundations of RFT, Lebesgue measure and intrinsic volumes. These geometric concepts are of central importance in RFT, because the probability distributions of topological features of Gaussian-related random fields (for example, the probability of the global maximum to exceed a given value) depend not only on the characteristics of the Gaussian-related random field, but also on the volume of the space it extends over. In RFT, this dual dependency is condensed in the notion of \textit{resel volumes}. In brief, resel volumes are smoothness-adjusted intrinsic volumes, which in turn are coefficients in a formula for the Lebesgue measure of \textit{tubes}. The treatment in \autoref{sec:volumetrics} is necessarily shallow and emphasizes intuition over formal rigour. 

\fullnameref{sec:theory} is devoted to the theoretical development of RFT. The central part of this section is the delineation of the parametric probability distributions used by RFT to capture the stochastic behaviour of topological features of a Gaussian-related random field's excursion set under the null hypothesis. The exceedance probabilities of these parametric distributions for observed data are the p-values reported in the SPM results table. We discuss the theory of RFT against the background of a continuous space, discrete data point model. This model is conceived as the theoretical analogue to the familiar mass-univariate GLM and is formally introduced at the outset of \fullref{sec:theory}. \fullnameref{sec:excursions_smoothness_resels} then introduces three foundational concepts of RFT: \textit{excursion sets} of Gaussian-related random fields are subsets of the field's domain on which the field exceeds a pre-specified value $u$. In the applied neuroimaging literature, this value $u$ (or, more specifically, its equivalent p-value) is known as the \textit{cluster-forming threshold} \citep{Cobidas2016} and has been at the center stage of the recent debate on the validity of parametric p-values in fMRI data analysis \citep[e.g.,][]{Flandin2017}. The topological constitution of excursion sets, and hence the probability distributions of their features, depends in a predictable manner on the value of $u$ and on the \textit{smoothness} of the underlying Gaussian-related random field. In the applied neuroimaging literature, the smoothness measure employed by RFT is known as ``FWHM'' \citep[e.g.,][]{Eklund2016}. In fact, the parameterization of a Gaussian-related random field's smoothness in terms of the full widths at half maximum (FWHMs) of isotropic Gaussian convolution kernels  applied to hypothetical white-noise Gaussian random fields conforms to a reparameterization of a more fundamental smoothness measure and was introduced early on in the RFT literature \citep{Friston1991, Worsley1992}. We introduce this more fundamental smoothness measure in the \textit{Smoothness} subsection of \fullref{sec:excursions_smoothness_resels}. To obtain an intuition about this measure, we show analytically in \ref{sec:proofs} that for Gaussian random fields with Gaussian covariance functions it evaluates to a scalar multiple of the covariance function's length parameter. We close \fullref{sec:excursions_smoothness_resels} by introducing the FWHM reparameterization of smoothness and finally conjoining the probabilistic and geometric threads of our exposition in the notion of \textit{resel volumes}. Equipped with these foundations, we then proceed to the core of RFT theory in \fullnameref{sec:probabilistic_properties}. More specifically, upon establishing a set of expected values, we review the probability distributions of the following excursion set features as evaluated by the SPM implementation of RFT: (1) the global maximum, (2) the number of clusters, (3) the cluster volume, (4) the number of clusters of a given volume, and (5) the maximal cluster volume. We supplement this discussion with number of proofs in \ref{sec:proofs}. The distributions of the global maximum and the maximal cluster volume are the distributions that endow the SPM implementation of RFT with FWER control at the peak- and cluster-level, respectively. To this end, we review the principal idea of controlling the FWER by means of maximum statistics in \ref{sec:hypothesis_testing_fwer_control}. With all theoretical aspects in place, we are then in the position to consider the application of RFT theory in a GLM-based fMRI data-analytical setting.

We thus begin \fullnameref{sec:application} by reconsidering the continuous space GLM introduced in \autoref{sec:theory} from the discrete spatial sampling perspective of fMRI. In \fullnameref{sec:parameter_estimation} we then discuss how RFT furnishes a data-adaptive null hypothesis model by estimating the smoothness and approximating the intrinsic volumes of the Gaussian-related random field that underlies the observed data. The estimated FWHM smoothness parameters and intrinsic volumes are combined in estimated resel volumes, which form the pivot point between the data-driven and theory-based aspects of RFT. We then close our treatment by considering the p-values reported in the SPM results table. As will become evident, the p-values reported for set-, cluster-, and peak-level inferences can actually be evaluated using a single (Poisson) cumulative distribution function that is adapted for different topological statistics and data-based null model characteristics \citep{Friston1996}. Finally, in \fullnameref{sec:discussion}, we briefly explore potential future avenues for the further refinement of RFT. 

In summary, we make the following novel contributions to the literature. From a scientific perspective, we provide a unified review of RFT, which is both mathematically comprehensive and computationally explicit. As such, our treatment allows for the detailed statistical interpretation of the experimental effects reported using RFT in the last decades. Further, given the recent discussions on the validity of SPM's cluster-level p-values, our treatment has the potential to serve as a readily accessible starting point for the further refinements of RFT. Finally, from an educational perspective, we provide a novel resource to introduce newcomers to the field of computational cognitive neuroscience to one of functional neuroimaging's equally most basic (univariate cognitive process mapping using the GLM) and sophisticated (random field-based modelling) analysis tools. All custom written Matlab code (The MathWorks, NA) implementing simulations and visualizations reported on herein is available from \url{https://osf.io/3dx9w/}. The code repository also contains the group-level fMRI data evaluated in \autoref{fig:rft_spm_table}, and documented and revised versions of SPM's parameter estimation and p-value evaluation routines (\textit{rft\_spm\_est\_smoothness.m} and \textit{rft\_spm\_table.m}, respectively). For details, please refer to the OSF project documentation.

\subsubsection*{Prerequisites and notation}

We assume throughout that the reader is familiar with the theoretical and practical aspects of GLM-based fMRI analysis as presented for example in \citet{ Kiebel2007, Monti2011, Poline2012} and \citet{Poldrack2011}. We presume that a familiarity with basic concepts of the multiple testing problem entailed by the mass-univariate GLM-based analysis of fMRI data, as covered for example in \citet{Brett2007, Nichols2012} and \citet{Ashby2011} is beneficial. As mentioned above, we provide a synopsis of essential aspects from the theory of multiple testing in  \ref{sec:hypothesis_testing_fwer_control}. In \autoref{tab:mathematical_symbols}, we list mathematical symbols that will be used throughout.

\begin{table}[]
\begin{small}
\renewcommand{\arraystretch}{1.3}
\begin{tabularx}{\textwidth}{ll} 
\toprule
Symbol & Meaning    																				\\ 
\hline
$\mathbb{N}^0$, $\mathbb{N}_{m}$, $\mathbb{N}_{m}^0$				
& The sets $\mathbb{N} \cup \{0\}$, $\{1,2,...,m\}$, and $\{0,1,...,m\}$, respectively
\\
$|S|$
& Cardinality of a set $S$
\\
$e_i$
& $i$th standard basis vector for $\mathbb{R}^d$
\\
$||x||_2$
& Euclidean norm of $x \in \mathbb{R}^d$															
\\
$|A|$
& Determinant of a matrix $A \in \mathbb{R}^{d \times d}$	
\\
$\mbox{p.d.}$
& Positive-definite
\\
$\mbox{inf}, \mbox{sup}$
& Infimum, supremum
\\
$\nabla f(x)$																								
& Gradient of a function $f : \mathbb{R}^d \to \mathbb{R}$ at $x$ 									
\\
$\frac{\partial}{\partial x_i} f(x)$ 
& $i$th partial derivative of a function $f : \mathbb{R}^d \to \mathbb{R}$ at $x$, $i = 1,...,d$ 	
\\
$\Gamma(x)$
& Gamma function evaluated at $x \in \mathbb{R}$
\\
$\mathbb{P}, \mathbb{E}, \mathbb{C}$
& Probability, expectation, covariance 																
\\
$N(\mu, \Sigma)$				
& Gaussian distribution
\\
$\bullet$
& End of definition
\\
$\Box$
& End of proof
\\
\bottomrule
\end{tabularx} 
\caption{List of mathematical symbols. Note that the meaning of the symbol $|\cdot|$ is usually clear from the context.}\label{tab:mathematical_symbols}
\end{small}
\end{table}

\section{Background}\label{sec:background}
\subsection{Real-valued random fields}\label{sec:random_fields}
We commence by discussing some basic aspects of real-valued random fields. For comprehensive introductions to the theory of random fields, see for example \citet{Christakos1992}, \citet{Abrahamsen1997}, \citet{Adler1981} and \citet{Adler2007}. We use the following definition: 

\begin{definition}[Real-valued random field]\label{def:random_field} 
A real-valued random field $\{X(x)|x \in S\}$ on a domain $S \subset \mathbb{R}^D$ is a set of random variables $X(x)$ on a probability space $(\Omega, \mathcal{A}, \mathbb{P})$, i.e., for each $x \in S, X(x) : \Omega \to \mathbb{R}$ is a random variable.

$\hfill \bullet$
\end{definition}

\noindent A variety of notations for real-valued random fields exist. In the RFT literature, random fields are most commonly denoted by ``$X(x),x \in S$'' \citep[e.g.,][]{Worsley1994, Taylor2007, Flandin2015}, a convention which we shall follow herein. With regards to this notation, it is important to realize that the symbol $X(x)$ denotes a random variable, while it may look suspiciously like the value of a function $X$ evaluated for an input argument $x$. This is of course intentional: because each $X(x)$ of a real-valued random field is a real-valued random variable, it can be written as
\begin{equation}\label{eq:rf_Xx}
X(x) : \Omega \to \mathbb{R}, \omega \mapsto X(x)(\omega).
\end{equation}
In expression \eqref{eq:rf_Xx}, $X(x)(\omega)$ denotes the value in $\mathbb{R}$ that $X(x)$ takes on for the input argument $\omega \in \Omega$. Because the double bracket notation $X(x)(\omega)$ is rather unconventional, an alternative notation mainly used in the mathematical literature is to write a real-valued random field as 
\begin{equation}\label{eq:rf_X_x_omega}
X: S \times \Omega \to \mathbb{R}, (x,\omega) \mapsto X(x,\omega),
\end{equation}
or to denote the constituent random variables of a real-valued random field by
\begin{equation}\label{eq:rf_Xx_omega}
X_x: \Omega \to \mathbb{R}, \omega \mapsto X_x(\omega).
\end{equation}
The three different notations of expressions \eqref{eq:rf_Xx}, \eqref{eq:rf_X_x_omega} and \eqref{eq:rf_Xx_omega} make the same point: for a fixed $\omega$, $X(x)$, $X$, or $X_x$ is a deterministic, real-valued function with domain $S$. The notation $X(x)$ suppresses the dependency of this function on the random elementary outcome $\omega$, while the other two notations do not. The clearest notation in this regard is perhaps offered by expression \eqref{eq:rf_X_x_omega}, but the notation $X(x),x \in S$ seems to be generally preferred. Crucially, and these notational subtleties aside, the domain $S$ of a random field is usually an uncountable infinite set (such as a $D$-dimensional interval), which implies that a random field usually comprises an uncountable infinite number of random variables.

Two further aspects of \defref{def:random_field} are worth discussing. First, we chose the same letter for both the random field (capital $X$) and the domain values (lowercase $x$), because on the one hand, we will need to define specific random fields later on, and $X$ is for the current purposes perhaps the most generic choice, while on the other hand, $x$ invokes a clearer notion of a spatial domain than, say, $t$. In fact, $t$ is a popular letter for the elements of the domain of a random field (e.g., \citet{Worsley1994}). However, because random fields will be used in the context of RFT for GLM-based fMRI data analysis as spatial models, we prefer $x$. Second, we defined the domain $S$ to be a subset of $\mathbb{R}^D$. We use the letter $S$, because in the context of RFT, the domains of random fields of interest are typically referred to as \textit{\underline{s}earch spaces}. Intuitively, the search spaces of interest in GLM-based fMRI data analysis correspond to the brain (in whole-brain analyses) or brain regions of interest (in small volume correction analyses). The $D$ we have primarily in mind is $D = 3$, i.e., the case of random fields over three-dimensional space, such as a volume of observed test statistics. For visualization purposes we will also consider the case $D = 2$. In the case of $D = 1$, random fields are commonly referred to as \textit{random processes}, or perhaps even more often as \textit{stochastic processes}. Finally, a linguistic remark: because, as the pinnacle of mass-univariate GLM-based fMRI data analysis, RFT concerns only univariate statistical values, we are in fact only concerned with real-valued (as opposed to, say, vector-valued) random fields. We will therefore use the terms ``real-valued random field'' and ``random field'' interchangeably henceforth.

\subsubsection*{Expectation, covariance, and variance functions}

Like random variables, random fields have certain analytical features that express their expected (or ``average'') behaviour over many realizations. These features, which generalize the concepts of a random variable's expectation and variance, are known as \textit{expectation}, \textit{covariance}, and \textit{variance functions}. We use the following definitions \citep[cf.][]{Abrahamsen1997}:

\begin{definition}[Expectation, covariance, and variance functions]
For a real-valued random field $X(x), x \in S$, the expectation function  is defined as
\begin{equation}\label{eq:exp_fun}
m : S \to \mathbb{R}, x \mapsto m(x) := \mathbb{E}(X(x)),
\end{equation}
the covariance function is defined as
\begin{equation}\label{eq:cov_fun}
c : S \times S \to \mathbb{R}, (x,y) \mapsto c(x,y) := \mathbb{C}(X(x),X(y))
\end{equation}\label{eq:var_fun}
and the variance function is defined as
\begin{equation}
s^2 : S \to \mathbb{R}_{\ge 0}, x \mapsto s^2(x) := c(x,x).
\end{equation}
$\hfill \bullet$
\end{definition}

\noindent Note that the expectation and covariance expressions \eqref{eq:exp_fun} and \eqref{eq:cov_fun} are evaluated for the random variables $X(x)$ and $X(x), X(y)$, respectively, with regards to the probability measure of the underlying probability space. Further note that the variance function is defined directly in terms of the covariance function.  

\subsubsection*{Stationarity}

An important property of real-valued random fields, and a fundamental assumption for the random fields considered in RFT, is \textit{stationarity}. We define  \textit{stationarity in the wide sense} as follows \citep[cf.][]{Abrahamsen1997}:

\begin{definition}[Wide-sense stationarity]
A real-valued random field $X(x), x \in S$ is called stationary in the wide sense, if the expectation function of $X(x)$ is a constant function, 
\begin{equation}
m : S \to \mathbb{R}, x \mapsto m(x) := \bar{m} \mbox{ for } \bar{m} \in \mathbb{R}
\end{equation}
and the covariance function of $X(x)$ is a function of the separation of its arguments only, i.e., for all $x,y \in S$ and $d := x - y$, the covariance function \eqref{eq:cov_fun} can be written as
\begin{equation}
c : S \to \mathbb{R}, d \mapsto c(d).
\end{equation}
$\hfill \bullet$
\end{definition} 

\noindent In addition to stationarity in the wide sense, there exists the notion of \textit{stationarity in the strict sense} \citep{Abrahamsen1997}. Stationarity in the strict sense means that all finite-dimensional distributions of a real-valued random field are invariant under arbitrary translations of their support arguments in the domain of the real-valued field. Because stationarity in the wide and the strict sense are equivalent for Gaussian and Gaussian-related random fields, which are our primary concern herein, and because the concept of wide-sense stationarity is more readily applicable in the computational simulation of random fields, we only formalize  stationarity in the wide sense. Note, however, that in general, stationarity in the strict sense implies stationarity in the wide sense, but not vice versa. Henceforth, we shall only use the term ``stationarity'' to mean stationarity in the wide sense. Finally, note that stationary random fields are also sometimes referred to as ``homogeneous'' random fields (e.g., \citet{Adler1981}, \citet{Worsley1995a}). 

The role of stationarity for RFT is fundamental, because, intuitively, it corresponds to the assumption that the null hypothesis holds true at every location of the search space $S$, and the null hypotheses over the entire search space are identical. More explicitly, defining $\bar{m} := 0$ for a stationary real-valued random field corresponds to the assumption that the expected value of the random variable at each location is $0$, which, from the perspective of statistical testing is identical with the expectation of no effect, i.e., a simple null hypothesis. 

\subsection{Gaussian and Gaussian-related random fields}\label{sec:gaussian_random_fields}
\subsubsection*{Gaussian random fields}

We next introduce a central class of random fields for RFT, Gaussian random fields, and use these to make the definitions of random fields, their associated expectation, covariance, and variance functions, as well as the concept of stationarity more concrete. 

\begin{definition}[Gaussian random field]
A Gaussian random field (GRF) on a domain $S \subset \mathbb{R}^D$ is a random field $G(x), x \in S$, such that every finite-dimensional vector of random variables \begin{equation}
g := (G(x_1), G(x_2), ..., G(x_n))^T
\end{equation}  is distributed according to a multivariate Gaussian distribution.

$\hfill \bullet$
\end{definition}

\noindent This definition implies that the random vector $g$ is distributed according to a multivariate Gaussian distribution for every choice of the $x_i, i = 1,...,n$ and $n \in \mathbb{N}$, such that we can write
\begin{equation}\label{eq:grf_g}
g \sim N(\mu,\Sigma) \mbox{ for } \mu \in \mathbb{R}^n, \Sigma \in \mathbb{R}^{n \times n} \mbox{ p.d. }.
\end{equation}
Moreover, if we denote the expectation function and the covariance function of a GRF by $m$ and $c$, respectively, the expectation parameter $\mu \in \mathbb{R}^n$ and covariance matrix parameter $\Sigma = (\Sigma_{ij}) \in \mathbb{R}^{n \times n}$, p.d. in expression \eqref{eq:grf_g} have the entries
\begin{equation}\label{eq:grf_g_parameters}
\mu_i = m(x_i) \mbox{ and } \Sigma_{ij} = c(x_i, x_j)  \mbox{ for } i,j = 1,...,n. 
\end{equation}
Notably, expressions \eqref{eq:grf_g} and \eqref{eq:grf_g_parameters} achieve two things: first, they  allow for reducing the fairly abstract concept of a random field to the multivariate Gaussian distribution as an arguably more familiar entity. Second, they immediately imply a computational approach to obtain realizations from a GRF: if one defines a set of discrete support points $x_1,...,x_n$ in the domain of a GRF of interest and evaluates the expectation and covariance functions of the GRF at these points, one obtains an expectation parameter and a covariance matrix that can be used as parameters for a multivariate Gaussian distribution random vector generator. Of course, this requires that the expectation and covariance functions of the GRF are defined as well. 

\begin{figure}[htbp]
\centerline{\includegraphics[width=15 cm]{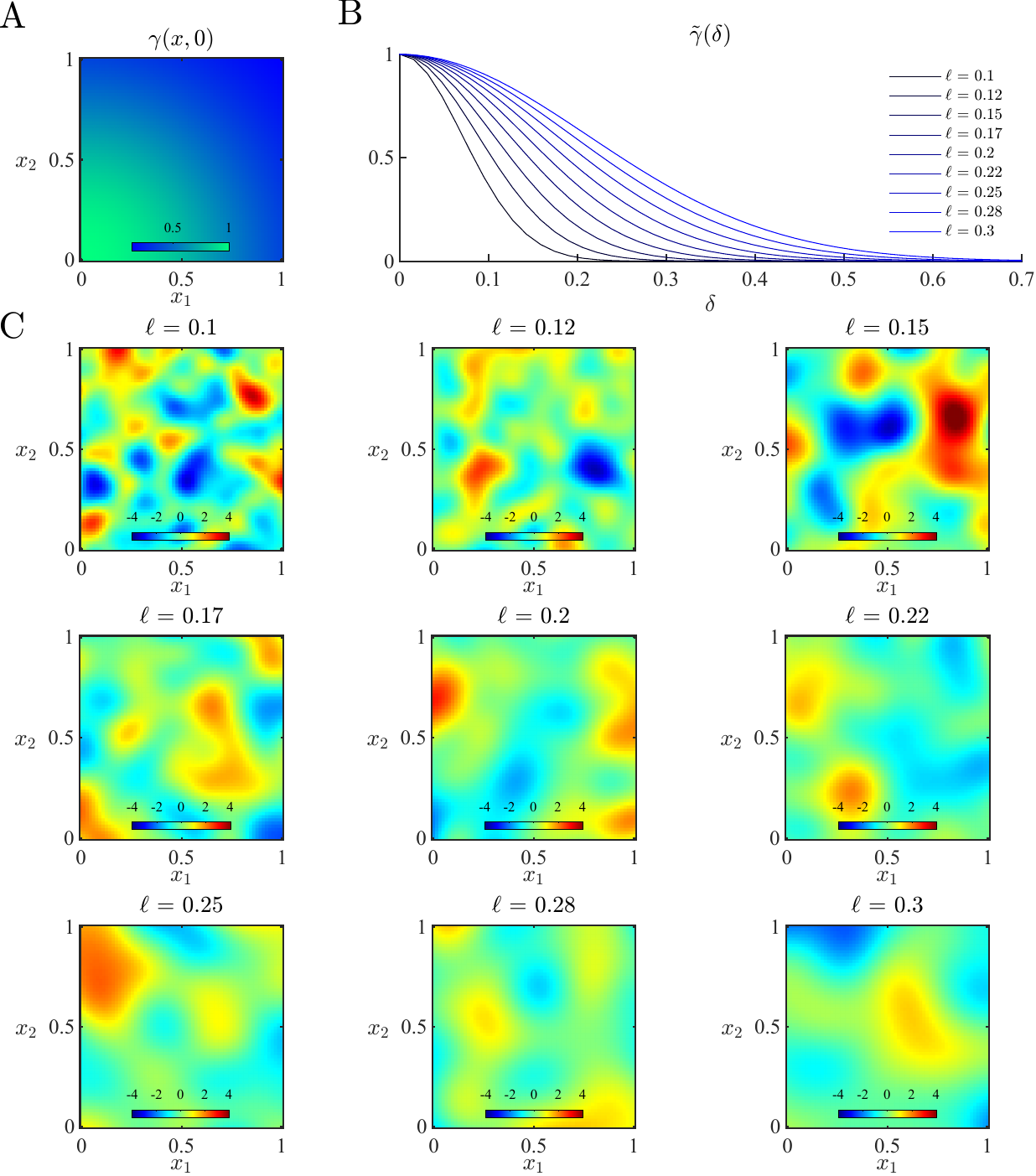}}
\caption{Realizations of Gaussian random fields with Gaussian covariance functions (GCFs).  (\textbf{A}) Panel A visualizes a GCF for a random field with two-dimensional domain $D = 2$ of the form of expression \eqref{eq:gauss_cf_1} with $y = (0,0)^T$ with parameters $v = 1$ and $\ell = 1$. (\textbf{B}) Panel B visualizes GCFs  for a random field with two-dimensional domain $D = 2$ of the form of expression \eqref{eq:gauss_cf_2} with parameter $v = 1$ and varying length constants $\ell$. Note that the height of $\tilde{\gamma}$ at a distance $\delta$ indicates the amount of covariation between two random variables at distance $\delta$ in the random field. (\textbf{C}) Panel C depicts nine realizations of a Gaussian random field with domain $S = [0,1] \times [0,1]$ with $2^6$ support points in each dimension, corresponding to spatially arranged samples from a $2^6$ dimensional multivariate Gaussian distribution of the form given by expressions \eqref{eq:grf_g} and \eqref{eq:grf_g_parameters}. Each realization is based on a Gaussian random field with Gaussian covariance function of varying length parameter $\ell$, as indicated in the subpanel titles. Note that for larger values of $\ell$, and hence larger covariation of two arbitrary random variables of the field at distance $\delta$, the realizations show less variability over space, and appear ``smoother'' than for smaller values of $\ell$. For the full implementational details of these simulations, please see \textit{rft\_2.m}.}
\label{fig:rft_grfs}
\end{figure}

Defining appropriate covariance functions and deciding whether a given function is actually a suitable covariance function, such that the corresponding covariance matrix is positive-definite, is a mathematical problem on its own (see e.g., \citet[][Chapter 4]{Rasmussen2006}, \citet{Abrahamsen1997}), and we will not delve into this issue here. We do, however, provide an important example of a valid covariance function, the so-called \textit{Gaussian covariance function} (GCF)  given by (e.g., \citet[][p. 71]{Christakos1992} and \citet[][p. 5]{Powell2014})
\begin{equation}\label{eq:gauss_cf_1}
\gamma : S \times S \to \mathbb{R}_{>0}, (x,y) \mapsto 
\gamma(x,y) := v \exp\left(-\frac{||x - y ||_2^2}{\ell^2} \right) \mbox{ with } v,\ell > 0.
\end{equation}
Note that the variance function of a GRF with a GCF evaluates to 
\begin{equation}
s^2(x) = \gamma(x,x) = v,
\end{equation}
and that the GCF is de-facto a function of the scalar Euclidean distance 
\begin{equation}
\delta := ||x - y ||_2 \in \mathbb{R}_{\ge 0}
\end{equation}
between $x$ and $y$ only. It may thus equivalently be expressed as
\begin{equation}\label{eq:gauss_cf_2}
\tilde{\gamma} : \mathbb{R}_{\ge 0} \to \mathbb{R}_{>0}, \delta \mapsto 
\tilde{\gamma}(\delta) := v \exp \left(- \frac{\delta^2}{\ell^2} \right) \mbox{ with } v,\ell > 0,
\end{equation}
for which $v = \tilde{\gamma}(0)$. The parameter $\ell$ in eqs. \eqref{eq:gauss_cf_1} and \eqref{eq:gauss_cf_2} is a length constant and determines  the strength of the covariation of the random variable $X(x)$ and $X(y)$ at a distance $\delta = ||x - y ||_2$ (\autoref{fig:rft_grfs}A and B). Intuitively, the larger the value of $\ell$, the stronger the covariation at a given distance, and hence the less variable the profile of GRF realizations  over space. In \autoref{fig:rft_grfs}C, we visualize the dependency of realizations of GRFs with GCF on the parameter $\ell$, for GRFs with domain $S = [0,1] \times [0,1] \subset \mathbb{R}^2$ and constant zero mean function. Note that the GCF is called ``Gaussian'' because of its functional form, and not because we consider it in the context of Gaussian random fields. Further note that for a constant mean function, GRFs with GCF are stationary.  Finally, note that GRFs with $D = 1$ are known as \textit{Gaussian processes} and enjoy some popularity in the machine learning community \citep{Rasmussen2004, Rasmussen2006}. 

\subsubsection*{Gaussian-related random fields}

Three additional random fields that derive from Gaussian random fields are of central importance in RFT. These fields, often referred herein under the umbrella term \textit{Gaussian-related random fields}, can be considered the random field analogues of the $Z\text{-},T\text{-},$ and $F\text{-}$distributions \citep{Shao2003}. Following \citet{Worsley1994} and \citet{Adler2017}, we define these fields as follows:

\begin{definition}[Gaussian-related random fields: $Z\text{-}$,$T\text{-}$, and $F\text{-}$fields]\label{def:tzf_fields}
A $Z\text{-}$field on a domain $S \subset \mathbb{R}^D$ is a stationary Gaussian random field with constant expectation function $m(x) := 0$ and variance function $s^2(x) = 1$. Let
\begin{equation}
Z_1(x), ...,Z_n(x), x \in S
\end{equation}  be $n$ independent $Z\text{-}$fields, i.e., 
\begin{equation}\label{eq:z_field_independence}
\mathbb{C}(Z_i(x), Z_j(x)) = 0 \mbox{ for all } 1 \le i,j\le n, i \neq j,x \in S
\end{equation} 
and let 
\begin{equation}\label{eq:z_field_grad_cov}
\mathbb{C}(\nabla Z_i(x)) = \mathbb{C}(\nabla Z_j(x)) \mbox{ for all }1 \le i,j\le n, i \neq j,x \in S.
\end{equation}
Then
\begin{equation}\label{eq:t_field}
T(x) := \frac{Z_1(x)\sqrt{n-1}}{\sqrt{\sum_{i=2}^n Z_i^2(x)}}, x \in S
\end{equation}
is called a $T\text{-}$field with $n-1$ degrees of freedom. Finally, let 
\begin{equation}
Z_1(x), ...,Z_n(x), Z_{n+1}(x), ..., Z_{n+m}(x), x \in S
\end{equation}
be $n+m$ independent $Z$-fields, i.e., 
\begin{equation}
\mathbb{C}(Z_i(x), Z_j(x)) = 0 \mbox{ for all } 1 \le i,j\le n+m, i \neq j,x \in S
\end{equation}
and let 
\begin{equation}
\mathbb{C}(\nabla Z_i(x)) = \mathbb{C}(\nabla Z_j(x)) \mbox{ for all } 1 \le i,j\le n+m, i \neq j,x \in S.
\end{equation}
Then
\begin{equation}\label{eq:F_field}
F(x) := \frac{m\sum_{i=1}^n Z_i^2(x)}{n\sum_{i=n+1}^{n+m} Z_i^2(x)}, x \in S
\end{equation}
is called an $F\text{-}$field with $n,m$ degrees of freedom.

$\hfill \bullet$
\end{definition}

\noindent In the context of RFT, we conceive of $Z\text{-},T\text{-}$ and $F\text{-}$fields as the theoretical, continuous-space analogues of \textit{statistical parametric maps}, i.e., discrete-space spatial maps of realized $Z\text{-}$, $T\text{-}$ and $F\text{-}$statistics \citep[cf.][]{Friston2007a}.

\subsection{Volumetrics}\label{sec:volumetrics}
A fundamental aspect of RFT is the fact that the probability distributions of topological features of random fields, such as the maximum of a random field to exceed a given threshold value, depend on (1) the stochastic characteristics of the random field and (2) the size of the search space over which the random field extends. Intuitively, the larger the space over which the random field extends and the more variable the random field per unit space, the higher the probability that, for example, the maximum of the field exceeds a given threshold value. Because random fields are mathematically developed over continuous space, the question of the size of a subset of a random field's domain is not trivial and rests on a large body of mathematical work that falls into the realms of differential and integral geometry \citep[][Part II]{Adler2007}. We here provide a bare minimum of terminology for describing volumes in continuous space by focussing on the notions of \textit{Lebesgue measure} and \textit{intrinsic volumes}. Lebesgue measure is a fundamental volume measure in mathematical measure theory and forms the basis for the concept of intrinsic volumes. Intrinsic volumes in turn are the foundation for the concept of resel volumes as discussed in \fullnameref{sec:excursions_smoothness_resels}.

\subsubsection*{Lebesgue measure}
Lebesgue measure is a fundamental building block of mathematical measure theory. Intuitively, Lebesgue measure can be understood as deriving from the desire to allocate a meaningful notion of volume to geometric objects that are modelled as subsets of $\mathbb{R}^D$. Specifically, the measure should (a) be compatible with the usual notion of length, area, and volume for lines, rectangles, and cuboids, respectively, (b) be smaller for an object A than for an object B, if A can be fit into object B, (c) be invariant under translations of the object, and (d) be additive in the sense that if two objects do not overlap, their joint volume is the sum of their individual volumes \citep{Meisters1997}. Lebesgue measure is a measure that fulfills these properties for a large class of subsets of $\mathbb{R}^D$. A full development of Lebesgue measure from first principles is beyond our scope and can be found in many books \citep[e.g.,][]{Billingsley1978, Cohn1980, Stein2009}. Instead, we here provide a general definition of Lebesgue measure based on \citet{Hunter2011}, discuss some of the intuition associated with this definition, and finally list some values for the Lebesgue measure of some familiar geometric objects. 

\begin{definition}[Lebesgue outer measure, Lebesgue measure]\label{def:lebesgue_measure}
Let 
\begin{equation}
R = [a_1,b_1] \times \dots \times [a_D,b_D]  \subset \mathbb{R}^D, \mbox{ with } -\infty < a_d \le b_d < \infty,\, d = 1,...,D
\end{equation}
denote a $D$-dimensional closed rectangle with sides oriented parallel to the coordinate axes, let $\mathcal{R}\big(\mathbb{R}^D\big)$ denote the set of all such rectangles in $\mathbb{R}^D$, and let
\begin{equation}\label{eq:rectangle_volume}
\varphi  : \mathcal{R}\big(\mathbb{R}^D\big) \to [0,\infty[, R \mapsto \varphi(R) := \prod_{d=1}^D (b_d - a_d)
\end{equation}
denote the volume of a rectangle in $\mathbb{R}^D$. Then the Lebesgue outer measure is defined as 
\begin{equation}\label{eq:lebesgue_outer_measure}
\lambda^* : \mathcal{P}\big(\mathbb{R}^D\big) \to [0,\infty], 
S \mapsto \lambda^*(S) := 
\inf \left\lbrace \sum_{i=1}^{\infty}\varphi(R_i)|S\subset \cup_{i=1}^{\infty} R_i, R_i \in \mathcal{R}\big(\mathbb{R}^D\big) \ \right\rbrace,
\end{equation}
where the infimum is taken over all countable collections of rectangles whose union contains $S$, a set $S \subset \mathbb{R}^D$  is called Lebesgue measurable, if for every $A \subset \mathbb{R}^D$
\begin{equation}\label{eq:lebesgue_measurability}
\lambda^*(A) = \lambda^*(A \cap S) + \lambda^*(A \textbackslash S),
\end{equation}
and the Lebesgue measure is defined as the function
\begin{equation}\label{eq:lebesgue_measure}
\lambda : \mathcal{P}\big(\mathbb{R}^D\big) \to [0,\infty], 
S \mapsto \lambda(S) := \lambda^*(S).
\end{equation}
$\hfill \bullet$
\end{definition}

\noindent The definition of Lebesgue measure thus rests on a familiar concept: the measure of a $D$-dimensional rectangle corresponds to the product of its side lengths. For $D=1$, this corresponds to the length of a line, for $D=2$ to the area of a rectangle (\autoref{fig:rft_volumetrics}A), and for $D = 3$ to the volume of a cuboid. This familiar measure of the content of $D$-dimensional rectangles is used in \defref{def:lebesgue_measure} to define an approximation of the measure of arbitrary geometric objects modelled by a subset $S \subset \mathbb{R}^D$ in expression \eqref{eq:lebesgue_outer_measure}. Specifically, the Lebesgue outer measure of $S$ is given by finding the smallest possible value of the countable sum of rectangle volumes, the rectangles of which cover the set $S$ of interest  (\autoref{fig:rft_volumetrics}B). Lebesgue measure itself is then defined as the Lebesgue outer measure restricted to a subset of subsets of $\mathbb{R}^D$ which fulfil the condition of eq. \eqref{eq:lebesgue_measurability}. 

The definition of Lebesgue measure by \defref{def:lebesgue_measure} is fairly abstract. For concrete subsets of $\mathbb{R}^D$ of geometrical interest, for example a circle in $\mathbb{R}^2$ or a cuboid in $\mathbb{R}^3$, it is not immediately clear how to compute their Lebesgue measure based on this definition. Lebesgue measure has, however, many mathematically desirable properties. For example, it can be shown that Lebesgue measure indeed fulfils the desiderata (a) - (d) discussed above, and, for $D$-dimensional rectangles, simplifies to the volume of rectangles \eqref{eq:rectangle_volume}. Some other noteworthy values of Lebesgue measure are
the Lebesgue measures of intervals in $\mathbb{R}$, 
\begin{equation}
\lambda(]a,b[) = \lambda([a,b[) = \lambda(]a,b]) = \lambda([a,b]) = b-a, 
\end{equation}
and the Lebesgue measure of a closed  ball $B$ with radius $r > 0$ in $\mathbb{R}^D$,
\begin{equation}\label{eq:ball_volume}
\lambda(B) 
= \lambda\left(\{x \in \mathbb{R}^D \vert \, ||x||_2 \le r\} \right) 
= \frac{\pi^{D/2}}{\Gamma(D/2 + 1)} r^D. 
\end{equation}
Finally, Lebesgue measure forms the foundation for the notion of the intrinsic volumes of a subset of $\mathbb{R}^D$ as discussed next.

\subsubsection*{Intrinsic volumes}

In general, intrinsic volumes can be thought of as measures of the $0$- to $D$-dimensional size of a set $S \subset \mathbb{R}^D$. Historically, \citet{Worsley1992} introduced the RFT framework for three-dimensional excursion sets only, for which it was assumed that the probability of them to touch the search space boundary was negligible. For these excursion sets, Lebesgue measure as a measure of volume was sufficient. \citet{Worsley1996} then introduced corrections for these ``boundary effects''  which require more refined measures of volume. These more refined measures of volume are furnished by intrinsic volumes, which appear in the mathematical literature under a variety of names, such as \textit{Quermassintegrale, Minkowski functionals, Dehn-Steiner functionals, integral curvatures}, and \textit{Euclidean Lipschitz-Killing curvatures} \citep{Adler2007, Adler2015, Adler2017}. As for the Lebesgue measure, a development of the concept of intrinsic volumes from first principles is beyond our scope. We thus content with a general definition of intrinsic volumes, for which we attempt to provide some intuition, and with listing some values for the intrinsic volumes of familiar geometric entities. 
As a general definition of the intrinsic volumes of a $D$-dimensional search volume $S \in \mathbb{R}^D$, we use the following implicit definition provided by \citet[][Appendix A.1.1]{Taylor2007} and \citet[][pp. 99 - 101]{Adler2015}):

\begin{definition}[Tube, intrinsic volumes]
Let $S \subset \mathbb{R}^D$ and let a tube $T_{S,r}$ of radius $r$ around $S$ be defined as 
\begin{equation}\label{eq:tube}
T_{S,r} := \{x \subset \mathbb{R}^D| \delta(x,S) \le r\}
\mbox{ with }
\delta(x,S) := \inf_{y \in S} ||x - y||_2.
\end{equation}
Then, for a convex set $S \in \mathbb{R}$ the Lebesgue measure of $T_{S,r}$ is given by 
\begin{equation}\label{eq:steiners_formula}
\lambda(T_{S,r}) = \sum_{i=0}^D b_{D-i} r^{D-i}\mu_d(S) 
\mbox{ with } 
b_j := \frac{\pi^{j/2}}{\Gamma(j/2 + 1)} \mbox{ for } j = D-i \mbox{ and }i = 0,...,D.
\end{equation}
and $\mu_0(S), \mu_1(S),..., \mu_D(S)$ are the intrinsic volumes of $S$.

$\hfill \bullet$
\end{definition}

\begin{figure}[htbp!]
\centerline{\includegraphics[width=\textwidth]{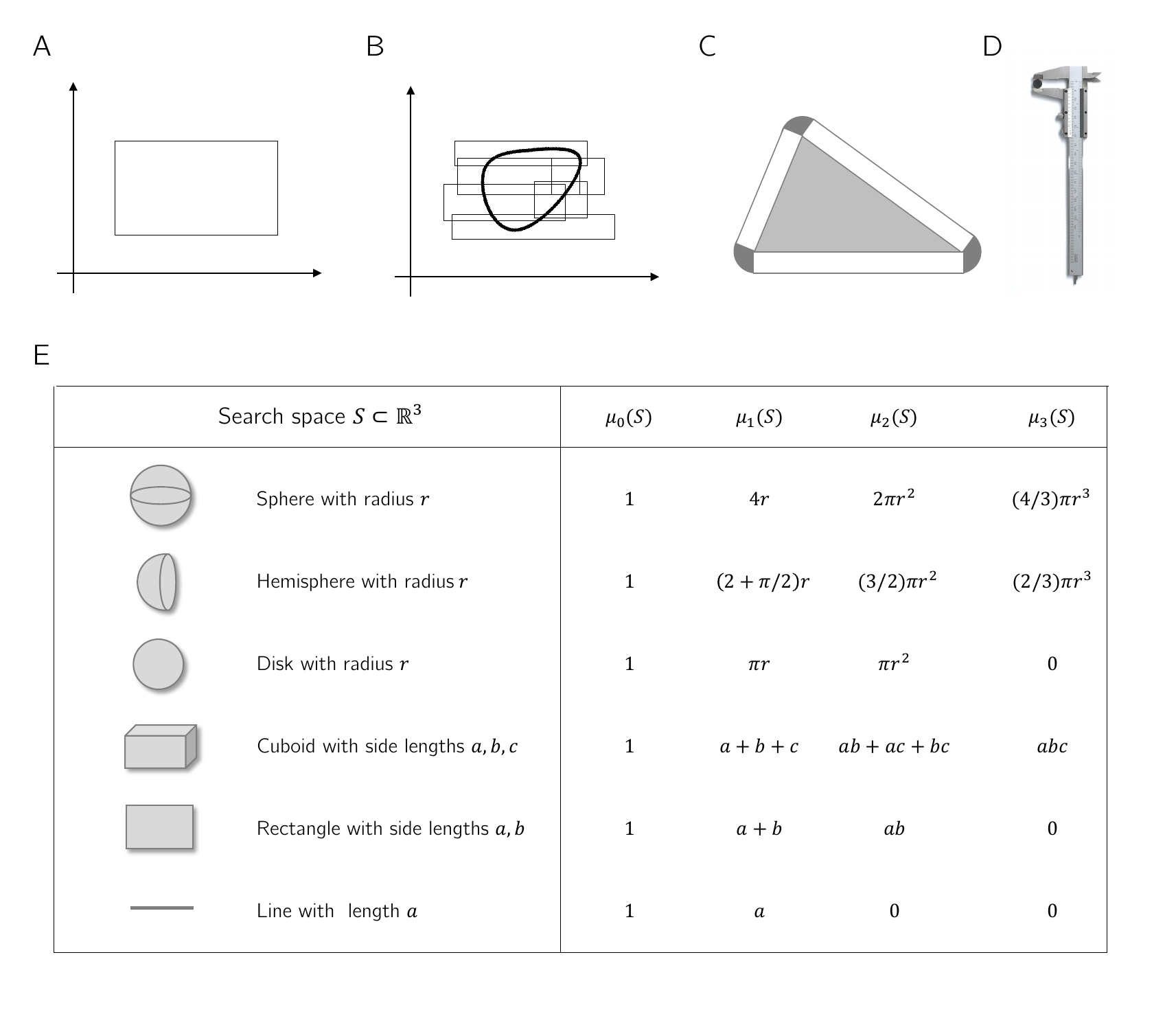}}
\caption{Lebesgue measure and intrinsic volumes. (\textbf{A}) A rectangle $R$ with sides oriented parallel to the coordinate axes in $\mathbb{R}^2$. (\textbf{B}) A visualization of Lebesgue outer measure. Consider the central object with the strong outline as a subset $S \subset \mathbb{R}^2$.  Then $S$ can be covered by a (possibly infinite) union of rectangles, such that $S \subset \cup_{i=1}^{\infty}R_i$. The panel shows a coverage of $S$ afforded by a finite number of rectangles. Clearly, in general the volume of the coverage overestimates the volume of the object of interest. (\textbf{C}) A two-dimensional tube around a triangle. The central grey area shows the triangle. The union of the triangle, the white rectangles, and the dark grey disc sections forms the two-dimensional tube around the triangle. (\textbf{D}) A caliper. (\textbf{E}) Intrinsic volumes for a selected set of basic geometric shapes as listed in \citet[][Table I]{Worsley1996}). Note that the intrinsic volumes of orders higher than the dimensionality of the geometric object are zero.}
\label{fig:rft_volumetrics}
\end{figure}

\noindent Note that the definition of the intrinsic volume $\mu_0(S),...,\mu_D(S)$ of a set $S \subset \mathbb{R}^D$ is implicit here in the sense that the quantities $\mu_0(S),...,\mu_D(S)$ appear as coefficients in the sum formula for the Lebesgue measure of a tube, and no direct way of how to evaluate these coefficients is provided. Also note that the $b_j$ correspond to the Lebesgue measures of unit balls in $\mathbb{R}$ (cf. \eqref{eq:ball_volume} with $r := 1$). To gain some insight into the meaning of this implicit definition of intrinsic volumes, we follow \citep[][pp. 100-101]{Adler2015} and discuss its application to a two-dimensional tube around a triangle (\autoref{fig:rft_volumetrics}C). Here, the triangle corresponds to the set $S \subset \mathbb{R}^2$ in eq. \eqref{eq:tube}. A two-dimensional tube $T_{S,r}$ of radius $r$ around $S$ is then formed by considering all points $x \in \mathbb{R}^2$ for which the distance $\delta(x,S)$ between the point $x$ and the subset $S$ is smaller than $r \ge 0$. The characteristic shape of a tube is afforded by the measure of distance $\delta(x,S)$:  for its evaluation, all points $y \in S$ are considered, their Euclidean distance to $x$ evaluated, and the distance between $x$ and $S$ then corresponds to the smallest distance between any $y \in S$ and $x$. This can be made more concrete by considering the tube around the triangle in  \autoref{fig:rft_volumetrics}C. First, consider the three sides of the triangle. Here, the smallest distance between a point $x\in \mathbb{R}^2$ and a point $y \in S$ is given by moving away from the triangle in a perpendicular manner. However, at the three corners of the triangle, the smallest distance from the cornerpoint is given by the corresponding arc of the circle of radius $r$ centred on the triangle's corner. By inspection of \autoref{fig:rft_volumetrics}C it can be inferred that the Lebesgue measure of the tube $T_{S,r}$, which in this two-dimensional scenario corresponds to the area of the tube, comprises three principal contributions: the area of the original triangle, the areas of the three rectangles at the sides of the triangle, and the contributions of the disc sections at the three corners of the triangle. With a little geometric intuition, it can also be inferred that these three disc sections in fact together form a full circle of radius $r$. Steiner's formula  of expression \eqref{eq:steiners_formula} in the current context then states
\begin{align}\label{eq:lm_triangle_tube}
\begin{split}
\lambda(T_{S,r}) 
& = \pi r^2 \mu_0(S) + 2  r \mu_1(S) +   \mu_2(S). \\
\end{split} 
\end{align}
The first term in eq. \eqref{eq:lm_triangle_tube} corresponds to the area of a circle with radius $r$, if $\mu_0(S) = 1$. Thus, the $0$th-order intrinsic volume of a triangle appears to be 1. More generally, the $0$th-order intrinsic volume corresponds to the Euler characteristic of a subset $S \subset \mathbb{R}^D$, which will be discussed in \fullnameref{sec:probabilistic_properties}. The second term in eq. \eqref{eq:lm_triangle_tube} corresponds to the area of the three rectangles of the tube, if $\mu_1(S)$ measures the perimeter of the triangle (i.e., the sum of its side lengths) and is multiplied by $1/2$. Finally, it follows that $\mu_2(S)$ must measure the area of the triangle. Note that it is also possible to form a three- or higher dimensional tube around a triangle. More generally, it can be shown that the intrinsic volumes of two-dimensional subsets $S \subset \mathbb{R}^2$ are given by 
\begin{align}
\begin{split}
\mu_0(S)	& : \mbox{Euler characteristic of } S			\\
\mu_1(S)	& : 0.5 \cdot \mbox{Circumference of } S	\\
\mu_2(S)	& : \mbox{Area of } S,				
\end{split}
\end{align}
and for three-dimensional subsets $S \subset \mathbb{R}^3$ by
\begin{align}\label{eq:intrinsic_vol_3d}
\begin{split}
\mu_0(S)	& : \mbox{Euler characteristic of } S			\\
\mu_1(S)	& : 2 \cdot \mbox{Caliper diameter of } S		\\
\mu_2(S)	& : 0.5 \cdot \mbox{Surface area of } S		 	\\
\mu_3(S)	& : \mbox{Volume of } S.					
\end{split}
\end{align}
Here, the \textit{caliper diameter} of a convex object is evaluated  by placing the solid between two parallel planes (or calipers), measuring the distance between the planes, and averaging over all rotations of object. This kind of measurement can be formed using a caliper, a tool that is shown in \autoref{fig:rft_volumetrics}D. As mentioned above, the concept of the Euler characteristic will be elaborated on below. As for Lebesgue measure, it should be evident by now that evaluating the intrinsic volumes for a given geometric object that is modelled as a subset of $\mathbb{R}^D$ is not trivial. In  \autoref{fig:rft_volumetrics}E, we collect the intrinsic volumes of a number of elementary geometrical objects. 

Finally, we note that in the context of RFT, the intrinsic volumes of the search space of interest are evaluated numerically based on an algorithm proposed by \citep{Worsley1996}. This algorithm implies that the first and second order intrinsic volumes in the case $S \subset \mathbb{R}^3$ can be decomposed into contributions from one- and two-dimensional subspaces along the cardinal axes, i.e., that
\begin{alignat}{3}\label{eq:intrinsic_vol_3d_decomp}
\mu_1(S) &	= \mu_1^{x_1}(S) 		&& \, + \mu_1^{x_2}(S)  	&& \, + \mu_1^{x_3}(S)  \\
\mu_2(S) &	= \mu_2^{x_1 x_2}(S)  	&& \, + \mu_2^{x_1 x_3}(S) 	&& \, + \mu_2^{x_2 x_3}(S),	
\end{alignat}
where $\mu_1^{x_d}(S) \in \mathbb{R},\, d = 1,2,3$ denotes a contribution from the respective one-dimensional subspace of $\mathbb{R}^3$, and $\mu_2^{x_i x_j}(S),  1 \le i,j \le 3, i \neq j$ denotes a contribution from the respective two-dimensional subspace of $\mathbb{R}^3$. Inspection of the analytical intrinsic volumes in \autoref{fig:rft_volumetrics}E indicates that this decomposition clearly holds for box-shaped geometric objects.

\section{Theory}\label{sec:theory}
We develop the theory of RFT against the background of the continuous space, discrete data point model
\begin{equation}\label{eq:continuous_glm}
Y_i(x) = m_i \beta(x) + \sigma Z_i(x) \quad \mbox{  for  } i = 1,...,n \mbox{ and } x \in S \subseteq \mathbb{R}^3.
\end{equation}
Here, $Y_i(x)$ denotes a random variable that models the $i$th of $n$ data observations at spatial location $x \in \mathbb{R}^3$, $m_i \in \mathbb{R}^p$ is a known space-independent vector of $p$ model constraints (commonly the $i$th row of a design matrix), $\beta(x)$ is an unknown data point-independent value of a space-dependent effect size parameter function $\beta: \mathbb{R}^3 \to \mathbb{R}^p$, $\sigma > 0$ is an unknown standard deviation parameter, and $Z_i(x)$ is a $Z$-field modelling observation error. In line with  \defref{def:tzf_fields} we assume that the $Z_i(x), i = 1,...,n$ are independent and of identical gradient covariances. The model equation \eqref{eq:continuous_glm} should thus be read as the generalization of the familiar mass-univariate GLM-based fMRI data analysis equation \citep[e.g.,][eq. 8.6]{Kiebel2007} to continuous space, which entails that the data observations $Y_i(x)$ are (usually non-stationary) random fields. In the context of GLM-based fMRI data analysis, eq. \eqref{eq:continuous_glm} may represent a first-level model, in which case the $m_i$ are typically derived from the convolution of condition-specific trial onset functions with a haemodynamic response function and it is assumed that temporal error correlations have been accounted for by \textit{pre-whitening} \citep[e.g.,][]{Glaser2007}. Equivalently,  eq. \eqref{eq:continuous_glm} may represent a second-level model in the summary statistics approach \citep[e.g.][]{Mumford2006}, in which case the $m_i$ typically represent categorical statistical designs and comprise primarily ones and zeros, and the assumption of independent error contributions is natural.  

The fundamental aim of RFT is to evaluate the probabilities of topological features of statistical parametric maps under the null hypothesis of no activation. With regards to the model of eq. \eqref{eq:continuous_glm}, such a null hypothesis corresponds to  $\beta(x) = 0$ for all $x \in \mathbb{R}^3$. The null hypothesis thus entails that the standardized data observations $\sigma^{-1}Y_i(x), i = 1,...,n$ are $Z$-fields. Consequently, the evaluation of location-specific $T\text{-}$ or $F\text{-}$ratios (cf. the right-hand sides of eqs. \eqref{eq:t_field} and \eqref{eq:F_field}, respectively) results in $T$- or $F$-fields. In the current section, we are concerned with the probabilistic theory of topological features of these Gaussian-related random fields, which we will denote generically by $X(x), x \in S$. As outlined in the Introduction, we first consider the fundamental RFT concepts of excursion sets, smoothness, and resel volumes of Gaussian-related random fields. Building on these concepts, we then consider the theory of the probabilistic properties of excursion set features. It should be noted that throughout this section we assume a continuous model of space and the validity of the null hypothesis.

\subsection{Excursion sets, smoothness, and resel volumes}\label{sec:excursions_smoothness_resels}
\subsubsection*{Excursion sets}\label{sec:excursion_sets}

Excursion sets of Gaussian-related random fields above a level $u \in \mathbb{R}$ are an elementary building block in the theory of RFT, because the topological features of Gaussian-related random fields that RFT is concerned with are topological features of its excursion sets. We use the following definition \citep[][Section 2, p. 14]{Worsley1994}:

\begin{definition}[Excursion set]
Let $X(x), x \in \mathbb{R}^D$ denote a Gaussian-related random field, let $S\subset \mathbb{R}^D$ denote a subset of $\mathbb{R}^D$ referred to as the search space, and let $u \in \mathbb{R}$ denote a level. Then the excursion set of $X(x)$ inside the search space $S$ above the level $u$ is defined as
\begin{equation}
E_u := \{x \in S| X(x) \ge u\}.
\end{equation}
$\hfill \bullet$
\end{definition}

\begin{figure}[htbp]
\centerline{\includegraphics[width=15 cm]{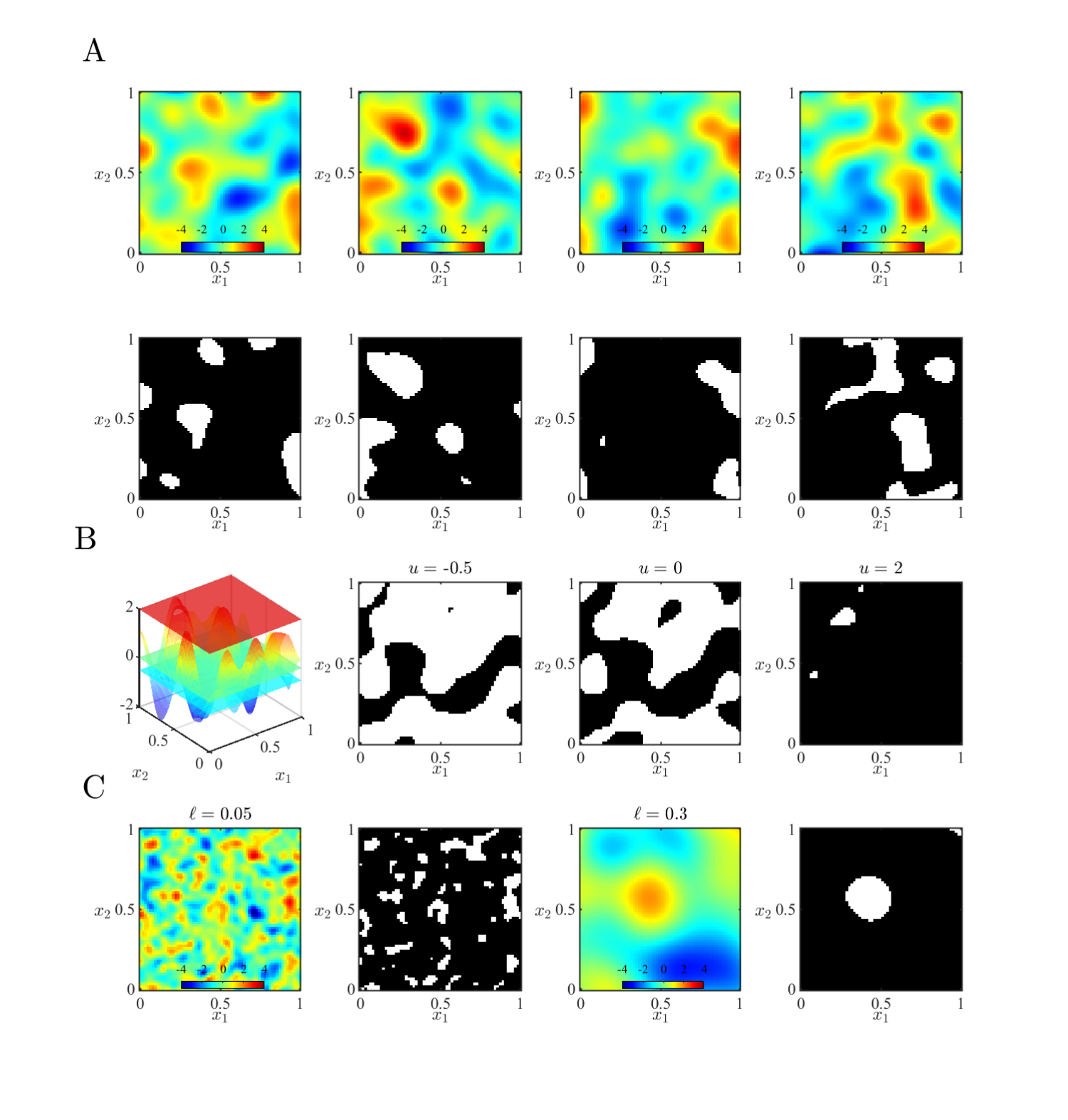}}
\caption{Excursion sets.  (\textbf{A}) This panel visualizes the random nature of excursion sets. The upper row of  four panels depicts four realizations of a $Z\text{-}$field with a Gaussian covariance function with parameters $v = 1, \ell = 0.15$. The lower four panels depict the realization-specific excursion set for a fixed level of $u=1$. Points included in the excursion sets are marked white, points not included in the excursion sets are black. (\textbf{B}) This panel visualizes the topological properties of the excursion set of a single $Z\text{-}$field realization (leftmost subpanel) as a function of the level value $u$. As $u$ increases from $-0.5$ to $2$, the topological constitution of the excursion set becomes less complex, and, in the limit of high levels $u$ comprises a single cluster comprising the global maximum of the Gaussian-related random field's realization or no points at all. (\textbf{C}) This panel visualizes the dependency of the topological properties of the excursion set on the smoothness of the Gaussian-related random field. The left two subpanels show a $Z\text{-}$field realization and its corresponding excursion set for $u=1$ for a rough (non-smooth) field. The excursion set comprises many small clusters. The right two subpanels show the same entities for a smoother field. Here, the excursion set comprises a smaller number of clusters, which are larger than in the case of the rough $Z\text{-}$ field. For implementational details of these simulations, please see \textit{rft\_4.m}.} 
\label{fig:rft_excursion_sets}
\end{figure}

\noindent In words, the excursion set $E_u$ comprises all points $x$ in the search space $S$ for which the Gaussian-related random field $X$ takes on values larger or equal to $u$.  In the neuroimaging literature, the level $u$ is referred to as the \textit{cluster-forming} (or \textit{cluster-defining}) \textit{threshold} \citep{Nichols2017, Flandin2017, Eklund2016}. Clearly, because $X$ is a random entity, the excursion set is also a random entity. From a probabilistic perspective, this entails that the characterization of excursion sets is concerned with expectations and distributions of excursion set features, because the precise properties of an excursion set vary from one realization of a Gaussian-related random field to another. \autoref{fig:rft_excursion_sets}A visualizes the excursion set above a level of $u := 1$ for four realizations of a $Z\text{-}$field with Gaussian covariance function. 

The probabilistic properties of topological features of excursion sets depend on many things. One observation, which is exploited repeatedly in the theory of RFT, is the following insight: as the level $u$ increases, the constitution of an excursion set follows a predictable path (\autoref{fig:rft_excursion_sets}B): at low values of $u$, excursion sets typically display complex topological shapes, comprising multiple \textit{clusters}, which in turn may have parts that are not part of the excursion set and appear as cluster holes. At higher values of $u$, holes tend to disappear and the number of clusters decreases. Finally, at very high values of $u$ around the level of the Gaussian-related random field realization's global maximum, the excursion set comprises only a single cluster (for $u$ slightly smaller than the realization's global maximum), a single point (for $u$ being equal to the global maximum), or no points at all (for all values of $u$ that are larger than the global maximum). As will be discussed below, this observation is used frequently in the approximation of probabilistic properties of excursion sets. A second observation is that, naturally, the topological properties of the excursion set at a fixed level $u$ depend on the characteristics of the underlying Gaussian-related random field. Clearly, for general nonstationary random fields, regions characterized by higher values of the expectation function than others have a higher probability of being contained in the excursion set. For the stationary Gaussian-related random fields that are of interest in RFT, the topological properties of excursion sets depend strongly on the field's smoothness: smoother random fields tend to have a more slowly and shallowly varying profile, while less smooth (rougher) random fields  typically display a more ``peaky'' profile. This results in the observation that excursion sets of smooth Gaussian-related random fields either contain few, but larger clusters, while excursion sets of rough Gaussian-related random fields contain more, but smaller clusters (\autoref{fig:rft_excursion_sets}C).

\subsubsection*{Smoothness}\label{sec:smoothness}
As visualized in \autoref{fig:rft_grfs} and \autoref{fig:rft_excursion_sets}, realizations of random fields can be \textit{smooth}, i.e., varying little over a given spatial distance, or \textit{rough}, i.e., varying strongly over a given spatial distance. This intuitive notion of smoothness is an important determinant of the probabilistic behaviour of topological features of random fields and of central importance in RFT \citep{Friston1991}. Thus, in order to establish quantitative relationships between a Gaussian-related random field's smoothness and the probabilistic behaviour of its topological features, a quantitative measure of smoothness is required. The principle measure used to quantify smoothness in RFT is the reciprocal unit-square \textit{Lipschitz-Killing curvature} \citep[][eq. 3]{Taylor2007}, defined as
\begin{equation}\label{eq:int_det_var_nabla_X}
\varsigma := \frac{1}{\int_{[0,1]^D} |\mathbb{V}\left(\nabla X(x) \right)|^{\frac{1}{2}} \,dx},
\end{equation}
where
\begin{small}
\begin{equation}\label{eq:var_nabla_X}
\mathbb{V}(\nabla X(x)) := 
\left(\mathbb{C} \left(\frac{\partial}{\partial x_i} X(x), \frac{\partial}{\partial x_j} X(x) \right) \right)_{1 \le i,j \le D} 
\end{equation}
\end{small}

\noindent denotes the $D \times D$ variance matrix of the gradient components (i.e., partial derivatives) of the random field  at $x \in [0,1]^D$. Note that we refer to $\mathbb{V}(\nabla X(x))$ as a \textit{variance matrix} (as opposed to a \textit{covariance matrix}) despite the fact that entries of $\mathbb{V}(\nabla X(x))$ are given by covariances. This is motivated by the fact that these covariances refer to the partial derivatives of the identical random variable $X(x)$ and not to partial derivatives of covariances of two different random variables \citep{Friston1991, Worsley1992, Taylor2007}. In the following, we first attempt to elucidate in which sense eqs. \eqref{eq:int_det_var_nabla_X} and \eqref{eq:var_nabla_X} capture the intuitive notion of a random field's smoothness. We then discuss, how eq. \eqref{eq:int_det_var_nabla_X} is reformulated in RFT to give rise to the ``full width at half maximum'' parameterization of smoothness. Throughout, we make the assumption that  the random field's gradient exists at all locations of the random field's domain, i.e., that the random field is differentiable with respect to space.

To obtain a first intuition in which sense eqs. \eqref{eq:int_det_var_nabla_X} and \eqref{eq:var_nabla_X} provide a measure of smoothness, we consider the case of a stationary GRF on a one-dimensional domain ($D = 1$) with a GCF $\gamma$ and parameters $v := 1$ and $\ell > 0$. In this case, the gradient of $X$ at the location $x\in [0,1]$ simplifies to the derivative of $X$ with respect to $x$, which, in line with the conventions of spatial statistics, we denote by $\dot{X}(x)$. Further, the variance matrix of the gradient components simplifies to the variance of this spatial derivative, and the determinant to the absolute value, which in turn is redundant for a non-negative variance. We are thus led to consider 
\begin{equation}\label{eq:int_v_dX}
\varsigma = \frac{1}{\int_{[0,1]} \mathbb{V}(\dot{X}(x))^{\frac{1}{2}}\,dx}.
\end{equation}  
We may first note that, regardless of the type of random field, the reciprocal of the square of the variance of the spatial derivative of the random field at a location $x$ is a sensible measure of the field's smoothness, if we assume that this value is constant over space: naturally, the spatial derivative quantifies the rate of change of the values of the random field, and if this is high, the random field changes a lot, if it is low, the random field does not change much. In addition, the variance describes how variable this spatial variability is over realizations of the random field, and, if it is small, most realizations of the random field with small spatial derivatives will appear smooth. Furthermore, for the current example of a one-dimensional GRF with a GCF, one may analytically evaluate eq. \eqref{eq:int_v_dX} and, as shown in \ref{sec:proof_gcf_smoothness}, obtains 
\begin{equation}\label{eq:smoothness_1d_grf_gauss_cov}
\varsigma =  \frac{\ell}{\sqrt{2}}.
\end{equation} 
Thus, for the current scenario, the smoothness measure defined by eqs. \eqref{eq:int_det_var_nabla_X} and \eqref{eq:var_nabla_X} is directly related to the spatial covariation parameter of the covariance function of the GRF. This is, of course, a special case. An important aspect of eq. \eqref{eq:int_det_var_nabla_X} is that it is defined as long as the partial derivatives of the random field of interest are defined, but for many random fields it may not directly map onto a single parameter of the field's covariance function. This is analogous to the difference between a variance of a random variable and the variance parameter of a univariate Gaussian distribution: while the former concept applies to any random variable, the case that it directly maps onto a parameter of the probability density function of a random variable, like the variance parameter of the univariate Gaussian, is rather an exception. 

More generally, the smoothness measure defined by eqs. \eqref{eq:int_det_var_nabla_X} and \eqref{eq:var_nabla_X} involves the determinants of the spatial gradients at locations $x \in [0,1]^D$. Assuming again that the gradients are constant over space, we next consider the two-dimensional scenario ($D = 2$). In this case, the determinant is given by

\begin{small}
\begin{equation}\label{eq:2d_smoothness}
|\mathbb{V}(\nabla X(x))| 
= \mathbb{V}\left(\frac{\partial}{\partial x_1}X(x) \right)
  \mathbb{V}\left(\frac{\partial}{\partial x_2}X(x) \right)
- \mathbb{C}\left(\frac{\partial}{\partial x_1}X(x),\frac{\partial}{\partial x_2}X(x)\right)^2.
\end{equation}
\end{small}

\noindent As is evident from eq. \eqref{eq:2d_smoothness}, the variances and covariances of the field's partial derivatives make opposing contributions to the field's smoothness: the variances of the partial derivatives contribute to roughness, while the covariances of the partial derivatives contribute to the field's smoothness. The former was already observed for the one-dimensional scenario. The latter can be interpreted as follows: a systematic simultaneous change of values in the direction of both the $x_1$ and $x_2$ ordinates indicates smoothness. Finally, an intuition about the three-dimensional case is afforded by the intuitive view of the determinant of a $3 \times 3$ matrix as the magnification factor of the volume of a unit cuboid under the linear transform furnished by the matrix \citep{Hannah1996}. In this case, large values for the off-diagonal elements with respect to the diagonal elements imply a smaller magnification and vice versa, thus implicating large variances of the partial derivatives in low smoothness and large correlations of the partial derivatives in high smoothness. In summary, if a random field is differentiable with respect to space, then eqs.  \eqref{eq:int_det_var_nabla_X} and \eqref{eq:var_nabla_X} provide a sensible scalar measure of the intuitive notion of a random field's smoothness.

\subsubsection*{Smoothness reparameterization}

While $\varsigma$ thus constitutes a perfectly fine scalar measure of a Gaussian-related random field's smoothness, \citet{Friston1991} and \citet{Worsley1992} proposed to reparameterize $\varsigma$ in the three-dimensional case ($D = 3$) as
\begin{equation}\label{eq:varsigma_fwhm}
\varsigma = (4 \ln 2)^{-\frac{3}{2}} f_{x_1}f_{x_2}f_{x_3} \mbox{ with }  f_{x_1},f_{x_2},f_{x_3} > 0.
\end{equation}
In the context of RFT, the values $f_{x_1},f_{x_2},f_{x_3}$ are referred to as the \textit{full widths at half maximum (FWHMs) in direction $x_1, x_2$ and $x_3$}, respectively. The reparameterization of smoothness in terms of FWHMs oriented along the cardinal axes of three-dimensional space was motivated by assuming that the realization of a Gaussian-related random field of interest had been created by convolving a GRF with a white-noise covariance function with a Gaussian convolution kernel (see \ref{sec:proof_smoothness_reparameterization} for formal definitions of these entities). In this case, the smoothness of the resulting Gaussian-related random field depends on the spatial width of the Gaussian convolution kernel. The spatial width of a Gaussian convolution kernel, in turn, can be described in terms of the covariance matrix of the kernel. If it is assumed that the covariance matrix of the Gaussian convolution kernel is diagonal, then the width of the convolution kernel is determined by the three diagonal parameters $\sigma_{x_1}^2, \sigma_{x_2}^2$ and $\sigma_{x_3}^2$, which govern the widths of the respective one-dimensional marginal Gaussian functions of the convolution kernel. An alternative representation of the width of a Gaussian function in turn is afforded by its FWHM: in general, the FWHM of a function $f$ is the value $x$ for which $f(-x/2) = f(x/2) =  f(0)/2$. As shown in \ref{sec:proof_smoothness_reparameterization}, for a univariate Gaussian function extending over the $x_1$ domain with variance parameter $\sigma^2_x$, this value is given by 
\begin{equation}\label{eq:gauss_fwhm}
f_x = \sqrt{8 \ln 2}\sigma_x.
\end{equation}
In other words, the FWHM of a univariate Gaussian function is a scalar multiple of the square root of its variance parameter. Eq. \eqref{eq:varsigma_fwhm} was then motivated by approximating the variance matrix of partial derivatives $\mathbb{V}(\nabla X(x))$ of a given Gaussian-related random field $X(x), x \in S$ by the variance matrix of partial derivatives of a hypothetical random field $Y(x), x \in S$ that was imagined to have been created by the convolution of a white-noise GRF with an isotropic Gaussian convolution kernel parameterized in terms of its FWHMs $f_{x_1}, f_{x_2}$, and $f_{x_3}$. For such a field, the variance matrix of partial derivatives was proposed to be  independent of space and of the form \citep[][eq. 2]{Worsley1992}
\begin{equation}\label{eq:Lambda}
\mathbb{V}(\nabla Y(x)) =
4 \ln 2 \left(\begin{matrix}
f_{x_1}^{-2}				& 	0 						& 	0						\\
0							&	f_{x_2}^{-2} 				& 	0						\\
0							&	0							& 	f_{x_3}^{-2}
\end{matrix}\right) =: \Lambda.
\end{equation}
For this hypothetical field, it then follows directly that 
\begin{equation}\label{eq:smoothness_fwhm}
\varsigma 
= \left(\int_{[0,1]^D} |\Lambda|^{\frac{1}{2}}\,dx \right)^{-1}
= \left((4 \ln 2)^3 (f_{x_1} f_{x_2} f_{x_3})^{-2} \right)^{-\frac{1}{2}}
= (4 \ln 2)^{-\frac{3}{2}} f_{x_1}f_{x_2}f_{x_3}.
\end{equation}
In \ref{sec:proof_smoothness_reparameterization} we show that the diagonal elements of the variance matrix of gradient components of a three-dimensional white-noise GRF convolved with an isotropic Gaussian convolution kernel with FWHMs $f_{x_1}, f_{x_2}$,  and $f_{x_3}$ are indeed of the form implied by the right-hand side of eq. \eqref{eq:Lambda} (see \citet{Holmes1994} and \citet{Jenkinson2000} for similar work). Note, however, that this only (at least partially) validates the construction of the approximation, but not the approximation 
\begin{equation}
\mathbb{V}(\nabla X(x)) \approx \mathbb{V}(\nabla Y(x)),
\end{equation}
which is implicit in the reparameterization of the smoothness parameter of a given Gaussian-related random field $X(x),x\in S$ in terms of FWHMs, \textit{perse}. 

\subsubsection*{Resel volumes}\label{sec:resel_volumes}

Building on the parameterization of smoothness in terms of FWHMs, the concept of \textit{resel volumes} was introduced by \citet{Worsley1992}. Intuitively, resel volumes are the smoothness-adjusted intrinsic volumes of subsets of domains of random fields. Stated differently, the smoother a random field, the larger the resel volumes of subsets of its domain at constant intrinsic volumes. Based on the general definition of intrinsic volumes for the three-dimensional case \eqref{eq:intrinsic_vol_3d}, the decomposition of the first- and second-order intrinsic volumes \eqref{eq:intrinsic_vol_3d_decomp}, and the FWHM parameterization of a random field's smoothness \eqref{eq:smoothness_fwhm}, the resel volumes of a set $S \subset \mathbb{R}^D$ in the domain of a random field are then defined as follows \citep[][pp. 63, eq. (3.2)]{Worsley1996}:

\begin{definition}[Resel volumes]\label{def:resel_volumes}
Let $X(x),x \in \mathbb{R}^3$ denote a Gaussian-related random field with 
smoothness 
\begin{equation}
\varsigma = (4 \ln 2)^{-\frac{3}{2}} f_{x_1}f_{x_2}f_{x_3}
\end{equation}
and let $S\subset \mathbb{R}^3$ denote a subset of the domain of the Gaussian-related random field. Then the $0$th- to $3$rd-order resel volumes of $S$ are defined in terms of the intrinsic volumes of $S$ and the smoothness parameters $f_{x_1},f_{x_2}$ and $f_{x_3}$ by
\begin{alignat}{3}
R_0(S) 	& := 	\mu_0(S) 																															\\
R_1(S) 	& := 	
\frac{1}{f_{x_1}} \mu_1^{x_1}(S) 				&& \, + 	
\frac{1}{f_{x_2}} \mu_1^{x_2}(S) 			 	&& \, + 	
\frac{1}{f_{x_3}} \mu_1^{x_3}(S) 				\\
R_2(S) 	& := 	
\frac{1}{f_{x_1}f_{x_2}} \mu_2^{x_1x_2}(S) 	&& \, +  	
\frac{1}{f_{x_1}f_{x_3}} \mu_2^{x_1x_3}(S) 	&& \, +  	
\frac{1}{f_{x_2}f_{x_3}} \mu_2^{x_2 x_3}(S) 	\\
R_3(S) 	& := 	\frac{1}{f_{x_1} f_{x_2} f_{x_3}} \mu_3(S) 																
\end{alignat}
$\hfill \bullet$
\end{definition}

\noindent The resel volumes of a given set $S \subset \mathbb{R}^3$ can thus be readily evaluated if the intrinsic volumes of $S$ and their subspace contributions, as well as the FWHM smoothness parameters $f_{x_1}$,$f_{x_2}$, $f_{x_3}$ of the Gaussian-related random field extending over $S$, are known. 

\subsection{Probabilistic properties of excursion set features}\label{sec:probabilistic_properties}

The probabilistic properties of the following five topological features of excursion sets of Gaussian-related random fields form the core of RFT theory: (P1) the global maximum of the Gaussian-related random field, (P2) the number of clusters within an excursion set, (P3) the volume of clusters within an excursion set, (P4) the number of clusters of a given volume within an excursion set, and (P5) the maximal volume of a cluster within an excursion set. In the theory of RFT, these topological features are described by random variables, the probability distributions of which are governed by the stochastic and geometric characteristics of the underlying Gaussian-related random field and the level $u$ of the respective excursion sets. The dependencies of the respective distributions on the properties of the underlying Gaussian-related random field and the excursion set level are mitigated by four expected values: (E1) the expected volume of an excursion set, (E2) the expected number of local maxima within an excursion set, (E3) the expected number of clusters within an excursion set, and (E4) the expected volume of a cluster within an excursion set. In the following, we first discuss the parametric dependencies of these expectations on the type of Gaussian-related random field and its resel volumes, i.e., its smoothness and intrinsic volumes. We then discuss the probability distributions of the five topological features that form the core of RFT theory. As will become evident throughout this section, a defining characteristic of RFT theory is that the parametric distributional forms of virtually all topological features of interest derive from approximations, rather than analytical transforms of the probability density functions describing the Gaussian-related random fields under  functions that instantiate the topological features. For the presentation of RFT theory herein, this entails that we will most often only report the parametric distributional form of a given topological feature under RFT and discuss some background about its motivation, rather than analytically deriving the respective parametric distributional form from first principles. The approximative-parametric nature is inherent to RFT and is rooted in the seminal works by \citet{Adler1976a} and \citet{Adler1976b} and has more recently seen major advances by \citet{Taylor2005, Taylor2006} and \citet{Adler2007} (see also \citet{Adler2000} for a historical perspective).

\subsubsection*{Expected values}

\noindent \textbf{(E1) The expected volume of an excursion set}
\vspace{2mm}

\begin{figure}[htp!]
\centerline{\includegraphics[width= 14.5 cm]{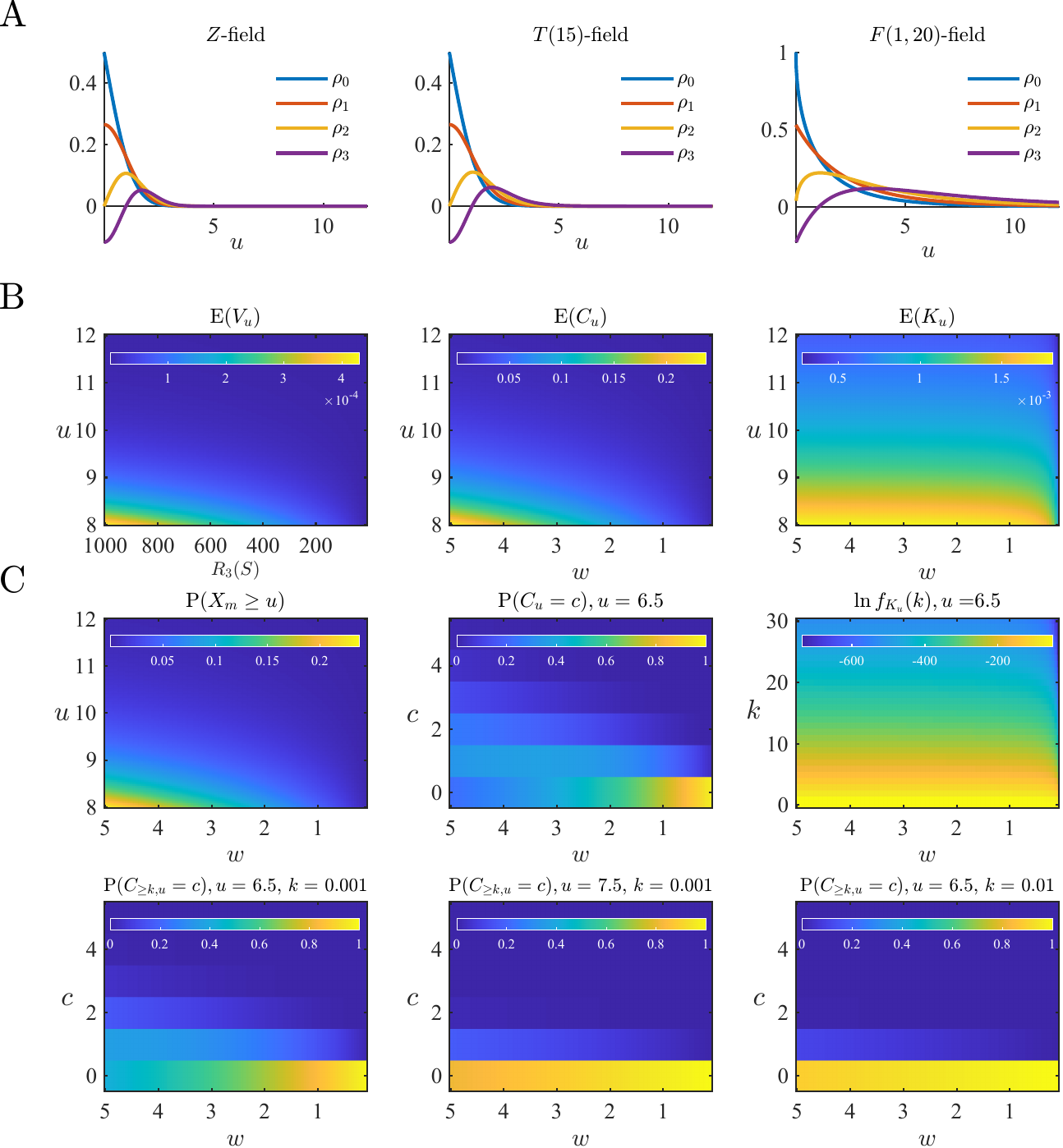}}
\caption{Probabilistic properties of excursion set features. (\textbf{A}) Euler characteristic densities. The panels depict, from left to right, the $Z$-field Euler characteristic (EC) densities, the $T$-field EC densities for a $T$-field with 15 degrees of freedom and the $F$-field EC densities for an $F$-field with (1,20) degrees of freedom. (\textbf{B}) Excursion set feature expectations for a $T$-field with $15$ degrees of freedom. From left to right, the panels show  the expected excursion set volume $\mathbb{E}(V_u)$ in units of resels, the expected number of clusters $\mathbb{E}(C_u)$, and the expected cluster volume $\mathbb{E}(K_u)$ in resels as a function of the excursion set level $u$ and a parameterization of smoothness. For the expected excursion set volume, this smoothness parameter is given by the third-order resel volume. For the remaining panels, including the panels in (C), we selected the resel volumes of the exemplary data set (cf. \autoref{fig:rft_spm_table}) given by $R_0(S) = 6.0, R_1(S) = 32.8, R_2(S) = 353.6,$ and $R_3(S) = 704.6$ and evaluated \eqref{eq:gkf} over a range of scalar multiples $wR_d, d = 0,1,2,3$. The value of $w \in [0,5]$ is displayed on the $x$-axis. Note that smoothness increases with decreasing resel volumes. (\textbf{C}) Excursion set feature probability distributions for a $T$-field with $15$ degrees of freedom as functions of smoothness. From the left to right, the panels show the probability for the maximum of the Gaussian-related random field to exceed a value $u$, $\mathbb{P}(X_m \ge u)$, the probability for the number of clusters within an excursion set to assume a value $c$, $\mathbb{P}(C_u = c)$, the log probability density for a cluster to assume a volume $k$, $f_{K_u}(k)$. (\textbf{D}) The distribution of the number of clusters of a given volume within an excursion set. The panels depict $\mathbb{P}(C_{\ge k,u} = c)$ as a function of $u$ and $k$.}
\label{fig:rft_excursion_set_properties}
\end{figure}

Recall that an excursion set $E_u$ is a subset of the search space $S$ for which a Gaussian-related random field takes on values larger or equal to $u \in \mathbb{R}$. As discussed above, the standard analytical measure that assigns a notion of volume to subsets of $\mathbb{R}^D$ is the Lebesgue measure $\lambda$, which returns the \textit{Lebesgue volume} $\lambda(E_u)$ of the excursion set. As discussed by \citet[][Section 1.7, pp. 18 - 19]{Adler1981}, the expected value of the Lebesgue volume of an excursion set is given by
\begin{equation}\label{eq:exp_volume}
\mathbb{E}\left(\lambda(E_u) \right) = \lambda(S)(1 - F_X(u)),
\end{equation}
where $\lambda(S)$ denotes the Lebesgue volume of the search space and $F_X$ denotes the distribution function of the Gaussian-related random field at the origin (see also \citet[][eq. 4]{Hayasaka2003} and, for $Z\text{-}$fields, \citet[][eq. 3]{Friston1994}). Because for stationary random fields, this distribution function equals the distribution functions at all other locations of the random field, we denote it by $F_X$ without reference to its spatial position.
A brief derivation of eq. \eqref{eq:exp_volume} is provided in  \ref{sec:miscellaneous_proofs}. As it stands, eq. \eqref{eq:exp_volume} has strong intuitive appeal: the expected value of the excursion set $E_u$ is a proportion of the search space $S$, and the proportion factor is given by the probability of the Gaussian-related random field to take on values larger than $u$ at each point in space. For example, for small values of $u$, the probability that the Gaussian-related random field takes on values larger than $u$ is always higher than for larger values of $u$. Accordingly, the expected value of the volume of the excursion set is larger in the former than in the latter case. From eq. \eqref{eq:exp_volume}, it follows directly that if the three-dimensional volume of the search space is expressed in terms of the third-order resel volume $R_3(S)$, then the expected third-order resel volume of the excursion set is given by
\begin{equation}\label{eq:exp_volume_res}
\mathbb{E}(R_3(E_u)) = R_3(S) (1 - F_X(u)).
\end{equation}
In the following, we will denote the random variable modelling the expected volume of an excursion set independent of its unit of measurement by $V_u$, where $V_u := \lambda(E_u)$ or $V_u := R_3(E_u)$ depending on the specified context. We visualize the expected volume of the excursion set as a function of $u$ and $R_3(S)$ in \autoref{fig:rft_excursion_set_properties}B. As is evident from this visualization, the expected volume of an excursion set $\mathbb{E}(V_u)$ decreases with higher cluster-forming thresholds $u$ and higher Gaussian-related random field smoothness, i.e., decreasing third-order resel volume $R_3(S)$.

\vspace{2mm}
\noindent \textbf{(E2) The expected number of local maxima within an excursion set}
\vspace{2mm}

Let $M_u$ denote the number of local maxima of a Gaussian-related random field $X(x), x \in \mathbb{R}^D$ above $u$ inside a set $S \subset \mathbb{R}^D$. Then $M_u$ is also the number of local maxima of the excursion set $E_u$. In RFT, the expected number of local maxima of a Gaussian-related random field is approximated by the expected \textit{Euler characteristic} $\chi(\cdot)$ of the excursion set \citep[][Section 2, p. 15]{Worsley1994}, i.e.
\begin{equation}\label{eq:exp_m_u_approx_chi}
\mathbb{E}(M_u) \approx \mathbb{E}(\chi(E_u)). 
\end{equation}
To unpack expression \eqref{eq:exp_m_u_approx_chi}, we first discuss the notion of the Euler characteristic as a topological measure. We then consider the motivation for the approximation of expression \eqref{eq:exp_m_u_approx_chi}, and finally discuss the parametric closed form for the expected Euler characteristic used in RFT.
 
The Euler characteristic is a function and is referred to as a \textit{topological invariant}. Topological invariants measure the geometrical characteristics of geometrical objects irrespective of the way they are bent. A formal description of the Euler characteristic is beyond our current scope, because, as \citet[][p.16]{Adler2000} puts it, the Euler characteristic ``(...) is one of those unfortunate cases in which what is easy for the human
visual system to do quickly and effectively requires a lot more care when
mathematicised''. This care is not warranted here, because, as will be seen below, the expected Euler characteristic of the excursion set $E_u$ will itself be approximated by a function of $u$ without recourse to an actual evaluation of 
$\chi(E_u)$. We thus only provide a brief intuition for the Euler characteristic itself: for a three-dimensional volume of interest, the Euler characteristic counts, in an alternating sum, the number of three types of topological features: (1) connected components, (2) visible open holes, often referred to as handles, and (3) invisible voids. Semi-formally, the Euler characteristic of an excursion set $E_u$ is hence given by
\begin{align}\label{eq:euler_characteristic}
\begin{split}
\chi(E_u)	=  \quad	& \, \mbox{Number of connected components of } E_u \\
 				-  		& \, \mbox{Number of handles of } E_u \\
 				+ 		& \, \mbox{Number of voids in } E_u.
\end{split}
\end{align}  
The following examples for familiar geometric objects provided by \citet{Worsley1996b} are instructive:

\vspace{2mm}
\begin{footnotesize}
\begin{itemize}[leftmargin = *]
\item[] $\chi(E_u) = 1$ for a single solid excursion set 
\tab (1 connected component - 0 handles + 0 voids),
\item[] $\chi(E_u) = 0$ for a doughnut-shaped excursion set 
\tab (1 connected component - 1 handles + 0 voids),
\item[] $\chi(E_u) = -2$ for a pretzel-shaped excursion set 
\tab (1 connected component - 3 handles + 0 voids), 
\item[] $\chi(E_u) = 2 $ for a tennisball-shaped excursion set 
\tab (1 connected component - 0 handles + 1 voids). 
\end{itemize}
\end{footnotesize}
\vspace{2mm}

\noindent More qualitatively, if an excursion set comprises many disconnected components, each with very few holes (in astrophysics referred to as a ``meatball'' topology), the Euler characteristic is positive; if the clusters of an excursion set are connected by many bridges,  thus creating many holes (in astrophysics referred to as a ``sponge'' topology), the Euler characteristic is negative, and if the excursion set comprises many surfaces that enclose hollows (in astrophysics referred to as a ``bubble'' topology), the Euler characteristic is again positive.

In the theory of RFT, it is, however, not the value of the Euler characteristic for a specific realization of an excursion set that is of primary interest, but rather its expected value - as evident from the right-hand side of expression \eqref{eq:exp_m_u_approx_chi}. This approximation (or ``heuristic'') is based on the following intuition: as the threshold value $u$ that defines the excursion set $E_u$ increases, voids and handles of the excursion set tend to disappear, and what remains are isolated connected components of the excursion set (in the neuroimaging literature referred to as clusters). In this scenario, $\chi(E_u)$ thus counts the number of connected components (cf. eq. \eqref{eq:euler_characteristic}). Each of the clusters in the excursion set naturally contains a local maximum of the excursion set, which thus motivates to approximate the expected number of local maxima by the expected Euler characteristic over realizations of the Gaussian-related random field $X(x)$. As for the accuracy of this approximation, it has been demonstrated that it becomes exact in the limit of very high values of $u$, in which $\chi(E_u)$ evaluates to either 1 (one remaining cluster) or 0 (no remaining clusters) \citep{Taylor2005}. For lower values of $u$, an analytical validation does not seem to exist so far. Crucially though, the approximation in expression \eqref{eq:exp_m_u_approx_chi} is helpful from a mathematical perspective, because there exists a parametric closed form for the expected Euler characteristic. This parametric closed form was introduced by \citep[][eq. 3.1]{Worsley1996} and is given by
\begin{equation}\label{eq:gkf}
\mathbb{E}(\chi(E_u)) = \sum_{d = 0}^D R_d(S)\rho_d(u),
\end{equation}
where the $R_d(S), d = 0,...,D$ are the $d$-dimensional resel volumes of the search space and the $\rho_d(u)$ are referred to as \textit{Euler characteristic (EC) densities}. The formula on the right-hand side of eq. \eqref{eq:gkf} is powerful because it decomposes the expected value of the Euler characteristic into (1) smoothness-adjusted volumetric contributions from each of the $d = 0$ to $d = D$ dimensions of the search space, and (2) unit volume-smoothness densities $\rho_d$ that only depend on the value of $u$ and the type of Gaussian-related random field on the other hand. We have already discussed the meaning of the $R_d$ in \fullref{sec:excursions_smoothness_resels} and hence focus on the EC densities in the following. Intuitively, the functions 
\begin{equation}
\rho_d : \mathbb{R} \to \mathbb{R}, u \mapsto \rho_d(u) \mbox{ for } d = 0,...,D
\end{equation}
describe the contribution of unit smoothness resel volumes to the expected value of the Euler characteristic. The values taken by these functions depend on $u$, and in general, as the value of $u$ increases (i.e., the excursion set is formed based on a higher threshold), these values decrease. For constant resel volumes, this implies that the value of the expected Euler characteristic, and thus the approximated expected number of local maxima, decreases. The exact parametric forms of the function $\rho_d$ depend on the type of Gaussian-related random field and have been analytically evaluated by \citet{Worsley1994}. We list their functional forms in  \autoref{tab:ecd} and visualize their form based on their SPM implementation in \autoref{fig:rft_excursion_set_properties}A. Note that the $0$th order EC density of each field corresponds to $1 - F_X(u)$, where $F_X$ denotes the distribution function of the respective field. This allows for evaluating the expected volume of the excursion set as expressed in eq. \eqref{eq:exp_volume_res} in terms of the third order resel volume and the $0$th order EC density. In summary, the expected number of local maxima of an excursion set can thus be approximated for a given Gaussian-related random field in a straight-forward manner, if the field's resel volumes are known (or have been estimated). 

\begin{table}[h!]
\renewcommand{\arraystretch}{2.5}
\begin{footnotesize}
\begin{tabularx}{\textwidth}{l}
$Z$-field Euler characteristic densities			 												\\
\hline  
$\rho_0(u) = 
\int_u^{\infty}\frac{1}{(2\pi)^{1/2}}
\exp \left(-t^2/2 \right) dt$		
\\
$\rho_1(u) = 
\frac{(4 \ln 2)^{1/2}}{2\pi} 
\exp \left(-u^2/2 \right)$ 				
\\
$\rho_2(u) = 
\frac{4 \ln 2}{(2\pi)^{3/2}} 
\exp \left(-u^2/2 \right)u$ 	 			
\\				
$\rho_3(u) = \frac{(4 \ln 2)^{3/2}}{(2\pi)^2} 
\exp \left(-u^2/2 \right)(u^2 - 1)$ 		
\\
$T$-field Euler characteristic densities			 												\\
\hline  
$\rho_0(u;\nu) = 
\int_u^{\infty}
\frac{\Gamma\left(\frac{\nu+1}{2}\right)}{\sqrt{\nu \pi}\Gamma\left(\frac{\nu}{2}\right)
								 \left(1+\frac{t^2}{\nu}\right)^{-1/2(\nu+1)}}dt$ 					\\
$\rho_1(u;\nu) = 
\frac{(4 \ln 2)^{1/2}}{2\pi} 
\left(1 + \frac{u^2}{\nu}\right)^{-1/2(\nu-1)}$ 		
\\
$\rho_2(u;\nu) = 
\frac{(4 \ln 2)\Gamma\left(\frac{\nu+1}{2}\right)}
{(2\pi)^{3/2}\left(\frac{\nu}{2}\right)^{1/2}\Gamma\left(\frac{\nu}{2}\right)}
 \left(1+\frac{u^2}{\nu}\right)^{-1/2(\nu-1)}u$									
\\
$\rho_3(u;\nu) =
\frac{(4 \ln 2)^{3/2}}{(2\pi)^{2}}
\left(1+\frac{u^2}{\nu}\right)^{-1/2(\nu-1)} 
\left(\frac{\nu-1}{\nu}u^2 -1 \right)$	 											
\\

$F$-field  Euler characteristic densities															
\\
\hline  	
														
$\rho_0(u;\nu_1, \nu_2) = 	
\int_u^{\infty} 
\frac{\Gamma((\nu_1 + \nu_2)/2)}{\Gamma(\nu_1/2)\Gamma(\nu_2/2)}
\frac{\nu_1}{\nu_2}
\left(\frac{\nu_1}{\nu_2} t\right)^{1/2(\nu_1-2)} 
\left(1 + \frac{\nu_1t}{\nu_2}\right)^{-1/2(\nu_1 + \nu_2)}
\,dt$ 
\\
$\rho_1(u;\nu_1, \nu_2) = 
\frac{(4 \ln 2)^{1/2}}{(2\pi)^{1/2}}
\frac{\Gamma((\nu_1 + \nu_2 - 1)/2)2^{1/2}}{\Gamma(\nu_1/2)\Gamma(\nu_2/2)}
\left(\frac{\nu_1 u}{\nu_2} \right)^{1/2(\nu_1 -1)}
\left(1 + \frac{\nu_1u}{\nu_2} \right)^{-1/2(\nu_1 + \nu_2 - 2)}$ 
\\
$\rho_2(u;\nu_1, \nu_2) = 
\frac{4 \ln 2}{2\pi}
\frac{\Gamma((\nu_1 + \nu_2 - 2)/2)}{\Gamma(\nu_1/2)\Gamma(\nu_2/2)}
\left(\frac{\nu_1 u}{\nu_2} \right)^{1/2(\nu_1 - 2)}
\left(1 + \frac{\nu_1u}{\nu_2} \right)^{-1/2(\nu_1 + \nu_2 - 2)} 
\left( (\nu_2 -1)\frac{\nu_1 u}{\nu_2} - (\nu_1 - 1)\right)$
\\
$\rho_3(u;\nu_1, \nu_2) = 
\frac{(4 \ln 2)^{3/2}}{(2\pi)^{3/2}}
\frac{\Gamma((\nu_1 + \nu_2 - 3)/2)2^{-1/2}}{\Gamma(\nu_1/2)\Gamma(\nu_2/2)}
\left(\frac{\nu_1 u}{\nu_2} \right)^{1/2(\nu_1 - 3)}
\left(1 + \frac{\nu_1u}{\nu_2} \right)^{-1/2(\nu_1 + \nu_2 - 2)} 
$
\\
$\quad \quad  \quad \quad  \quad \quad  \quad \quad \quad \times 
\left((\nu_2 - 1)(\nu_2 - 2)
\left(\frac{\nu_1 u}{\nu_2}\right)^2 
- (2\nu_1\nu_2 - \nu_1 - \nu_2 - 1)
\frac{\nu_1 u}{\nu_2} 
+ (\nu_1 - 1)(\nu_1 - 2)\right)
$
\\

\end{tabularx}
\end{footnotesize}
\caption{Euler characteristic densities for $Z\text{-}$fields and for $T\text{-}$ and $F\text{-}$fields with $\nu$ and $\nu_1,\nu_2$ degrees of freedom, respectively. Note that the $0$th-order Euler characteristic densities correspond to complementary distribution functions of $Z,T$ and $F$ random variables. Reproduced from Table II in \citet{Worsley1996}.}\label{tab:ecd}
\end{table}

\vspace{2mm}
\noindent \textbf{(E3) The expected number of clusters within an excursion set}
\vspace{2mm}

In the SPM implementation of RFT, the expected number of clusters is approximated by the expected Euler characteristic \citep[][Chpt. 18, pp. 233-234, Chpt. 19, pp. 240-241]{Friston2007}. This means that the approximations of the expected number of local maxima and the approximation of the expected number of clusters are identical. This identity entails the assumption that each cluster contains only a single local maximum of the excursion set. Cast more formally,  let $M_u$ denote the number of local maxima of an excursion set $E_u$, and let 
\begin{equation}
\mathcal{M} := \{ m_1, ..., m_{M_u}\}  \subset E_u
\end{equation}
denote the set of local maxima of $E_u$. Clusters are unconnected components of the excursion set and thus form a partition $\mathcal{P}_u$ of $E_u$, i.e.,
\begin{equation}
\mathcal{P}_u = \{\mathcal{C}_1, ..., \mathcal{C}_{C_u} \} \mbox{ with }
\mathcal{C}_i \cup \mathcal{C}_j = \emptyset \mbox{ for } i \neq j \mbox{ and }
\cup_{i=1}^{C_u} \mathcal{C}_i = E_u,
\end{equation}
where 
\begin{equation}
C_u := |\mathcal{P}_u|
\end{equation}
denotes the number of clusters. While each cluster contains at least one local maximum, in general it is very well possible that a given cluster comprises multiple local maxima. This is also a common finding in the actual analysis of fMRI data using SPM, in which case the local maxima for large clusters are indeed listed in the SPM results table based on a distance selection criterion of typically larger than 8 mm apart. 
The validity of the approximation chain
\begin{equation}\label{eq:exp_c_u_approx}
\mathbb{E}(C_u) 
\approx \mathbb{E}(M_u) 
\approx \mathbb{E}(\chi(E_u))
=  \sum_{d = 0}^D R_d(S)\rho_d(u),
\end{equation}
thus strongly rests on the assumption of high values for the excursion set-defining threshold $u$: in the case that only the global maximum of the random field is contained in the single connected component of the field's excursion set, the number of local maxima of the excursion set and the number of clusters are identical with probability one. In Figure \ref{fig:rft_excursion_set_properties}B, we visualize the expected number of clusters $\mathbb{E}(C_u)$ within an excursion set over a range of values for $u$ and interpolations of the resel volumes $R_0(S), ...,R_3(S)$ of the exemplary data set underlying \autoref{fig:rft_spm_table}. As is evident from this visualization, the expected number of clusters within an excursion set decreases with increasing values for the cluster-forming threshold and increasing smoothness of the Gaussian-related random field, corresponding to decreasing resel volumes.

\vspace{2mm}
\noindent \textbf{(E4) The expected volume of clusters within an excursion set}
\vspace{2mm}

The expected volume of a single cluster is evaluated in RFT based on the expressions for (1) the expected volume of the excursion set, and (2) the expected number of clusters in such an excursion set. For a value $u\in \mathbb{R}$, let the \textit{cluster volume} be defined by the random variable
\begin{equation}\label{eq:k_u_definition}
K_u := \frac{V_u}{C_u}.
\end{equation}
Then, if $V_u$ and $C_u$ are independent random variables with finite expected values, the expected cluster volume is given by
\begin{equation}\label{eq:k_u_expectation}
\mathbb{E}(K_u) = \frac{\mathbb{E}(V_u)}{\mathbb{E}(C_u)}
\end{equation}
(e.g., \citet[][eq. 2]{Friston1996}, \citet[][eq. 3]{Hayasaka2003}). Naturally, if the expected volume of the excursion set is expressed in terms of Lebesgue measure, i.e., $V_u = \lambda(E_u)$, eq. \eqref{eq:k_u_expectation} denotes the expected cluster Lebesgue volume, and if the expected volume of the excursion set is expressed in terms of third-order resel volume, i.e., $V_u = R_3(E_u)$, eq. \eqref{eq:k_u_expectation} denotes the expected cluster resel volume. We visualize the expected cluster resel volume as a function of the cluster-forming threshold and the smoothness of the Gaussian-related random field in 
\autoref{fig:rft_excursion_set_properties}B. As is evident from this visualization, the expected cluster resel volume decreases with increasing values of the cluster-forming threshold and increasing Gaussian-related random field smoothness.

\subsubsection*{Probability distribution of excursion set features}
\vspace{2mm}
\noindent \textbf{(P1) The distribution of the global maximum of a Gaussian-related random field}
\vspace{2mm}

Historically, the distribution of the global maximum of a Gaussian-related random field has been one of the seeds of the RFT framework \citep{Worsley1992}. This is because maximum statistics afford FWER control in multiple testing scenarios, as we show in \ref{sec:hypothesis_testing_fwer_control}. To introduce the distribution of the global maximum used in RFT, let 
\begin{equation}
X_{m} := \max_{x \in S} {X(x)}
\end{equation}
denote the global maximum of the Gaussian-related random field $X(x)$ on a set $S$. In RFT, the probability distribution of the global maximum is specified in terms of its distribution function, which is assumed to be of the following form (e.g., \citet[][eqs. 1,6]{Worsley1992}, \citet[][eq. 3.1]{Worsley1996})
\begin{equation}\label{eq:maximum_distribution}
F_{X_{m}} : \mathbb{R} \to [0,1], u \mapsto F_{X_{m}}(u) := 1 - \mathbb{E}(\chi(E_u)).
\end{equation}
The reasoning behind this form is similar to that justifying the approximation of the expected number of local maxima by the expected Euler characteristic: first, if the excursion set defining value $u$ is high enough, $E_u$ will not comprise any handles and voids any more. Second, in the limit of even higher values $u$, there will remain only one connected component comprising the global maximum or no connected component at all. In this case, the Euler characteristic of the excursion set can assume only the values $1$ or $0$: if the global maximum is equal to or larger than the excursion set defining value, i.e., if $X_m \ge u$, the Euler characteristic will assume the value $1$, i.e., $\chi(E_u) = 1$. If the global maximum is smaller than the excursion set defining value, i.e., if $X_m < u$, the Euler characteristic will assume the value $0$, i.e., $\chi(E_u) = 0$. That is, in the limit of high values for $u$, the Euler characteristic $\chi(E_u)$ of the excursion set can be conceived as a random variable taking on values in the set $\{0,1\}$ with a probability mass function defined in terms of the mutually exclusive event probabilities 
\begin{equation}
\mathbb{P}(\chi(E_u) = 0)  := \mathbb{P}(X_m < u)		\mbox{ and }
\mathbb{P}(\chi(E_u) = 1)  := \mathbb{P}(X_m \ge u).
\end{equation}
Evaluating the expected value $\mathbb{E}(\chi(E_u))$ in this scenario then yields
\begin{equation}
\mathbb{E}(\chi(E_u)) = \mathbb{P}(\chi(E_u) = 0) \cdot 0 +  \mathbb{P}(\chi(E_u) = 1) \cdot 1 = \mathbb{P}(X_m \ge u).
\end{equation}
With the definition of the cumulative distribution function of a continuous random variable, expression \eqref{eq:maximum_distribution} then follows directly. 

It is instructive to consider the dependency of the probability for the global maximum to exceed a value $u$ as a function of the Gaussian-related random field's smoothness: an increase in the degree of smoothness at constant intrinsic volumes results in a decrease of the Gaussian-related random field's resel volumes (cf. \defref{def:resel_volumes}). A decrease in resel volumes at constant value $u$ in turn leads to a decrease in the value of the expected Euler characteristic (cf. eq. \eqref{eq:gkf}). A decrease in the expected Euler characteristic in turn results in an increase of the probability $\mathbb{P}(X_m < u)$ (cf. eq. \eqref{eq:maximum_distribution}), which in turn is equivalent to a decrease of the probability $\mathbb{P}(X_m \ge u)$. In summary, an increase in a Gaussian-related random field's smoothness thus yields a lower probability for the field's global maximum to exceed a given value $u$. We visualize this dependency in \autoref{fig:rft_excursion_set_properties}C. Note that $\mathbb{E}(C_u)$ and $\mathbb{P}(X_m \ge u)$ are identically equal to $\mathbb{E}(\chi(E_u))$. 

\vspace{2mm}
\noindent \textbf{(P2) The distribution of the number of clusters within an excursion set}
\vspace{2mm}

The RFT approximation for the distribution of the number of clusters within an excursion set was introduced by \citet[][eq. 8]{Friston1994} and is of the form
\begin{equation}\label{eq:c_u_distribution}
C_u \sim \mbox{Poiss}(\lambda_{C_u}),
\end{equation}
where
\begin{equation}\label{eq:c_u_distribution_lambda}
\lambda_{C_u} := \mathbb{E}(C_u).
\end{equation}
That is, the number of clusters is assumed to be distributed according to a Poisson distribution with parameter $\mathbb{E}(C_u)$, which in turn is available from eq. \eqref{eq:exp_c_u_approx}. \citet{Friston1994} base the approximation for the distribution of the number of clusters in an excursion set on \citet[][Theorem 6.9.3, p. 161]{Adler1981} and the fact that a Poisson distribution is fully specified in terms of its expectation parameter. Theorem 6.9.3 in \citet{Adler1981} is concerned with the limit probability of an Euler characteristic number for high values of $u$ and remarks that ``Since this number is (...) essentially the same as the number of components of the excursion set (...) and the number of local maxima above $u$, these should have the same limiting distribution (...). This, however, has never been rigorously proven''. Intuitively, eq. \eqref{eq:c_u_distribution} can be understood as reflecting the fact that the occurrence of clusters in an excursion set under the null hypothesis should not depend on spatial location, but only depend on an ``average spatial occurrence rate'' given by $\mathbb{E}(C_u)$. By means of the approximation \eqref{eq:exp_c_u_approx}, an increase in the smoothness of a Gaussian-related random field results in a decrease of the expected number of clusters in an excursion set at a fixed level $u$. In \autoref{fig:rft_excursion_set_properties}D, we visualize the dependency of $\mathbb{P}(C_u = c)$ as a function of the Gaussian-related random field's smoothness for a fixed level $u = 6.5$. As evident from the visualization, increases in smoothness result in a shift of probability mass towards smaller values of $c$.

\vspace{6mm}
\noindent\textbf{(P3) The distribution of the volume of clusters within an excursion set}
\vspace{2mm}

As in definition \eqref{eq:k_u_definition}, let $K_u$ denote a real-valued random variable that models the volume of clusters within an excursion set measured either in terms of Lebesgue measure or third-order resel volume. In the SPM implementation of RFT, the probability distribution of this random variable is assumed to be specified in terms of the probability density function (cf. \citet[][p. 213, eq. (12)]{Friston1994})
\begin{equation}\label{eq:k_u_distribution}
f_{K_u} : \mathbb{R}_{\ge 0} \to \mathbb{R}_{\ge 0},
k \mapsto f_{K_u}(k) 
:= \frac{2\kappa}{D} k^{\frac{2}{D} - 1} \exp\left(-\kappa k^{\frac{2}{D}}\right),
\end{equation} 
where 
\begin{equation}\label{eq:k_u_distribution_kappa}
\kappa := \left(\Gamma\left(\frac{D}{2} + 1 \right) \frac{1}{\mathbb{E}(K_u)} \right)^{\frac{2}{D}}.
\end{equation}
These expressions are motivated as follows. First, eq. \eqref{eq:k_u_distribution} has been derived by \citet{Friston1994} based on a result by \citet{Nosko1969, Nosko1969a, Nosko1970}. Nosko's result, which is reported in \citet[][p. 158]{Adler1981}, states that the size of the distribution of connected components of a GRF's excursion set to the power of $2/D$ has an exponentional distribution with expectation parameter $\kappa$.  As shown in \ref{sec:miscellaneous_proofs}, eq. \eqref{eq:k_u_distribution} then follows directly from Nosko's result by the change of variables theorem (e.g., \citet[][p. 136]{Fristedt2013}). Note that \citet[][p. 158]{Adler1981} remarks that ``no detailed proofs are given'' for Nosko's result. Second, eq. \eqref{eq:k_u_distribution_kappa} follows from the expression for the expected cluster volume eq. \eqref{eq:k_u_expectation} and the fact that an exponential distribution is fully specified in terms of the reciprocal value of its expectation. It should be noted that the probability density function eq. \eqref{eq:k_u_distribution} applies to GRFs and, in the limit of high degrees of freedom, to $T$-fields (cf. comments in \textit{spm\_P\_RF.m}). Refined results for the distribution of cluster volumes in excursion sets of $T$- and $F$-fields were derived by \citet{Cao1999}, but are not implemented in SPM.

For later reference, we note two formal consequences of the parametric form eq. \eqref{eq:k_u_distribution} for the probability distribution of the cluster volume $K_u$. First, as shown in \ref{sec:miscellaneous_proofs}, eq. \eqref{eq:k_u_distribution} implies that the cumulative density function of the cluster volume is given by
\begin{equation}\label{eq:k_u_cdf}
F_{K_u} : \mathbb{R} \to [0,1], k \mapsto F_{K_u}(k) =  1 - \exp \left(-\kappa k^{\frac{2}{D}} \right).
\end{equation}
Second, from eq. \eqref{eq:k_u_cdf} it follows directly, that the probability of the cluster volume to exceed a constant $k$ under the RFT framework evaluates to (cf. \citet[][eq. 11]{Friston1994})
\begin{equation}\label{eq:k_u_ge_k}
\mathbb{P}(K_u \ge k) =  \exp \left(-\kappa k^{\frac{2}{D}} \right).
\end{equation}
In \autoref{fig:rft_excursion_set_properties}C, we visualize the log probability density function $f_{K_u}$ for a fixed value $u = 6.5$ as a function of the Gaussian-related random field's smoothness. The allocation of probability density mirrors the effect of smoothness on the expected volume of clusters within the excursion set $\mathbb{E}(K_u)$ in \autoref{fig:rft_excursion_set_properties}B.

\begin{small}
\vspace{2mm}
\noindent\textbf{(P4) The distribution of the number of clusters of a given volume within an excursion set}
\vspace{2mm}
\end{small}

Let $C_{\ge k,u}$ denote a random variable that models the number of clusters within an excursion set of level $u$ that have a volume equal to or larger than some constant $k$. Under the RFT framework, it is assumed that
(cf. \citet[][eq. 4]{Friston1996})
\begin{equation}\label{eq:q_ku_distribution}
C_{\ge k,u} \sim \mbox{Poiss}(\lambda_{C_{\ge k,u}}),
\end{equation}
where
\begin{equation}\label{eq:q_ku_distribution_lambda}
\lambda_{C_{\ge k,u}} := \mathbb{E}(C_u)\mathbb{P}(K_u \ge k).
\end{equation}
That is, the random variable $C_{\ge k,u}$ is assumed to be distributed according to a Poisson distribution, the expectation parameter of which is given by the product of the expected number of clusters and the probability of a cluster volume to assume a value equal to or larger than the constant $k$. Note again that parametric forms for $\mathbb{E}(C_u)$ and $\mathbb{P}(K_u \ge k)$ are available from eqs. \eqref{eq:exp_c_u_approx} and \eqref{eq:k_u_ge_k}, respectively. Further note that eq. \eqref{eq:q_ku_distribution} specifies probabilities of the form $\mathbb{P}(C_{\ge k,u} = i)$, which should be read as representing the probability that the number of clusters within an excursion set defined by the level value $u$ with a volume larger than or equal to $k$ is $i$. For later reference, we note that the sum $\sum_{i=0}^{c-1} \mathbb{P}(C_{\ge k,u} = i)$
thus represents the probability that the number of clusters within an excursion set defined by level $u$ with volume larger than or equal to $k$ is either smaller than or equal to a constant $c \in \mathbb{N}$ minus $1$, and hence the expression $1 - \sum_{i=0}^{c-1} \mathbb{P}(C_{\ge k,u} = i)$ represents the probability that the number of clusters within an excursion set defined by level $u$ having volume equal to or larger than $k$ is equal to or larger than $c$.

Expressions \eqref{eq:q_ku_distribution} and \eqref{eq:q_ku_distribution_lambda} can be motivated as follows. Assume the existence of a random variable $C_u$ that models the number of clusters within an excursion set at level $u$ and the existence of a  parametric expression for its distribution (provided by eq. \eqref{eq:c_u_distribution} and eq. \eqref{eq:c_u_distribution_lambda} in the theory of RFT). Assume further that one would like to specify the distribution $\mathbb{P}(C_{\ge k,u})$ of a second random variable $C_{\ge k,u}$ that models the number of clusters in an excursion set $E_u$ at level $u$ that have a volume equal to or larger than some constant $k$. By definition, this distribution is a marginal distribution of the joint distribution
\begin{equation}
\mathbb{P}(C_u,C_{\ge k,u}) = \mathbb{P}(C_u)\mathbb{P}(C_{\ge k,u}|C_u).
\end{equation}
Define the conditional distribution of $C_{\ge k,u}$ given $C_u$ by the binomial distribution
\begin{equation}
C_{\ge k,u}|C_u \sim \mbox{BN}_{j}(\mu), 
\mbox{ with } 
\mu := \mathbb{P}(K_u \ge k)
\end{equation}
such that by definition
\begin{equation}\label{eq:p_q_ku_giv_c}
\mathbb{P}(C_{\ge k, u} = i|C_u = j) 
= \left(\begin{matrix} j \\ i \end{matrix} \right)\mu^i (1 - \mu)^{j-i}, 
\mbox{ for } i \in \mathbb{N}_j^0 \mbox{ and } j \in \mathbb{N}.
\end{equation} 
Intuitively, the event that $i$ of $j$ realized clusters (i.e., $i = 0,1,...,j$) have a volume equal or larger to $k$ is thus conceived as one realization of $j$ independent Bernoulli events with ``success probability'' $\mathbb{P}(K_u \ge k)$. Based on eq. \eqref{eq:p_q_ku_giv_c}, it can then be shown that (cf. \citet[][Appendix]{Friston1996})
\begin{equation}\label{eq:p_q_ku_poiss}
\mathbb{P}(C_{\ge k,u} = i) 
= \sum_{j = 1}^{\infty} \mathbb{P}(C_u = i,C_{\ge k,u} = j)
= \frac{\lambda_{C_{\ge k,u}}^i \exp\left(-\lambda_{C_{\ge k,u}} \right)}{i!} ,
\end{equation}
i.e., that such a random variable $C_{\ge k,u}$ has a Poisson distribution with expectation parameter $\lambda_{C_{\ge k,u}}$.  We document the derivation of eq. \eqref{eq:p_q_ku_poiss} in \ref{sec:miscellaneous_proofs}. The importance of the random variable $C_{\ge k,u}$ in the SPM implementation of RFT will be elaborated on in \fullref{sec:spm_table}. Finally, in \autoref{fig:rft_excursion_set_properties}D, we visualize the distribution of $C_{\ge k,u}$ as a function of smoothness and the values of $u$ and $k$. At constant smoothness, increasing $u$ and $k$ leads to a shift of probability mass to lower values of $c$.

\vspace{2mm}
\noindent \textbf{(5) The distribution of the maximal cluster volume within an excursion set}
\vspace{2mm}

Finally, to afford FWER control in multiple testing scenarios relating to the volume of clusters within excursion sets, a parametric form of the respective maximum statistic is required (cf. \ref{sec:hypothesis_testing_fwer_control}). To this end, let, for a set of $c \in \mathbb{N}$ of clusters within an excursion set $E_u$ at level $u$, $k_i, i = 1,...,c$ denote the volumes of clusters $i = 1, ...,c$. Further, let
\begin{equation}
K_u^m := \max \{k_1,...,k_c\}
\end{equation}
denote the maximal cluster volume within the excursion set. Then, with the definition of the random variable $C_{\ge k,u}$, we have for $k\in \mathbb{R}$ 
\begin{align}
\begin{split}
\mathbb{P}(K_u^m < k) = \mathbb{P}(C_{\ge k,u} = 0) \mbox{ and }
\mathbb{P}(K_u^m \ge k) = \mathbb{P}(C_{\ge k,u} \ge 1). 
\end{split}
\end{align}
In words, first, the probability that the maximal cluster volume within an excursion set $E_u$ at level $u$ is smaller than a constant $k$ is the same as the probability that the number of clusters in the excursion set $E_u$ which have a volume equal to or larger than $k$ is zero. Second, the probability that the maximal cluster volume in an excursion set $E_u$ is equal to or larger than a constant $k$ is the same as the probability that the number of clusters in the excursion set $E_u$ which have a volume equal to or larger than $k$ is equal to or larger than $1$. The distribution of the maximal cluster volume $K_u^m$ is thus fully determined by the distribution of $C_{\ge k,u}$. For later reference, we note that with eqs. \eqref{eq:q_ku_distribution}, \eqref{eq:q_ku_distribution_lambda}, and \eqref{eq:p_q_ku_poiss} (cf. \citep[][eq. 14]{Friston1994}) 
\begin{align}\label{eq:p_k_u_m}
\begin{split}
\mathbb{P}(K_u^m \ge k) 
& = \mathbb{P}(C_{\ge k,u} \ge 1) \\
& = 1 - \sum_{i = 0}^0 \mathbb{P}(C_{\ge k,u} = i) \\
& = 1 - \exp \left(-\mathbb{E}(C_u) \mathbb{P}(K_u \ge k) \right).
\end{split}
\end{align}

\section{Application}\label{sec:application}
The general theory of random fields - and the theory of RFT in particular - model three-dimensional space by $\mathbb{R}^3$. In reality, however, fMRI data points are only acquired at a finite number of discrete spatial locations, commonly known as voxels. This discretization of space has to be accounted for when considering the application of RFT theory to GLM-based fMRI data analysis. In the current section, we thus consider the discrete space, discrete data-point linear model
\begin{equation}\label{eq:discrete_glm}
Y_{iv} = m_i \beta_v + \sigma Z_{iv}  \mbox{  for  } i = 1,...,n \mbox{ and } 
v \in \mathcal{V},
\end{equation}
where
\begin{equation}
\mathcal{V} := 
\left \lbrace 
v |
v = (v_1,v_2,v_3) \mbox{ with } 
v_1 \in \mathbb{N}_{m_{1}}, v_2 \in \mathbb{N}_{m_{2}}, v_3 \in  \mathbb{N}_{m_{3}} 
\right\rbrace
\end{equation}
denotes a \textit{voxel index set}. Eq. \eqref{eq:discrete_glm} should be read as the discrete-space analogue of eq. \eqref{eq:continuous_glm} for which the continuous space coordinates $x \in \mathbb{R}^3$ have been replaced by voxel indices $v \in \mathcal{V}$. We assume that these voxel indices correspond to the weighting parameters of three orthogonal, but not necessarily orthonormal, basis vectors $b_d \in \mathbb{R}^3, d = 1,2,3$ of $\mathbb{R}^3$, such that the set 
\begin{equation}\label{eq:lattice}
L := \left\lbrace \sum_{d = 1}^3 v_d b_d | v_1 \in 
\mathbb{N}_{m_{1}}, v_2 \in \mathbb{N}_{m_{2}},  v_3 \in \mathbb{N}_{m_{3}}
\right \rbrace
\end{equation}
is a finite lattice covering the subset of $\mathbb{R}^3$ of interest. Assuming an MNI space bounding box of -78 to 78 mm in the coronal, -112 mm to 76 mm in the sagittal, and  -70 mm to 95 mm in the axial direction and a 2 $\times$ 2 $\times$ 2 mm voxel size, typical values for the number of indices are $m_{1} = 79$, $m_2 = 95$, and $m_3 = 69$. With these definitions, we thus have the correspondences
\begin{equation}\label{eq:cont_discr_corres}
Y_{iv} = Y_i(\tilde{x}), 
\beta_{v} = \beta_i(\tilde{x}), \mbox{ and } 
Z_{iv} = Z_i(\tilde{x}) \mbox{ for } \tilde{x} := \sum_{d = 1}^3 v_d b_d \mbox{ and } v \in \mathcal{V} 
\end{equation}
between the discrete space and the continuous space linear model representations of eqs.  \eqref{eq:discrete_glm} and \eqref{eq:continuous_glm}, respectively. 

Building on these correspondences between the theoretical scenario of a continuous space and the applied scenario of discrete spatial sampling, we are now in a position to discuss  the application of the theoretical framework of the previous section to  GLM-based fMRI data analysis. As outlined in the Introduction, we first consider the \textit{parameter estimation} of the RFT null model, which comprises the estimation of the Gaussian-related random field's smoothness and the approximation of the search space's intrinsic and resel volumes. We then consider the evaluation of the parametric p-values reported in the SPM results table.

\subsection{Parameter estimation}\label{sec:parameter_estimation}

As discussed in \fullref{sec:probabilistic_properties}, the probability distributions of all excursion set features of a Gaussian-related random field depend on the resel volumes of the search space via the expected Euler characteristic (cf. eq. \eqref{eq:gkf}). The resel volumes in turn depend on the FWHM parameterization of the Gaussian-related random field's smoothness and the search space's intrinsic volumes (cf. Definition \eqref{def:resel_volumes}). Application of the parametric forms for the probability distributions of excursion set features discussed in \fullref{sec:probabilistic_properties} in a data-analytical context thus necessitates the estimation of the Gaussian-related random field's smoothness, the approximation of its search space's intrinsic volumes, and the subsequent evaluation of its resel volumes. In effect, the estimated resel volumes are intended to furnish a data-adaptive null hypothesis model against which the actually observed data can be evaluated. Because the data typically do not fully conform to the null hypothesis $\beta_v = 0$ for all $v \in \mathcal{V}$, the SPM implementation of RFT bases its parameter estimation scheme on the so-called \textit{standardized residuals}
\begin{equation}\label{eq:standardized_residuals}
r_{iv} := \frac{Y_{iv} - m_i \hat{\beta_v}}
{\sqrt{\frac{1}{n-p}\sum_{i=1}^n \left(Y_{iv} - m_i \hat{\beta_v} \right)^2}}, 
i = 1,...,n, v \in \mathcal{V},
\end{equation}
where $\hat{\beta}_v$ denotes the Gauss-Markov estimator for the effect size parameter at voxel $v$. Practically, these values are stored in \textit{ResI\_000*.nii} files, which are created and deleted by SPM's \textit{spm\_spm.m} function during SPM's GLM estimation. Note that the denominator of the standardized residuals in eq. \eqref{eq:standardized_residuals} corresponds to the estimate $\hat{\sigma}$ of the standard deviation parameter $\sigma$. Under the assumption that $m_i \hat{\beta_v}$ veridically reflects the expectation of $Y_{iv}$, the standardized residuals can thus be conceived as realizations of the $Z$-fields $Z_i(x)$  at discrete spatial locations (cf. eq. \eqref{eq:continuous_glm} and the introduction of \fullref{sec:theory}). Stated differently, the RFT framework assumes that the standardized residuals can be considered observations of the null hypothesis model used to evaluate the probabilities of observed excursion set features.  RFT parameter estimation as implemented in SPM's \textit{spm\_est\_smoothness.m} function then comprises three steps: first, the estimation of the FWHM smoothness parameters based on the standardized residuals, second, the approximation of the intrinsic volumes of the search space, and finally, the evaluation of the resel volumes.

\subsubsection*{FWHM smoothness parameter estimation}

\begin{figure}[t]
\centerline{\includegraphics[width=\textwidth]{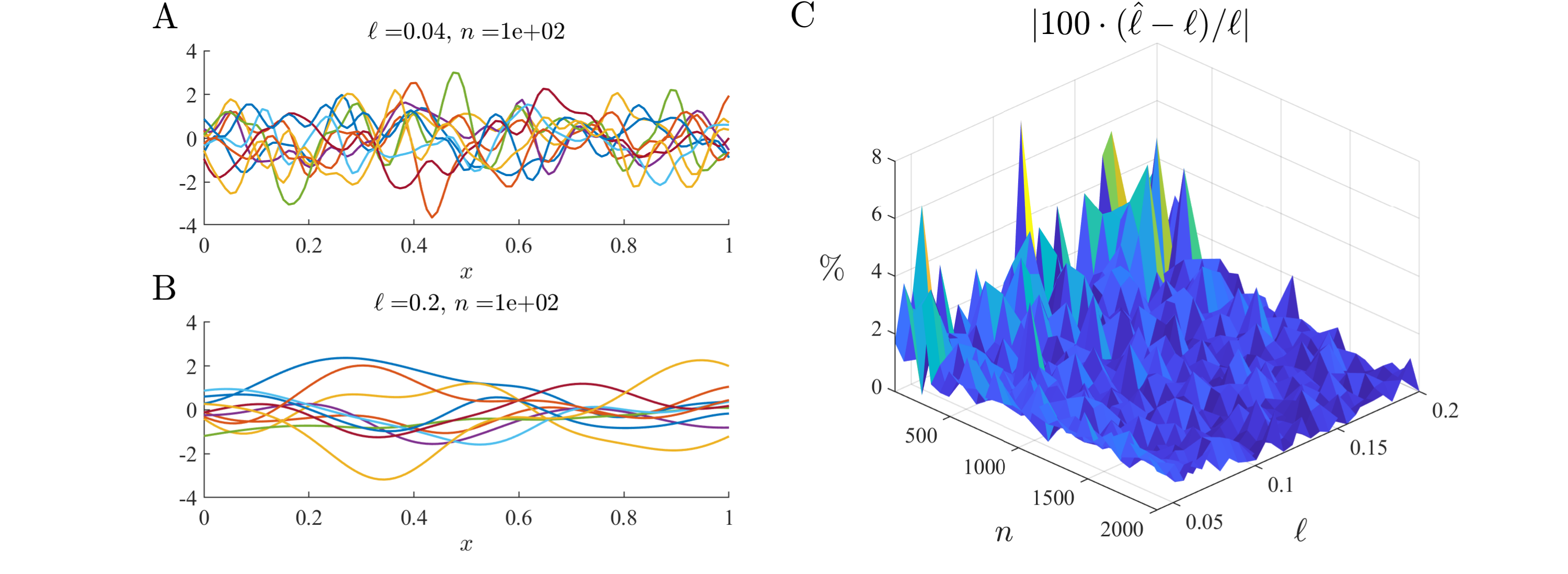}}
\caption{Smoothness parameter estimation. To exemplify the smoothness estimation scheme of SPM's RFT implementation, we capitalize on the fact that for one-dimensional GRFs with Gaussian covariance function, the reciprocal unit-square Lipschitz-Kiling curvature $\varsigma$ is identical to the Gaussian covariance function's length parameter $\ell$ up to a factor of $2^{-1/2}$ (cf. eq. \eqref{eq:smoothness_1d_grf_gauss_cov}). A straightforward estimator for $\ell$ is thus $\hat{l} := \sqrt{2}\varsigma$. For the simulations depicted in the figure, we evaluated $\varsigma$ using numerical differentiation and sample variances to estimate $\ell$ for one-dimensional GRFs on the interval $[0,1]$. We varied $\ell$ in the interval $[0.04, 0.2]$ and examined sample sizes in the range from $n = 100$ to $n = 2000$. (\textbf{A}) Panel A visualizes a sample of $n = 100$ rough GRF realizations with $\ell = 0.04$ . (\textbf{B})  Panel B visualizes a sample of $n = 100$ smooth  GRF realizations with $\ell = 0.2$. (\textbf{C}) Panel C visualizes the absolute percent relative estimation error $|100 \cdot (\hat{\ell} - \ell)/\ell|$ for the ensuing estimates. In general, this error is small ($\lesssim 3$ \% on average) and decreases for larger sample sizes uniformly over the space of $\ell$. For the full implementational details of these simulations, please see \textit{rft\_6.m}.}
\label{fig:rft_smoothness_estimation}
\end{figure}

SPM's estimation scheme for the FWHM smoothness parameters can be conceived as a two-step procedure. In the first step, an estimate of the smoothness parameter $\varsigma$ (cf. eq.  \eqref{eq:int_det_var_nabla_X}) of the $Z$-fields $Z_i(x)$ is constructed based on the standardized residuals.  In a second step, this estimate is used together with an estimate of the approximating matrix $\Lambda$ to afford the estimation of  FWHM smoothness parameters that conform to the equality of eq. \eqref{eq:smoothness_fwhm} and, in their relative magnitude, reflect the diagonal entries of the variance matrix of partial derivatives (cf. eq. \eqref{eq:var_nabla_X}). More specifically, the smoothness parameter estimate of the first step is evaluated as
\begin{equation}\label{eq:varsigma_hat}
\hat{\varsigma} := \frac{1}{\frac{1}{m}\sum_{v=1}^m |\frac{n}{\tilde{n}(n-p)}V_v|^{1/2}},
\end{equation}
where $V_v \in \mathbb{R}^{3 \times 3}$ denotes a scaled voxel-wise estimate of the variance matrix of gradient components with elements
\begin{equation}\label{eq:V_v}
(V_v)_{jk} 
:= \sum_{i = 1}^{\tilde{n}} {(r_{i(v + e_j)} - r_{iv})(r_{i(v + e_k)} - r_{iv})} \,\,\mbox{ for } j,k = 1,2,3 \mbox{ and } v \in \mathcal{V},  
\end{equation}
$m = m_1 + m_2 + m_3$ denotes the cardinality of the voxel index set, and $\tilde{n} \le n$ denotes the size of a data point subset potentially used by SPM for computational efficiency. The forms of eqs. \eqref{eq:varsigma_hat} and \eqref{eq:V_v} are motivated as follows \citep{Hayasaka2003, Hayasaka2004}: the $(V_v)_{jk}$ are products of the first-order finite differences between the standardized residuals of adjacent voxel indices. These first-order finite differences may be interpreted as scaled versions of numerical first-order derivatives. Notably, SPM thus foregoes the direct approximation of the gradient components of the Gaussian-related random field. The first order differences in the factors in eq. \eqref{eq:V_v} are evaluated by \textit{spm\_sample\_vol.c}. The formation of the sum of products of these first-order finite differences of standardized residuals in  eq. \eqref{eq:V_v} and its subsequent division by $\tilde{n}$ can be interpreted as an estimation of the (co)variances of the gradient components, if it is assumed that the sample average over available data points $\tilde{n}$ is zero. The subsequent multiplication by $n$ and division by the degrees of freedom $n-p$ is meant to rescale the respective estimates to the entire set of data points and render the estimate unbiased by effectively converting standardized to normalized residuals (cf. \textit{spm\_est\_smoothness.m}, comments ll. 144 - 169 and \citet{Worsley1996c}). In effect, the sum terms in the denominator of eq. \eqref{eq:varsigma_hat} can be interpreted as scaled voxel-specific numerical estimates of the square root of the determinants of $\mathbb{V}(\nabla X(x)), x \in S$. Finally, the summation of these terms and division by the number of voxel indices $m$ can be interpreted as the numerical integration over the unit cuboid for first-order finite differences, rather than numerical derivative approximations to the Gaussian-related random fields' gradient components. In \autoref{fig:rft_smoothness_estimation} we demonstrate that a numerically simplified version of SPM's smoothness parameter estimation approach indeed results in a valid recovery of the smoothness parameter of a two-dimensional Gaussian random field with Gaussian covariance function.

The reparameterization of smoothness in terms of FWHM smoothness parameters in the second step then is achieved by setting
\begin{equation}\label{eq:f_tilde_j}
\tilde{f}_{x_j} 
:= \frac{1}{4 \ln 2}
\left(\frac{1}{m}\sum_{v=1}^m \frac{n}{\tilde{n}(n-p)}(V_v)_{jj}\right)^{-\frac{1}{2}}
\end{equation}
and
\begin{equation}\label{eq:f_hat_j}
\hat{f}_{x_j} 
= \frac{\hat{\varsigma}^{\frac{1}{3}}}{ (4 \ln 2)^{-\frac{1}{2}}}
\frac{\tilde{f}_{x_j}}{\left(\prod_{k=1}^3 \tilde{f}_{x_j} \right)^{\frac{1}{3}}}
\end{equation}
for $j = 1,2,3$. Eqs. \eqref{eq:f_tilde_j} and \eqref{eq:f_hat_j} implement SPM's assumption that the variance matrix of gradient components is of diagonal form and thus that the smoothness parameter $\varsigma$ can be reparameterized in terms of the FWHM smoothness parameters $f_{x_j}$ (cf. eq. \eqref{eq:Lambda}). More specifically, these equations are motivated as follows (cf. \textit{spm\_est\_smoothness.m}, comments ll. 246 - 248): the $\tilde{f}_{x_j}$ can be interpreted as estimates of the diagonal elements of $\Lambda$ based on the $\mathbb{V}(\nabla X(x))$ matrix estimate discussed above. By normalizing these values using their geometric average in the second factor of eq. \eqref{eq:f_hat_j} and scaling the resulting normalized $\tilde{f}_{x_j}$'s by the first factor, it is then ensured that the $\hat{f}_{x_j}$'s reproduce the smoothness parameter estimate $\hat{\varsigma}$ if they are substituted as estimates of the $f_{x_j}$'s in the matrix $\Lambda$ (cf. eq. \eqref{eq:smoothness_fwhm}). The thus evaluated estimates are then used for the evaluation of the resel volumes and are reported in the footnote of the SPM results table as \textit{FWHM (voxels)}.

\subsubsection*{Intrinsic volume approximation}

To approximate the intrinsic volumes of the search space, SPM uses an algorithm proposed by \citet[][pp. 62 -63]{Worsley1996} and implemented in \textit{spm\_resels\_vol.c}. This algorithm corresponds to a summation of the contributions of zero- to  three-dimensional subspaces along the cardinal axes to the respective intrinsic volumes of the search space. More specifically, the algorithm  evaluates
\begin{itemize}
\item the total number of voxels
\begin{equation}
m := m_{x_1}m_{x_2}m_{x_3},
\end{equation}
\item the number of \textit{edges} in the $x_1,x_2$ and $x_3$ directions, i.e.,
\begin{align}
\begin{split}
m_{1}^e 	& := 	| \{(v, v + e_1) | v, v + e_1 \in \mathcal{V} \}|,	\\
m_{2}^e 	& := 	| \{(v, v + e_2) | v, v + e_2 \in \mathcal{V} \}|,	\\
m_{3}^e 	& := 	| \{(v, v + e_3) | v, v + e_3 \in \mathcal{V} \}|,	\\
\end{split}
\end{align}
 respectively,
\item  the number of \textit{faces} in the $x_1\text{-}x_2$-, $x_1\text{-}x_3$-, and $x_2\text{-}x_3$- directions, i.e.,  
\begin{align}
\begin{split}
m_{12}^f 	
& := 	| \{(v, v + e_1, v + e_2, v + e_1 +  e_2) | v, v + e_1, v + e_2, v + e_1 +  e_2 \in \mathcal{V} \}|,	\\
m_{13}^f  	
& := 	| \{(v, v + e_1, v + e_3, v + e_1 +  e_3) | v, v + e_1, v + e_3, v + e_1 +  e_3 \in \mathcal{V} \}|,	\\
m_{23}^f  	
& := 	| \{(v, v + e_2, v + e_3, v + e_2 +  e_3) | v, v + e_2, v + e_3, v + e_2 +  e_3 \in \mathcal{V} \}|,	\\
\end{split}
\end{align}
respectively, 
and 
\item the number of \textit{cubes}, i.e., 
\begin{footnotesize}
\begin{align}
\begin{split}
m^c 		:= 		
|\{( & v, v + e_1, v + e_2, v + e_3,v + e_1 +  e_2, v + e_2 +  e_3, v + e_1 +  e_3, v + e_1 + e_2 + e_3) \\
| &  v, v + e_1, v + e_2, v + e_3,v + e_1 +  e_2, v + e_2 +  e_3, v + e_1 + e_3, v + e_1 + e_2 + e_3 \in \mathcal{V} \}|.	\\
\end{split}
\end{align}
\end{footnotesize}
\end{itemize}
Based on these counts and the physical voxel sizes $\delta_1, \delta_2$ and $\delta_3$ in the $x_1$-, $x_2$-, and $x_3$-directions, respectively,  the intrinsic volumes of the search space are then approximated by 

\begin{small}
\begin{align}
\begin{split}
\tilde{\mu}_0(S) 	& 	= 	m 
							- (m_{1}^e + m_{2}^e + m_{3}^e) 
							+ (m_{12}^f + m_{13}^f + m_{23}^f) 
							- m^c, 													\\
\tilde{\mu}_1(S) 	& 	= \delta_1(m_{1}^e - m_{12}^f - m_{13}^f + m^c)	 		
							+ \delta_2(m_{2}^e - m_{12}^f - m_{23}^f + m^c) 			
							+ \delta_3(m_{3}^e - m_{13}^f - m_{23}^f + m^c),	\\		
\tilde{\mu}_2(S) 	& 	= 	  \delta_1\delta_2(m_{12}^f - m^c) 
							+ \delta_1\delta_3(m_{13}^f - m^c) 
							+ \delta_2\delta_3(m_{23}^f - m^c), 														\\
\tilde{\mu}_3(S) 	& 	= \delta_1\delta_2\delta_3 m^c.
\end{split}
\end{align}
\end{small}

\noindent Note, for example, that $\tilde{\mu}_3(S)$ corresponds to a summation of the volume $\delta_1\delta_2\delta_3$ of each of the $m^c$ voxels constituting the search space. For a whole-brain GLM-based fMRI data analysis, this quantity thus typically corresponds to an approximation of the physical brain volume by many small cuboids. 

\subsubsection*{Resel volume estimation}

De-facto, SPM does not evaluate an approximation of the search space's intrinsic volumes \textit{perse}. Instead, it capitalizes on the subspace decomposition assumption of eq. \eqref{eq:intrinsic_vol_3d_decomp} and directly estimates the resel volumes in \textit{spm\_resel\_vol.m} by
\begin{align}
\begin{split}
\hat{R}_0(S) 	& 		= 	\tilde{\mu}_0(S),								\\
\hat{R}_1(S)	& 		= 	
\left(\begin{matrix}
\frac{\delta_1}{\hat{f}_{x_1}} 
& \frac{\delta_2}{\hat{f}_{x_2}} 
& \frac{\delta_3}{\hat{f}_{x_3}} 
\end{matrix}\right)
\left(\begin{matrix}
m_{1}^e - m_{12}^f - m_{13}^f + m^c \\
m_{2}^e - m_{12}^f - m_{23}^f + m^c \\
m_{3}^e - m_{13}^f - m_{23}^f + m^c
\end{matrix}\right), 		
\\ \\
\hat{R}_2(S) 	& 		=	
\left(\begin{matrix}
\frac{\delta_1 \delta_2}{\hat{f}_{x_1}\hat{f}_{x_2}}	
& \frac{\delta_1 \delta_3}{\hat{f}_{x_1}\hat{f}_{x_3}}
& \frac{\delta_2 \delta_3}{\hat{f}_{x_2}\hat{f}_{x_3}}
\end{matrix}\right) 
\left(\begin{matrix}
m_{12}^f - m^c 				\\
m_{13}^f - m^c				\\
m_{23}^f - m^c				\\
\end{matrix} \right), \mbox{and } 	\\
\hat{R}_3(S) 	
& = \frac{1}{\hat{f}_{x_1}\hat{f}_{x_2}\hat{f}_{x_3}} \tilde{\mu}_3(S). 
\end{split}
\end{align}

The estimated resel volumes $\hat{R}_d(S), d = 0,1,2,3$ are the pivot points between the practical perspective of the current section and the theoretical perspective of \fullnameref{sec:theory}. As previously noted, virtually all probability distributions of the topological features of interest in RFT depend on the resel volumes via the expected Euler characteristic (cf. eq. \eqref{eq:gkf}). Substitution of $\hat{R}_d(S), d = 0,1,2,3$ for the resel volumes in the expected Euler characteristic formula thus results in a set of probability distributions that reflect a data-adapted Gaussian-related random field null model. The exceedance probabilities of the observed topological features of interest under this data-adapted null model are then reported in the SPM results table.

This concludes our discussion of the RFT parameter estimation scheme implemented in SPM. We provide a reproduction of this estimation scheme in a simplified and annotated version of \textit{spm\_est\_ smoothness.m} as \textit{rft\_spm\_est\_smoothness.m} in the accompanying OSF project. 

\subsection{The SPM results table}\label{sec:spm_table}
The majority of p-values in the SPM results table relate to the random variable $C_{\ge k, u}$ introduced in \fullref{sec:probabilistic_properties} \citep{Friston1996}. Recall that $C_{\ge k, u}$ models the number of clusters within a level $u$-dependent excursion set $E_u$ that have a volume equal to or larger than a constant $k \in \mathbb{R}$. As discussed in \fullref{sec:probabilistic_properties}, the random variable $C_{\ge k, u}$ takes on values in $\mathbb{N}^0$ and is distributed according to a Poisson distribution with parameter $\lambda_{C_{\ge k,u}} := \mathbb{E}(C_u)\mathbb{P}(K_u \ge k)$, where $\mathbb{E}(C_u)$ represents the expected number of clusters within an excursion set $E_u$ and is given by eq. \eqref{eq:exp_c_u_approx}, while $\mathbb{P}(K_u \ge k)$ denotes the probability for a cluster within an excursion set $E_u$ to assume a volume equal to or larger than the constant $k$ and  is given by eq. \eqref{eq:k_u_ge_k}. Formally, we have 
\begin{equation}
C_{\ge k,u} \sim \mbox{Poiss}(\lambda_{C_{\ge k,u}}), 
\end{equation}
and thus
\begin{equation}
\mathbb{P}(C_{\ge k,u} = c) = \exp\left(-\mathbb{E}(C_u)\mathbb{P}(K_u \ge k)\right)\frac{\left(\mathbb{E}(C_u)\mathbb{P}(K_u \ge k)\right)^c}{c!} 
\mbox{ for } c = 0,1,... \mbox{.}
\end{equation}
For ease of notation, we denote the distribution function of $C_{\ge k,u}$  in the current section by
\begin{equation}\label{eq:spm_Pcdf}
F_{\lambda_{C_{\ge k, u}}} : \mathbb{N}^0 \to [0,1], c \mapsto 
F_{\lambda_{C_{\ge k, u}}}(c) := \sum_{j = 0}^c \mathbb{P}(C_{\ge k,u} = j). 
\end{equation}
The majority of p-values in the SPM results table are then special cases of probabilities related to $C_{\ge k, u}$ for the values $c\in \mathbb{N}^0, k \in \mathbb{R}$, and $u \in \mathbb{R}$ and are evaluated using the implementation of the Poisson distribution function \eqref{eq:spm_Pcdf} in \textit{spm\_Pcdf.m}. To distinguish these special cases, we make the following assumptions and use the following notation. We assume that
\begin{enumerate}[label={(\arabic*)}, leftmargin = *]
\item a statistical parametric map, for example a three-dimensional voxel map comprising $T$ value realizations, has been obtained,
\item for this map, a \textit{cluster-forming threshold} value $\tilde{u}$ and a \textit{cluster-extent threshold} value $\tilde{k}$ in resel volume have been defined,
\item the number of clusters $\tilde{c}$ in the resulting excursion set $E_{\tilde{u}}$ has been evaluated, 
\item the volumes of these clusters $\tilde{k}_i, i = 1,...,\tilde{c}$ have been assessed, and
\item for each cluster, the maximum test statistic within this cluster has been evaluated and is denoted by $\tilde{t}_i, i = 1,...,\tilde{c}$.
\end{enumerate}
A few remarks from a practical perspective on these assumptions are in order. In SPM, statistical parametric maps for $T$ tests are evaluated based on GLM beta and variance parameter estimates for user-defined contrast weight vectors and are saved under the filename \textit{spmT\_000*.nii} or the like. When interacting with SPM's \textit{Results} graphical user interface, or specifying options in SPM's batch editor for the \textit{Results Report} module, the user is prompted to specify the cluster-forming threshold value $\tilde{u}$ by selecting a \textit{Threshold Type} with options ``None'' and ``FWE'' for p-value adjustment, and a \textit{Threshold Value} in the form of a p-value. The combination of these two options is evaluated by SPM to define the value $\tilde{u}$ as follows. In the case of ``p-value adjustment: none'' and the subsequent specification of a p-value (e.g. 0.001), SPM uses the function \textit{spm\_u.m} to evaluate the inverse cumulative density function for the probabilistic complement of the specified value and the respective statistic. For example, in the case of a statistical parametric map comprising $T$-values, $\tilde{u}$ thus corresponds to the ``critical $T$-value'', i.e., the value $t_c$ for which values of the $T$-statistic equal to or larger than $t_c$ have a probability of equal to or smaller than the specified p-value under the standard $T$- distribution with the appropriate degrees of freedom (e.g. for a $T$ distribution with 30 degrees of freedom and p-values of 0.05, 0.025, and 0.001, the respective $\tilde{u}$ values evaluate to 1.697,2.042, and 3.385, respectively). In the case of ``p-value adjustment: FWE'' and the subsequent specification of a p-value (e.g. 0.05), SPM uses the function \textit{spm\_uc.m} to evaluate a critical threshold using the RFT framework. This option leads, in general, to higher values for $\tilde{u}$. The cluster-extent threshold $\tilde{k}$ is specified  by the user in units of voxels and is transformed into units of resels by SPM by multiplying the user-specified value by $1/\prod_{d = 1}^3 \hat{f}_{x_d}$. Finally, the excursion set $E_{\tilde{u}}$, the number of clusters $\tilde{c}$, and the cluster-specific volume and maxima statistics $\tilde{k}_i$ and $\tilde{t}_i, i = 1,...,\tilde{c}$ are evaluated using low-level SPM functionality, such as \textit{spm\_bwlabel.c}, \textit{spm\_clusters.m}, and \textit{spm\_max.m}. For the computational details of these routines, the reader is referred directly to the documentation in the functions' headers. Based on an interaction between the \textit{spm\_list.m} and \textit{spm\_P\_RF.m} functions, SPM then documents p-values in the SPM results table for \textit{set-level, cluster-level} and \textit{peak-level inferences} \citep{Friston1996}. In the following, we specify for each level of inference which probabilities these p-values correspond to.

\subsubsection*{Set-level inference}
Assume we are interested in the probability
\begin{equation}\label{eq:set_level_p}
\mathbb{P}(C_{\ge \tilde{k},\tilde{u}} \ge \tilde{c}),
\end{equation}
i.e., the probability that for the specified cluster-forming threshold $\tilde{u}$, the number of clusters within the excursion set $E_{\tilde{u}}$ that have a volume equal to or larger than the specified cluster-extent threshold value $\tilde{k}$ is equal to or larger than the observed number of clusters $\tilde{c}$. Clearly, we have
\begin{align}
\begin{split}
\mathbb{P}(C_{\ge \tilde{k},\tilde{u}} \ge \tilde{c}) 
& = 1 - \mathbb{P}(C_{\ge \tilde{k},\tilde{u}} < \tilde{c}) \\
& = 1 - F_{\lambda_{C_{\ge \tilde{k}, \tilde{u}}}}(\tilde{c})
\end{split}
\end{align}
and eq. \eqref{eq:set_level_p} can thus be evaluated using the Poisson distribution function \eqref{eq:spm_Pcdf}. In the SPM results table, the probability $\mathbb{P}(C_{\ge \tilde{k},\tilde{u}} \ge \tilde{c})$ is reported as the ``set-level p value'' and the value $\tilde{c}$ is documented as the number of clusters c (\autoref{fig:rft_spm_table} and \autoref{tab:spm_p_values}). The idea behind using \eqref{eq:set_level_p} is to enable inferences about ``network-activations'' \citep[][pp. 229-230]{Friston1996}: if the number of actually observed clusters $\tilde{c}$ exceeds the number $\mathbb{E}(C_u)$ of clusters expected under the RFT null hypothesis model by a large margin, eq. \eqref{eq:set_level_p} evaluates to a small value. In this case, it may thus be concluded that the observed pattern of distributed clusters reflects the engagement of a distributed network of brain regions in the GLM contrast of interest. Note that the set-level inference probability \eqref{eq:set_level_p} is the only p-value of the SPM results table that directly depends on the user specified cluster-extent threshold $\tilde{k}$.

\subsubsection*{Cluster-level inference}
Assume that for a cluster of the excursion set $E_{\tilde{u}}$ indexed by $i$, where $i = 1,...,\tilde{c}$, we are interested in the probability
\begin{equation}\label{eq:cluster_level_p}
\mathbb{P}(C_{\ge \tilde{k_i},\tilde{u}} \ge 1),
\end{equation}
i.e., the probability that for the specified cluster-forming threshold $\tilde{u}$ the number of clusters within the excursion set that have a volume equal to or larger than the observed cluster volume $\tilde{k}_i$ is equal to or larger than $1$. For this probability, we have
\begin{align}\label{eq:cluster_level_p_cluster_max}
\begin{split}
\mathbb{P}(C_{\ge \tilde{k}_i,\tilde{u}} \ge 1) 
& = 1 - \mathbb{P}(C_{\ge \tilde{k_i},\tilde{u}} < 1) 	\\
& = 1 - F_{\lambda_{C_{\ge \tilde{k}_i, \tilde{u}}}}(0) \\
& = 1 - \mathbb{P}(C_{\ge \tilde{k}_i,\tilde{u}} = 0)    \\
& = 1 - \exp\left(-\mathbb{E}(C_{\tilde{u}})\mathbb{P}(K_{\tilde{u}} \ge \tilde{k}_i)\right) \\
& = \mathbb{P}\left(K_{\tilde{u}}^m \ge \tilde{k}_i \right).
\end{split}
\end{align}
From the second equality of eq. \eqref{eq:cluster_level_p_cluster_max}, we thus see that \eqref{eq:cluster_level_p} can be evaluated using the Poisson distribution function \eqref{eq:spm_Pcdf}. Moreover, from the last equality, we see that $\mathbb{P}(C_{\ge \tilde{k_i},\tilde{u}} \ge 1)$ corresponds to the probability for the maximal cluster volume within an excursion set $E_{\tilde{u}}$ to exceed the value $\tilde{k}_i$ (cf. eq. \eqref{eq:p_k_u_m}). Because the probability \eqref{eq:cluster_level_p} hence relates to the maximum statistic for cluster extent, declaring a cluster activation significant, if $\mathbb{P}(C_{\ge \tilde{k_i},\tilde{u}} \ge 1) < \alpha$ furnishes a statistical hypothesis test for spatial cluster extent that controls the FWER at level $\alpha \in [0,1]$ (cf. \ref{sec:hypothesis_testing_fwer_control}). Note that the family of statistical tests for which the Type I error rate is controlled here refers to the volumes of clusters $K_{\tilde{u}}$ within an excursion set. In the SPM results table, the probability $\mathbb{P}(C_{\ge \tilde{k}_i,\tilde{u}} \ge 1)$ is reported as the ``FWE-corrected p-value at the cluster level'' for each cluster  $i = 1, ..., \tilde{c}$ in the column labelled $\text{p}_\text{FWE-corr}$ and the voxel number equivalent $\tilde{k}_i \prod_{d = 1}^3 \hat{f}_{x_d}$  of the resel volume of each cluster $i = 1, ..., \tilde{c}$ is reported in the column labelled $\text{k}_\text{E}$ (\autoref{fig:rft_spm_table} and \autoref{tab:spm_p_values}). 

As ``uncorrected p-value at the cluster level'' for cluster $i = 1, ..., \tilde{c}$ SPM reports the  probability 
\begin{equation}\label{eq:uncorrected_cluster_level}
\mathbb{P}(K_{\tilde{u}} \ge \tilde{k}_i),
\end{equation}
which corresponds to the probability that the cluster volume in an excursion set $E_ {\tilde{u}}$ exceeds the observed volume of the $i$th cluster. Note that rejecting the null hypothesis based on $\mathbb{P}(K_{\tilde{u}} \ge \tilde{k}_i) <  \alpha$ for a significance level $\alpha \in [0,1]$ over a family of $\tilde{c}$ clusters entails a probability of $1 - (1 - \alpha)^{\tilde{c}}$ for at least one Type I error and hence offers no FWER control. The ``p-value correction'' procedure implied by the SPM results table header labelling can thus be interpreted as the transformation of \eqref{eq:uncorrected_cluster_level} by means of the right-hand side of the penultimate equation in eq. \eqref{eq:cluster_level_p_cluster_max}.

\subsubsection*{Peak-level inference}
Assume we are interested in the probability
\begin{equation}
\mathbb{P}(C_{\ge 0,\tilde{t}_i} \ge 1),
\end{equation}
i.e., the probability that for a hypothetical cluster-forming threshold corresponding to the maximum statistic $\tilde{t}_i$ of cluster $i$, there are one or more clusters within the excursion set that have a volume equal to or larger than $0$. We have
\begin{align}\label{eq:p_c_ge_0_ti_ge_1}
\begin{split}
\mathbb{P}(C_{\ge 0,\tilde{t}_i} \ge 1) 
& = 1 - \mathbb{P}(C_{\ge 0,\tilde{t}_i} < 1) 	\\
& = 1 - F_{\lambda_{C_{\ge 0, \tilde{t}_i}}}(0) \\
& = 1 - \mathbb{P}(C_{\ge 0,\tilde{t_i}} = 0)    \\
& = 1 - \exp\left(-\mathbb{E}(C_{\tilde{t}_i})\mathbb{P}(K_{\tilde{t}_i} \ge 0)\right)	\\
& = 1 - \exp\left(-\mathbb{E}(C_{\tilde{t}_i})\right) \\
& \approx \mathbb{P}(X_m \ge \tilde{t}_i).		
\end{split}
\end{align}
In eq. \eqref{eq:p_c_ge_0_ti_ge_1}, the penultimate equality follows with expression \eqref{eq:k_u_ge_k} for $k = 0$. As shown in \ref{sec:proofs}, the ultimate approximation in eq. \eqref{eq:p_c_ge_0_ti_ge_1} follows using an approximation of the exponential function by the first two terms in its series definition \citep[][eq. 5]{Friston1996}. Thus, $\mathbb{P}(C_{\ge 0,\tilde{t}_i} \ge 1)$ approximates the probability that the global maximum of the statistical field of interest is equal or larger to the observed cluster maximum
statistic value $\tilde{t}_i$. In the SPM table, $\mathbb{P}(C_{\ge 0,\tilde{t}_i} \ge 1)$  is reported as the ``FWE-corrected p-value at the peak level'' for cluster $i = 1,...,\tilde{c}$ (\autoref{fig:rft_spm_table} and \autoref{tab:spm_p_values}). Because this probability hence relates to the maximum statistic for the test statistics comprising the statistical parametric map, declaring a voxel test statistic significant, if $\mathbb{P}(X_m \ge \tilde{t}_i) < \alpha$ furnishes a statistical hypothesis test for voxel test statistics that controls the FWER at level $\alpha \in [0,1]$ (cf. \ref{sec:hypothesis_testing_fwer_control}). Note that the family of statistical tests for which the Type I error rate is controlled here refers to the values of voxel-wise test statistics.

Finally, the SPM results table also reports an ``uncorrected p-value at the peak level''. This p-value is the probability $\mathbb{P}(C_{\ge 0,\tilde{t}_i} \ge 1) $ under the assumption that the resel volumes of the search space are given by $R_0 = 1$ and $R_d = 0$ for $d = 1,2,3$. Hence, in light of eqs. \eqref{eq:maximum_distribution} \eqref{eq:exp_c_u_approx}, and \eqref{eq:p_c_ge_0_ti_ge_1}, this p-value is evaluated as
\begin{equation}\label{eq:peak_level_uncorr}
\mathbb{E}(\chi(E_{\tilde{t}_i})) = \rho_0 (\tilde{t}_i).
\end{equation}
Intuitively, this corresponds to the scenario of a Gaussian-related random field comprising only point volumes at the locations of voxels with no spatial dependencies. Because the zero-order Euler characteristic densities are equivalent to the cumulative density functions of the $Z$-,$T$-, and $F$-distributions, the uncorrected p-value at the peak level corresponds to the familiar p-value of single statistical testing scenarios. From this perspective, the ``p-value correction'' at the peak-level implied by the SPM results table header hence corresponds to an adjustment of eq.  \eqref{eq:peak_level_uncorr} in terms of the search space's intrinsic volumes (i.e., the multiple testing multiplicity) and the underlying Gaussian-related random field's smoothness (i.e., the correlative nature of the multiple tests). 

\subsubsection*{The SPM results table footnote}

The footnote of the SPM results table documents a number of statistics that contextualize the reported p-values: the `Height threshold' and `Extent threshold' entries indicate the user-defined cluster-forming threshold $\tilde{u}$ and the voxel number equivalent of the cluster-extent threshold, i.e.,  $\tilde{k}\prod_{d = 1}^3\hat{f}_{x_d}$, respectively. In addition to these values, the height threshold entry lists the uncorrected and corrected peak-level p-value equivalents of $\tilde{u}$, and the extent threshold entry lists the uncorrected and corrected cluster-level p-values of $\tilde{k}$. The `Expected voxels per cluster <c>' and `Expected number of clusters <c>' entries report the voxel equivalent of the expected cluster volume, and the expected number of clusters with volume equal to or larger than the user-specified cluster-extent threshold $\tilde{k}$, respectively. The `FWEp' and 'FWEc' entries correspond to the critical $\tilde{u}$ value for a corrected peak-level p-value equal to or less than 0.05 and the smallest observed cluster size with a cluster-level corrected p-value less than 0.05, respectively. The `Degrees of freedom' entry reports the degrees of freedom of the underlying GLM design in an $F$-test format, such that for a one-sample $T$-test, the first number is set to 1, and the second number corresponds to the degrees of freedom of the $T$-test, $n-p$. The `FWHM' entry reports the estimated FWHM parameters in units of mm, i.e., $\delta_d\hat{f}_{x_d}, d = 1,2,3$ and in units of resels, i.e., $\hat{f}_{x_d}, d = 1,2,3$. The `Volume' entry indicates the search space volume in terms of Lebesgue measure $\lambda(S) = m\delta$ in units of mm\textsuperscript{3}, where $\delta := \delta_1\delta_2\delta_3$ is the Lebesgue volume of a single voxel in mm\textsuperscript{3}, in terms of the number of voxels $m$, and in terms of the third-order resel volume $R_3(S)$. Finally, the `Voxel size` entry reports the voxels sizes $\delta_1, \delta_2, \delta_3$ in units of mm and the volume of a ``cuboid resel'' $\hat{f} := \prod_{d=1}^3 \hat{f}_{x_d}$ in voxels. 

\begin{table}[]
\renewcommand{\arraystretch}{1.4}
\begin{footnotesize}
\begin{tabularx}{\textwidth}{lllllllllll}

set-level 													
& 				
& cluster-level			
& 
&			
& peak-level	

\\
\hline
p 															
& c 			
& p\textsubscript{FWE-corr}				
& k\textsubscript{E} 				
& p\textsubscript{uncorr}
& p\textsubscript{FWE-corr}				
& T
& p\textsubscript{uncorr} 
\\
$\mathbb{P}(C_{\ge \tilde{k},\tilde{u}} \ge \tilde{c})$ 	
& $\tilde{c}$	
& $\mathbb{P}(C_{\ge \tilde{k}_i,\tilde{u}} \ge 1)$     			
& $\tilde{k}_i \hat{f}$
& $\mathbb{P}(K_{\tilde{u}} \ge \tilde{k}_i)$
& $\mathbb{P}(C_{\ge 0, \tilde{t}_i} \ge 1)$
& $\tilde{t}_i$
& $\rho_0 (\tilde{t}_i)$
\\
\\
\hline
\\
Height threshold
&
& $\tilde{u}, \sum_{d = 0}^D \rho_d (\tilde{u}),\mathbb{P}(C_{\ge 0, \tilde{u}} \ge 1)$
&
& Degrees of freedom
& $1, n - p$
&
\\
Extent threshold
&
& $\tilde{k}\hat{f}, \mathbb{P}(K_{\tilde{u}} \ge \tilde{k}), \mathbb{P}(C_{\ge \tilde{k},\tilde{u}} \ge 1)$
&
& FWHM
& $\delta_d\hat{f}_{x_d},\hat{f}_{x_d}$
& 
& $d = 1,2,3$
\\
<k>
&
& $\mathbb{E}\left(K_{\tilde{u}}\right)\hat{f}$
&
& Volume
& $\lambda(S), m, R_3(S)$
\\
<c>
&
& $\mathbb{E}\left(C_{\tilde{u}}\right) \mathbb{P}(K_{\tilde{u}}\ge\tilde{k})$
&
& Voxel size
&  $\delta_1, \delta_2, \delta_3, \hat{f}$
&
\\
FWEp
&
& $u_c|\mathbb{P}(C_{\ge 0,u_c} \ge 1) = 0.05$
& 
&
&
&
&
\\
FWEc
&
& $\min \{\tilde{k}_i | \mathbb{P}(C_{\ge \tilde{k}_i,\tilde{u}} \ge 1) < 0.05 \}$ 
&
&
&
&
&
\end{tabularx}
\end{footnotesize}
\caption{p-values and other statistics reported in the SPM results table (cf. \autoref{fig:rft_spm_table}). Clusters are indexed by $i = 1,..., \tilde{c}$. We use the definitions $\hat{f} := \prod_{d = 1}^3 \hat{f}_{x_d}$  and $\delta := \delta_1\delta_2\delta_3$.}\label{tab:spm_p_values}
\end{table}

This concludes our documentation of the SPM results table. For an overview, please see \autoref{tab:spm_p_values}.  We provide a reproduction of SPM's p-value tabulation functionality in simplified and annotated versions of \textit{spm\_list.m} and \textit{spm\_P\_RF.m} as \textit{rft\_spm\_results\_table.m} in the accompanying OSF project.

\section{Discussion}\label{sec:discussion}

RFT is a highly sophisticated and computationally efficient mathematical framework and has repeatedly been validated empirically \citep[e.g.,][]{Hayasaka2003, Flandin2017, davenport_expected_2021}. In the field of statistics, RFT probably constitutes the most original and most advanced contribution of the neuroimaging community \citep{Adler2015}, while in the neuroimaging field itself, RFT likely constitutes the most routinely used statistical inference strategy \citep{Cobidas2016}.  Nonetheless, in the following, we will point to a few mathematical and statistical issues that may be elaborated on in the further refinement of RFT. 

First, from a mathematical perspective, it may be desirable to establish formal justifications for the distribution of the number of clusters within an excursion set and the distribution of the volume of clusters within an excursion set. As noted by \citet{Adler1981} and discussed in \autoref{sec:probabilistic_properties}, the currently used parametric forms appear to lack formal proofs. Similarly, it may be a worthwhile endeavour to mathematically model the conditional probabilistic structure of the number of clusters and their associated volumes more explicitly. At present, it is assumed that the expected number of clusters and the expected volume of a cluster are independent (cf. eq. \eqref{eq:k_u_expectation}). This appears to be a rather strong assumption, given the finite size of the search space. Finally, following \citet{Taylor2007}, it would be desirable to formally delineate some additional qualitative properties of the FWHM estimators implemented in SPM. The currently available reference in this regard, \citet{Worsley1996c}, appears to be somewhat outdated and superseded by the de-facto estimation of the FWHMs based on standardized, rather than normalized, residuals. Please note that given the approximative nature of the entire RFT framework and its repeated empirical validation, we do not mean to imply by these suggestions that RFT as it stands is fundamentally flawed, but that these foundations could be elaborated on to further ground the approach mathematically.  

Second, from a statistical perspective, a promising avenue for future research on RFT could be an update from the estimation of resel volumes to Lipschitz-Killing regression \citep{Adler2017}. In brief, Lipschitz-Killing regression foregoes an explicit approximation of the data's smoothness and intrinsic volumes and instead fits observed empirical Euler characteristics to the Gaussian kinematic formula \citep{Taylor2006} via generalized least squares. As such it potentially offers better analytical tractability, facilitated diagnostic, and higher computational efficiency in higher dimensions,  for example in the full spatiotemporal modelling of combined M/EEG and fMRI data.

Notwithstanding these potential future refinements, we hope that with our current documentation of RFT, we can contribute to the continuing efforts towards ever higher degrees of computational reproducibility and transparency in computational cognitive neuroscience \citep[][]{Nichols2017,Poldrack2017, Millman2018, Toelch2018}. Moreover, we hope that we can contribute to a factually grounded and technically precise discussion about the mathematical and computational aspects of fMRI data analysis.

\begin{small}
\paragraph{Data availability statement} The data and code that support the findings of this study are openly available from the Open Science Framework at \url{http://doi.org/10.17605/OSF.IO/3DX9W}.
\end{small}

\bibliographystyle{apalike}
\bibliography{Referenzen}

\renewcommand{\thesection}{Supplement \Alph{section}}
\renewcommand{\theequation}{S\arabic{section}.\arabic{equation}}
\setcounter{section}{0}
\setcounter{figure}{0}

\newpage
\renewcommand{\thesection}{Supplement S1}
\setcounter{definition}{0}
\setcounter{equation}{0}

 \section{Proofs}\label{sec:proofs}
\subsection{Gaussian covariance functions and smoothness}\label{sec:proof_gcf_smoothness}
We first collect a number of results on the relationships between a random field's expectation and covariance functions and the expectation and covariance functions of its gradient.  We retrieve these results from \citet[pp. 22 - 26]{Abrahamsen1997}, proofs of these results are available in \citet{Christakos1992}.

\subsubsection*{Derivatives of Gaussian random fields}
\noindent Let $X(x),x\in \mathbb{R}^n$ denote a GRF on a probability space $(\Omega, \mathcal{A}, \mathbb{P})$. Assume that $X(x), x \in \mathbb{R}^n$ has differentiable sample paths.  Then, for fixed $\omega \in \Omega$, the gradient field of $X(x)$ is defined as 
\begin{equation}
\nabla X(x) := \left(\begin{matrix}
\frac{\partial}{\partial x_i} X(x) 	\\
\end{matrix} \right)_{1 \le i \le n}
:= \left(\begin{matrix}
\lim_{h \to 0} \frac{X(x + h e_i) - X(x)}{h} 	\\
\end{matrix} \right)_{1 \le i \le n}
\in \mathbb{R}^n,
\end{equation}
where $e_i$ denotes the $i$th unit basis vector of $\mathbb{R}^n$. Note that like the GRF $X(x),x\in \mathbb{R}^n$, the gradient field is a function of both the spatial coordinates $x \in \mathbb{R}^n$ and the (random) elementary outcomes $\omega \in \Omega$. Let $m$ and $c$ denote the expectation and covariance functions of the GRF  $X(x), x \in \mathbb{R}^n$, respectively. Then the expectation function of the $i$th component of the gradient field is given by 
\begin{equation}\label{eq:exp_nabla_x}
\dot{m}_i : \mathbb{R}^n \to \mathbb{R},
x \mapsto \dot{m}_i(x) := \mathbb{E}\left(\frac{\partial}{\partial x_i} X(x)\right) 
= \frac{\partial}{\partial x_i} m(x) \mbox{ for } i = 1,...,n
\end{equation}
and the covariance function of the $i$th and $j$th components of the gradient field is given by
\begin{equation}\label{eq:cov_nabla_x}
\dot{c}_{ij} : \mathbb{R}^n \times \mathbb{R}^n  \to \mathbb{R},
(x,y) \mapsto \dot{c}_{ij}(x,y) 
:= \mathbb{C}\left(\frac{\partial}{\partial x_i} X(x), \frac{\partial}{\partial y_j} X(y) \right) 
= \frac{\partial^2}{\partial x_i \partial y_j} c(x,y)
\end{equation}
for $i,j = 1,...,n$. In words, the value of the expectation function of the $i$th component of a GRF's gradient field is given by the value of the $i$th partial derivative of the GRF's expectation function, and the value of the covariance function of the $i$th and $j$th components of a GRF's gradient field at locations $x$ and $y$ is given by the second-order partial derivative of the GRF's covariance function for $x$ and $y$ with respect to the $i$th component of $x$ and the $j$th component of $y$. Eq. \eqref{eq:cov_nabla_x} provides an explicit expression for the variance matrix of a GRF's gradient components in terms of the covariance function of the GRF. To see this, let $X(x), x\in \mathbb{R}^n$ with covariance function $c$. Let the variance matrix of the partial derivatives of $X$ at location $x \in \mathbb{R}^n$ be given by 
\begin{equation}
\mathbb{V}(\nabla X(x)) := 
\left(\begin{matrix}
\mathbb{C}\left(\frac{\partial}{\partial x_i}X(x), \frac{\partial}{\partial x_j}X(x)\right) 								\end{matrix}\right)_{1 \le i,j \le n}.
\end{equation}
Then, from \eqref{eq:cov_nabla_x}, we have
\begin{equation}
\mathbb{C}\left(\frac{\partial}{\partial x_i}X(x), \frac{\partial}{\partial x_j}X(x)\right) 
= \frac{\partial^2}{\partial x_i \partial x_j} c(x,x) \mbox{ for } 1 \le i,j \le n.
\end{equation}

\subsubsection*{Proof of eq. \eqref{eq:smoothness_1d_grf_gauss_cov}}

To now show eq. \eqref{eq:smoothness_1d_grf_gauss_cov}, we proceed in five steps. We first show that for a (mean-square differentiable) random process $X(x),x\in \mathbb{R}$ the covariance of its derivative at locations $x,y\in\mathbb{R}$ corresponds to the second derivative of its covariance function with arguments and $x$ and $y$ \citep[][pp. 43 - 47]{Christakos1992}, i.e.,
\begin{equation}\label{eq:varsigma_proof_1}
\mathbb{C}\left(\dot{X}(x), \dot{X}(y)\right) 
= \frac{\partial^2}{\partial x \partial y} c(x,y). 
\end{equation} 
We then consider the special case of a covariance function $c(x,y)$ that can be written as a function $\tilde{c}(\delta)$ of the difference $\delta = x - y$ between its input arguments and show that in this case \citep[][p. 71]{Christakos1992},
\begin{equation}\label{eq:varsigma_proof_2}
\frac{\partial^2}{\partial x \partial y} c(x,y) 
= - \frac{d^2}{d \delta^2}\tilde{c}(\delta).
\end{equation}
Next, we show that in this case
\begin{equation}\label{eq:varsigma_proof_3}
\mathbb{V}(\dot{X}(x)) = - \frac{d^2}{d \delta^2}\tilde{\gamma}(\delta).
\end{equation} 
We then consider the second-order derivative of the GCF $\tilde{\gamma}$ and show that
\begin{equation}\label{eq:varsigma_proof_4}
- \frac{d^2}{d\delta^2}\tilde{\gamma}(\delta) 
= - \frac{4\delta^2 - 2\ell^2}{\ell^4} \tilde{\gamma}(\delta). 	
\end{equation} 
Finally, we show that it then follows that
\begin{equation}\label{eq:varsigma_proof_5}
\varsigma = \frac{\ell}{\sqrt{2}}.
\end{equation}

\noindent (I) \eqref{eq:varsigma_proof_1} can be seen for a mean-square differentiable random process with constant zero mean function $m(x) = 0$ in general as follows. Recall that the $i$th partial derivative of a multivariate real-valued function
\begin{equation}
f : \mathbb{R}^n \to \mathbb{R}, x \mapsto f(x) 
\end{equation} 
at a point $a \in \mathbb{R}^n$ is defined as 
\begin{equation}
\frac{\partial}{\partial x_i}f(a) := 
\lim_{h \to 0} \frac{f(a_1 ..., a_i + h, ..., a_n) - f(a_1, ..., a_i, ...,a_n)}{h}.
\end{equation}
Then, the left-hand side of \eqref{eq:varsigma_proof_1} can be rewritten as
 
\begin{footnotesize}
\begin{align}
\begin{split}
\mathbb{C}\left(\dot{X}(x),\dot{X}(y)\right)
& = \mathbb{E}\left(\dot{X}(x)\dot{X}(y) \right) \\
& = \mathbb{E}\left(\lim_{h,r \to 0} \frac{1}{hr} \left((X(x + h) - X(x))(X(y +r) - X(y)\right)\right) \\
& = \mathbb{E}\left(\lim_{h,r \to 0} \frac{1}{hr} \left(X(x + h)X(y+r) - X(x+h)X(y) - X(x)X(y+r) + X(x)X(y))\right)\right) \\
& = \lim_{h,r \to 0} \frac{1}{hr} \mathbb{E}\left(X(x + h)X(y+r) - X(x+h)X(y) - X(x)X(y+r) + X(x)X(y)\right) \\
& = \lim_{h,r \to 0} \frac{1}{hr} \left(\mathbb{E}(X(x + h)X(y+r)) - \mathbb{E}(X(x+h)X(y)) - \mathbb{E}(X(x)X(y+r)) + \mathbb{E}(X(x)X(y))\right) \\
& = \lim_{h,r \to 0} \frac{1}{hr} \left(c(x + h,y+r) - c(x+h, y) - c(x, y+r) + c(x,y)\right) \\
& = \lim_{h\to 0} \frac{1}{h} \lim_{r \to 0} \frac{1}{r} \left(c(x + h,y+r) - c(x+h, y) - (c(x, y+r) - c(x,y))\right) \\
& = \lim_{h\to 0} \frac{1}{h} \left(\frac{\partial}{\partial y} c(x + h,y) - \frac{\partial}{\partial y} c(x,y)\right) \\
& = \frac{\partial^2}{\partial x \partial y} c(x,y), 
\end{split}
\end{align}
\end{footnotesize}

\noindent where we repeatedly assumed the appropriateness of exchanging limits and integrals, capitalized on the definition of the covariance function, and exploited the fact that $m(x) = 0, x \in \mathbb{R}$.

\vspace{2mm}

\noindent (II) \eqref{eq:varsigma_proof_2} can be seen by considering a covariance function 
\begin{equation}
c : \mathbb{R}^2 \to \mathbb{R}, (x,y) \mapsto c(x,y)
\end{equation}
which can be expressed as a function of the distance $\delta = x - y$ between $x$ and $y$, i.e., that with a function
\begin{equation}
\tilde{c} : \mathbb{R} \to \mathbb{R}, \delta \mapsto \tilde{c}(\delta),
\end{equation}
and the function
\begin{equation}
f : \mathbb{R}^2 \to \mathbb{R}, (x,y) \mapsto f(x,y) := x-y
\end{equation}
we can write
\begin{equation}
c(x,y) = \tilde{c}(f(x,y)).
\end{equation}
Then we have 

\begin{small}
\begin{align}
\begin{split}
\frac{\partial^2}{\partial x \partial y} c(x,y)
= & \frac{\partial^2}{\partial x \partial y} \tilde{c}(f(x,y)) \\
= & \frac{\partial}{\partial x} \left(\frac{\partial}{\partial y} \left(\tilde{c}(f(x,y)) \right)\right)  \\
= & \frac{\partial}{\partial x} \left(\frac{\partial}{\partial y} \tilde{c}(f(x,y)) \frac{\partial}{\partial y} f(x,y) \right)\\
= & \frac{\partial}{\partial x} \left(-\frac{\partial}{\partial y} \tilde{c}(f(x,y)) \right) \\
= & - \frac{\partial^2}{\partial x \partial y} \tilde{c}(f(x,y)) \\
= & - \frac{d^2}{d \delta^2} \tilde{c}(\delta), \\
\end{split}
\end{align}
\end{small}

\noindent where the last step follows with the definition of the total differential.

\vspace{2mm}

\noindent (III) With \eqref{eq:varsigma_proof_1} and \eqref{eq:varsigma_proof_2} we have for $x = y$:
\begin{equation}
\mathbb{V}\left(\dot{X}(x)\right) 
= \mathbb{C}\left(\dot{X}(x), \dot{X}(x)\right) 
= \frac{\partial^2}{\partial x \partial x}\gamma(x,x)
= - \frac{d^2}{d \delta^2}\tilde{\gamma}(0).
\end{equation}

\vspace{2mm}

\noindent(IV)
We have

\begin{small}
\begin{align}
\begin{split}
- \frac{d^2}{d \delta^2} \tilde{\gamma}(\delta) 
& = - \frac{d}{d \delta} \left(\frac{d}{d \delta} \tilde{\gamma}(\delta) \right) 										\\
& = - \frac{d}{d \delta} \left(\frac{d}{d \delta} \left( v \exp\left(-\frac{\delta^2}{\ell^2} \right) \right) \right) 	\\
& = - \frac{d}{d \delta} \left(\tilde{\gamma}(\delta) \left(-\frac{2\delta}{\ell^2} \right) \right) 						\\
& = - \frac{d}{d \delta}\tilde{\gamma}(\delta)\left(-\frac{2\delta}{\ell^2} \right) 
+\tilde{\gamma}(\delta)  \frac{d}{d \delta}\left(-\frac{2\delta}{\ell^2} \right) 										\\
& = - \tilde{\gamma}(\delta) \left(-\frac{2\delta}{\ell^2} \right)\left(-\frac{2\delta}{\ell^2} \right) 
+\tilde{\gamma}(\delta)  \frac{d}{d \delta}\left(-\frac{2\delta}{\ell^2} \right) 										\\
& = - \frac{4\delta^2}{\ell^4} \tilde{\gamma}(\delta) -\frac{2}{\ell^2} \tilde{\gamma}(\delta)   						\\
& = - \frac{4\delta^2 - 2\ell^2}{\ell^4} \tilde{\gamma}(\delta). 														 	\\
\end{split}
\end{align}
\end{small}

\noindent (V)
With $\delta = 0$, we obtain from \eqref{eq:varsigma_proof_4}
\begin{equation}\label{eq:var_x_and_ell}
\mathbb{V}\left(\dot{X}(x)\right)  = \frac{2}{\ell^2}.
\end{equation}
Finally, substitution of \eqref{eq:var_x_and_ell} in eq. \eqref{eq:int_v_dX} then yields
\begin{equation}
\varsigma 
= \int_{[0,1]} \left(\frac{2}{\ell^2} \right)^{-\frac{1}{2}} \,dx 
= \int_{[0,1]} \frac{\ell}{\sqrt{2}} \,dx 
= \frac{\ell}{\sqrt{2}}.
\end{equation}
$\hfill \Box$

\subsection{Elements of smoothness reparameterization}\label{sec:proof_smoothness_reparameterization}

\subsubsection*{Proof of eq. \eqref{eq:gauss_fwhm}}

The \textit{full width at half maximum} $f_x$ of a univariate real-valued function $h$ with a maximum at $x_m$ is defined as the distance
\begin{equation}
f_x := |x_1 - x_2|
\end{equation}
of two points $x_1 \le x_m \le x_2$, such that
\begin{equation}\label{eq:gauss_FWHM_p1}
h(x_1) = g(x_2) = \frac{1}{2} h(x_m).
\end{equation}
Let
\begin{equation}\label{eq:gauss_FWHM_p3}
g : \mathbb{R} \to \mathbb{R}, x \mapsto g(x)
:= \frac{1}{\sqrt{2\pi \sigma^2}}\exp\left(-\frac{1}{2\sigma^2} (x - \mu)^2 \right)
\end{equation}
denote the Gaussian function with parameters $\mu \in \mathbb{R}$ and $\sigma^2>0$. Then we have

\begin{footnotesize}
\begin{align}
\begin{split}
g(x) 
& = \frac{1}{2}g(x_m) \\
\Leftrightarrow
\frac{1}{\sqrt{2\pi \sigma^2}}\exp\left(-\frac{1}{2\sigma^2} (x - \mu)^2 \right)
& =\frac{1}{2} \frac{1}{\sqrt{2\pi \sigma^2}}\exp\left(-\frac{1}{2\sigma^2} (x_m - \mu)^2 \right) \\
\Leftrightarrow
\exp\left(-\frac{1}{2\sigma^2} (x - \mu)^2 \right)
& = \frac{1}{2}\exp\left(-\frac{1}{2\sigma^2} (x_m - \mu)^2 \right) \\
\Leftrightarrow
\exp\left(-\frac{1}{2\sigma^2} (x - \mu)^2 \right)
& = \frac{1}{2}\exp\left(-\frac{1}{2\sigma^2} (\mu - \mu)^2 \right) \\
\Leftrightarrow
\exp\left(-\frac{1}{2\sigma^2} (x - \mu)^2 \right)
& = \frac{1}{2} \\
\Leftrightarrow
-\frac{1}{2\sigma^2} (x - \mu)^2 
& = - \ln 2 \\
\Leftrightarrow
(x - \mu)^2 
& = 2 \ln 2 \, \sigma^2 \\
\end{split}
\end{align}
\end{footnotesize}

\noindent Assuming that $g$ is centered on zero, i.e., $\mu = 0$, then yields
\begin{equation}
x^2 = 2 \ln 2 \sigma^2 \Rightarrow x_{1,2} = \pm \sqrt{2 \ln 2 } \, \sigma \\
\end{equation}
such that
\begin{equation}
f_x = |x_2 - x_1| = 2 \sqrt{2 \ln 2 } \, \sigma = \sqrt{8 \ln 2 } \, \sigma.
\end{equation}
$\hfill \Box$

\subsubsection*{Diagonal elements of $\Lambda$}

Here we show that the diagonal elements of the variance matrix of gradient components of a random field that results from the convolution of a white-noise GRF with an isotropic Gaussian convolution kernel parameterized by FWHMs are of the form specified in eq. \eqref{eq:Lambda}. To this end, we first collect a number of results on the relationship between a Gaussian random field's expectation and covariance functions and the expectation and covariance functions of a weighted integral of the Gaussian random field.  Again, we retrieve these results from \citet[pp. 22 - 26]{Abrahamsen1997}, with proofs delegated to \citet{Christakos1992}.

\subsubsection*{Integrals of random fields}

\noindent Let $X(x),x\in \mathbb{R}^n$ denote a GRF with an everywhere continuous correlation function, and let
\begin{equation}
w : \mathbb{R}^n \times \mathbb{R}^n, (x,y) \mapsto w(x,y)
\end{equation}
denote a piecewise continuous, bounded, and everywhere differentiable function. Then the Riemann integrals
\begin{equation}
Y(x) := \int_S X(s)w(x,s)\,ds \mbox{ for } x \in \mathbb{R}^n
\end{equation}
define a GRF on $S \subset \mathbb{R}^n$ . Let $m_X$ and $c_X$ denote the expectation and covariance functions of $X(x),x\in \mathbb{R}^n$, respectively. Then the expectation and covariance functions of $Y(x),x\in S$ are given in terms of the expectation and covariance functions of $X(x),x \in \mathbb{R}^n$ and $w$ by 
\begin{equation}
m_{\small Y} : S \to \mathbb{R}, 
x \mapsto m_{\small Y}(x) := \mathbb{E}(Y(x))
 = \int_S m_X(s)w(x,s)\, ds
\end{equation}
and
\begin{equation}
c_{\small Y} : S \times S \to \mathbb{R}, 
(x,y) \mapsto c_{\small Y}(x,y) := \mathbb{C}(Y(x), Y(y))
 = \int_S \int_S c_X(s,t) w(x,t) w(y,s) \, ds \,dt.
\end{equation}
In words, the expectation and covariance functions of the GRF resulting from the weighted integration of a GRF $X(x),x\in \mathbb{R}^n$ are given by the weighted integrals of the expectation and covariance functions of the GRF $X(x),x\in \mathbb{R}$, respectively. Furthermore, the gradient field of $Y$ is given by 
\begin{equation}
\nabla Y(x) := \left(\begin{matrix}
\frac{\partial}{\partial x_i} Y(x) 	\\
\end{matrix} \right)_{1 \le i \le n}
=  \left(\begin{matrix}
\int_S X(s)\frac{\partial}{\partial x_i} w(x,s) \,ds 	\\
\end{matrix} \right)_{1 \le i \le n},
\end{equation}
the  expectation function of the $i$th component of the gradient field of $Y(x),x \in S$ is given by 
\begin{equation}
\dot{m}_{Y_i} : \mathbb{R}^n \to \mathbb{R},
x \mapsto \dot{m}_{Y_i}(x) := 
\mathbb{E}\left(\frac{\partial}{\partial x_i} Y(x)\right) 
= \int_S m^{X}(s)\frac{\partial}{\partial x_i} w(x,s) \,ds 
\end{equation}
for $i = 1,...,n$, and the covariance function of the $i$th and $j$th components of the gradient field of $Y(x),x\in \mathbb{R}^n$ is given by
\begin{multline}\label{eq:cov_nabla_y}
\dot{c}_{Y_{ij}} : \mathbb{R}^n \times \mathbb{R}^n  \to \mathbb{R},
(x,y) \mapsto
\dot{c}_{Y_{ij}}(x,y) 
:= \mathbb{C}\left(\frac{\partial}{\partial x_i} Y(x), \frac{\partial}{\partial y_j} Y(y) \right) \\
= \int_S \int_S c_X(s,t) \frac{\partial}{\partial x_i} w(x,t)
						 \frac{\partial}{\partial y_j} w(y,s)
 							\,ds \,dt
\end{multline}
for $i,j = 1,...,n$. In words, the expectation and covariance functions of the GRF $Y(x),x \in S$ can be evaluated based on the expectation and covariance functions of the GRF $X(x), x \in S$ and the partial derivatives of the weighting function $w$. As above, \eqref{eq:cov_nabla_y} provides an explicit expression for the variance matrix of the GRF $Y(x),x\in S$'s gradient components for the case that the GRF is constructed by means of a weighted integral of a GRF $X(x),x\in \mathbb{R}^n$. Specifically, let  the variance matrix of the partial derivatives of $Y$ at location $x \in S$ be given by 
\begin{equation}
\mathbb{C}(\nabla Y(x)) := 
\left(\begin{matrix}
\mathbb{C}\left(\frac{\partial}{\partial x_i}Y(x), \frac{\partial}{\partial x_j}Y(x)\right) 								
\end{matrix}\right)_{1 \le i,j \le n}.
\end{equation}
Then, from \eqref{eq:cov_nabla_y}, we have 
\begin{equation}\label{eq:var_nabla_y}
\mathbb{C}\left(\frac{\partial}{\partial x_i}Y(x), \frac{\partial}{\partial x_j}Y(x)\right) 
= \int_S \int_S c_X(s,t) \frac{\partial}{\partial x_i} w(x,t)
						 \frac{\partial}{\partial x_j} w(x,s)
 							\,ds \,dt.
\end{equation}.

\subsubsection*{Proof for the diagonal elements in eq. \eqref{eq:Lambda}}

Let $w_g$ denote an isotropic \textit{Gaussian convolution kernel}, i.e.,
\begin{equation}\label{eq:gauss_conv_kern}
w_g : \mathbb{R}^3 \times \mathbb{R}^3  \to \mathbb{R}_{>0}, 
(x,y) \mapsto w_g(x,y)
:= \prod_{k=1}^3 g_{\sigma_k}(x_k, y_k)
\end{equation}
where
\begin{equation}
g_{\sigma_k} : \mathbb{R} \times \mathbb{R} \to \mathbb{R}_{>0},
(x_k,y_k) \mapsto g_{\sigma_k}(x_k,y_k)
:= (2\pi)^{-\frac{1}{2}}(\sigma_k)^{-1}
\exp\left(-\frac{1}{2\sigma_k^2}(x_k - y_k)^2\right). 
\end{equation}
Further, let $W(x),x \in \mathbb{R}^3$ denote a stationary \textit{white-noise} GRF, i.e.,  GRF with expectation function
\begin{equation}
m_W(x) := 0  
\end{equation}
and
\begin{equation}
c_W(x,y) :=
\begin{cases}
\sigma_W^2 := \frac{1}{2}(4 \pi)^{\frac{3}{2}}\sigma_1 \sigma_2 \sigma_3 & \mbox{ for } x = y \\
0 & \mbox{ for } x \neq y.
\end{cases}
\end{equation}
Finally, let
\begin{equation}
Y(x) = \int_{\mathbb{R}^3} W(s)w_g(x,s) \,ds
\end{equation}
denote the GRF that results from the convolution of $W(x),x \in \mathbb{R}^3$ with the Gaussian convolution kernel $w_g$. With \eqref{eq:var_nabla_y}, the entries of the variance matrix of the gradient field of $Y(x),x\in \mathbb{R}^3$ are then given by
\begin{equation}\label{eq:var_nabla_wn_gauss_conv}
\mathbb{C}\left(\frac{\partial}{\partial x_i}Y(x), \frac{\partial}{\partial x_j}Y(x)\right) 
= \int_{\mathbb{R}^3} \int_{\mathbb{R}^3} c_W(s,t) \frac{\partial}{\partial x_i} w_g(x,t)
						 \frac{\partial}{\partial x_j} w_g(x,s)
 							\,ds \,dt.
\end{equation}
To show that the diagonal entries of $\mathbb{V}(\nabla Y(x))$ are of the form specified in eq. \eqref{eq:Lambda} we are thus led in step (I) to evaluate the partial derivatives of the convolution kernel $w_g(x,y)$ with respect to its first argument and in step (II) to evaluate the integral eq. \eqref{eq:var_nabla_wn_gauss_conv}. We show below that this results in 
\begin{equation}
\mathbb{V}\left(\frac{\partial}{\partial x_d}Y(x)\right)  
= \frac{1}{2 \sigma_d^2} \mbox{ for } d = 1,2,3. 
\end{equation}
In step (III), we then substitute the variance parameter of $w_g$ by its FWHMs parameterization and see that indeed, 
\begin{equation}
\mathbb{V}\left(\frac{\partial}{\partial x_d}Y(x)\right)  
= 4 \ln 2 f_{x_d}^{-2}.
\end{equation}

\vspace{2mm} 
\noindent (I) We have

\begin{footnotesize}
\begin{align}\label{eq:partial_g_w}
\begin{split}
\frac{\partial}{\partial x_i} w_g(x,y)
& = (2\pi)^{-\frac{3}{2}}(\sigma_1 \sigma_2 \sigma_3)^{-1}
\frac{\partial}{\partial x_i}  \prod_{k=1}^3 \exp\left(-\frac{1}{2\sigma_k^2}(x_k - y_k)^2\right) \\
& = (2\pi)^{-\frac{3}{2}}(\sigma_1 \sigma_2 \sigma_3)^{-1}
\frac{\partial}{\partial x_i} \exp\left(-\frac{1}{2\sigma_k^2} \sum_{k=1}^3(x_k - y_k)^2\right) \\
& = (2\pi)^{-\frac{3}{2}}(\sigma_1 \sigma_2 \sigma_3)^{-1}  \exp\left(-\frac{1}{2\sigma_k^2} \sum_{k=1}^3(x_k - y_k)^2\right)
\frac{\partial}{\partial x_i} \left(-\frac{1}{2\sigma_i^2} (x_i - y_i)^2\right) \\
& = - w_g(x,y)\left(\frac{x_i - y_i}{\sigma_i^2} \right).
\end{split}
\end{align}
\end{footnotesize}

\vspace{2mm} 
\noindent (II) Substitution of \eqref{eq:partial_g_w} in eq. \eqref{eq:var_nabla_wn_gauss_conv} then yields

\begin{footnotesize}
\begin{align}
\begin{split}
\mathbb{C}\left(\frac{\partial}{\partial x_i}Y(x), \frac{\partial}{\partial x_j}Y(x)\right) 
& = \int_{\mathbb{R}^3} \int_{\mathbb{R}^3} c_W(s,t) 
						\left(\frac{x_i - t_i}{\sigma_i^2}\right) w_g(x,t)
						\left(\frac{x_j - s_j}{\sigma_j^2}\right) w_g(x,s)
 						\,ds \,dt \\
& =  \int_{\mathbb{R}^3} c_W(t,t) 
						\left(\frac{x_i - t_i}{\sigma_i^2}\right) w_g(x,t)
						\left(\frac{x_j - t_j}{\sigma_j^2}\right) w_g(x,t)
 						\,dt \\
& =  \sigma_W^2 \int_{\mathbb{R}^3}  
						\left(\frac{x_i - t_i}{\sigma_i^2}\right) w_g(x,t)
						\left(\frac{x_j - t_j}{\sigma_j^2}\right) w_g(x,t)
 						\,dt. \\
\end{split}
\end{align}
\end{footnotesize}

\noindent We next consider the case $i = j = 1$, i.e., the first  diagonal elements of the variance matrix of the partial derivatives of $Y(x),x\in \mathbb{R}^3$. We have

\begin{footnotesize}
\begin{align}
\begin{split}
\mathbb{V}\left(\frac{\partial}{\partial x_1}Y(x)\right) 
& =  \sigma_W^2 \int_{\mathbb{R}^3}  
						\left(\frac{x_1 - t_1}{\sigma_1^2}\right)^2 w_g^2(x,t)
 						\,dt \\
& =  \sigma_W^2 \int_{\mathbb{R}^3}  
						\left(\frac{x_1 - t_1}{\sigma_1^2}\right)^2 w_g^2(x,t)
 						\,dt \\
& =  \sigma_W^2 \int_{\mathbb{R}^3}  
						\left(\frac{x_1 - t_1}{\sigma_1^2}\right)^2 \prod_{i=k}^3g_{\sigma_k}^2(x_k,t_k)
 						\,dt \\
& =  \sigma_W^2 
\int_{\mathbb{R}}\left(\frac{x_1 - t_1}{\sigma_1^2}\right)^2 g_{\sigma_1}^2(x_1,t_1) \,dt_1 
\int_{\mathbb{R}} g_{\sigma_2}^2(x_2,t_2) \,dt_2 
\int_{\mathbb{R}} g_{\sigma_3}^2(x_3,t_3) \,dt_3. 
\end{split}
\end{align}
\end{footnotesize}

\noindent We consider the remaining integrals in reverse order. For the latter two integrals, we have with $i = 2,3$ (cf. \citet[][eqs. 9- 13]{Jenkinson2000})

\begin{footnotesize}
\begin{align}
\begin{split}
\int_{\mathbb{R}} g_{\sigma_i}^2(x_i,t_i) \,dt_i 
& = \int_\mathbb{R} \left(\frac{1}{\sqrt{2\pi \sigma_i^2}}\exp\left(-\frac{1}{2\sigma_i^2}(x_i - t_i)^2\right)\right)^2\,dt_i \\
& = \frac{1}{2\pi \sigma_i^2} \int_\mathbb{R} \exp \left(-\frac{1}{\sigma_i^2}(x_i - t_i)^2 \right)  dt_i \\
& = \frac{1}{2}(\pi \sigma_i^2)^{-1} \int_\mathbb{R} \exp \left(-\frac{1}{\sigma_i^2}(t_i - x_i)^2 \right)  dt_i \\
& = \frac{1}{2}(\pi \sigma_i^2)^{-1}(\pi \sigma_i^2)^{\frac{1}{2}} \\
& = \frac{1}{\sqrt{4 \pi \sigma_i^2}}.
\end{split}
\end{align}
\end{footnotesize}

\noindent For the first integral, we have

\begin{footnotesize}
\begin{align}
\begin{split}
\int_{\mathbb{R}}\left(\frac{x_1 - t_1}{\sigma_1^2}\right)^2 g_{\sigma_1}^2(x_1,t_1) \,dt_1
& = \frac{1}{\sigma_1^4} \int_{\mathbb{R}}(t_1 - x_1)^2 g_{\sigma_1}^2(x_1,t_1) \,dt_1 \\
& = \frac{1}{\sigma_1^4} \left(\int_{\mathbb{R}}t_1^2 g_{\sigma_1}^2(x_1,t_1) \,dt_1
							-  2x_1\int_{\mathbb{R}}t_1 g_{\sigma_1}^2(x_1,t_1) \,dt_1 
							+  x_1^2\int_{\mathbb{R}} g_{\sigma_1}^2(x_1,t_1) \,dt_1 \right) \\
& = \frac{1}{\sigma_1^4} \left(\int_{\mathbb{R}}t_1^2 g_{\sigma_1}^2(x_1,t_1) \,dt_1
							-  2x_1\int_{\mathbb{R}}t_1 g_{\sigma_1}^2(x_1,t_1) \,dt_1 
							+  \frac{x_1^2}{\sqrt{4 \pi \sigma_1^2}} \right).
\end{split}
\end{align}
\end{footnotesize}

\noindent For the remaining terms, we have

\begin{footnotesize}
\begin{align}
\begin{split}
2x_1\int_{\mathbb{R}}t_1 g_{\sigma_1}^2(x_1,t_1) \,dt_1
& = 2x_1\int_\mathbb{R} t_1 \left(\frac{1}{\sqrt{2\pi \sigma_1^2}}\exp\left(-\frac{1}{2\sigma_1^2}(x_1 - t_1)^2\right)\right)^2\,dt_1 \\
& = 2x_1\frac{1}{2\pi \sigma_1^2}
\int_\mathbb{R} t_1 \exp\left(-\frac{1}{\sigma_1^2}(t_1 - x_1)^2\right)\,dt_1 \\
& = 2x_1\frac{1}{2\pi \sigma_1^2}
\frac{\sqrt{\pi \sigma_1^2}}{\sqrt{\pi \sigma_1^2}}
\int_\mathbb{R} t_1 \exp\left(-\frac{1}{\sigma_1^2}(t_1 - x_1)^2\right)\,dt_1 \\
& = 2x_1\frac{\sqrt{\pi \sigma_1^2}}{2\pi \sigma_1^2}
\int_\mathbb{R} t_1 \frac{1}{\sqrt{ \pi \sigma_1^2}} \exp\left(-\frac{1}{\sigma_1^2}(t_1 - x_1)^2\right)\,dt_1 \\
& = \frac{1}{\sqrt{4\pi \sigma_1^2}} 2x_1^2 \\
\end{split}
\end{align}
\end{footnotesize}

\noindent and

\begin{footnotesize}
\begin{align}
\begin{split}
\int_{\mathbb{R}}t_1^2 g_{\sigma_1}^2(x_1,t_1) \,dt_1
& = \int_\mathbb{R} t_1^2 \left(\frac{1}{\sqrt{2\pi \sigma_1^2}}\exp\left(-\frac{1}{2\sigma_1^2}(x_1 - t_1)^2\right)\right)^2\,dt_1 \\
& = \frac{1}{2\pi \sigma_1^2} \int_\mathbb{R} t_1^2 \exp\left(-\frac{1}{\sigma_1^2}(x_1 - t_1)^2\right)\,dt_1 \\
& = \frac{1}{2\pi \sigma_1^2}\frac{\sqrt{\pi \sigma_1^2}}{\sqrt{ \pi \sigma_1^2}} \int_\mathbb{R} t_1^2 \exp\left(-\frac{1}{\sigma_1^2}(x_1 - t_1)^2\right)\,dt_1 \\
& = \frac{\sqrt{\pi \sigma_1^2}}{2\pi \sigma_1^2}\int_\mathbb{R} t_1^2 \frac{1}{\sqrt{ \pi \sigma_1^2}} \exp\left(-\frac{1}{\sigma_1^2}(x_1 - t_1)^2\right)\,dt_1 \\
& = \frac{1}{\sqrt{4\pi \sigma_1^2}}(\sigma_1^2 + x_1^2). \\
\end{split}
\end{align}
\end{footnotesize}

\noindent Substitution then yields

\begin{footnotesize}
\begin{align}
\begin{split}
\int_{\mathbb{R}}\left(\frac{x_1 - t_1}{\sigma_1^2}\right)^2 g_{\sigma_1}^2(x_1,t_1) \,dt_1
& = \frac{1}{\sigma_1^4} 
\left(\frac{1}{\sqrt{4\pi \sigma_1^2}}(\sigma_1^2 + x_1^2)
 	-  \frac{1}{\sqrt{4\pi \sigma_1^2}} (2 x_1^2)
	+  \frac{}{\sqrt{4 \pi \sigma_1^2}} x_1^2 \right) \\
& = \frac{1}{\sigma_1^4} 
\left(\frac{1}{\sqrt{4\pi \sigma_1^2}}(\sigma_1^2 + x_1^2 - 2x_1^2 + x_1^2)\right) \\
& = \frac{1}{\sigma_1^4} 
\left(\frac{\sigma_1^2}{\sqrt{4\pi \sigma_1^2}}\right) \\
& = \frac{1}{\sigma_1^2} 
\frac{1}{\sqrt{4\pi \sigma_1^2}}.
\end{split}
\end{align}
\end{footnotesize}

\noindent We thus obtain

\begin{footnotesize}
\begin{align}
\begin{split}
\mathbb{V}\left(\frac{\partial}{\partial x_1}Y(x)\right) 
& =  \sigma_W^2 
\frac{1}{\sigma_1^2} 
\frac{1}{\sqrt{4\pi \sigma_1^2}}
\frac{1}{\sqrt{4\pi \sigma_2^2}}
\frac{1}{\sqrt{4\pi \sigma_3^2}}\\
& =  \sigma_W^2 
\frac{1}{\sigma_1^2} 
\frac{1}{(4\pi)^{\frac{3}{2}} \sigma_1 \sigma_2 \sigma_3}\\
& = 
\frac{1}{2 \sigma_1^2} 
\frac{(4\pi)^{\frac{3}{2}} \sigma_1 \sigma_2 \sigma_3}{(4\pi)^{\frac{3}{2}} \sigma_1 \sigma_2 \sigma_3}\\
& = 
\frac{1}{2 \sigma_1^2}.
\end{split}
\end{align}
\end{footnotesize}

\noindent The equivalent line of reasoning yields
\begin{small}
\begin{align}
\begin{split}
\mathbb{V}\left(\frac{\partial}{\partial x_2}Y(x)\right) 
& = 
\frac{1}{2 \sigma_2^2} 
\end{split}
\end{align}
\end{small}
and
\begin{small}
\begin{align}
\begin{split}
\mathbb{V}\left(\frac{\partial}{\partial x_3}Y(x)\right) 
& = 
\frac{1}{2 \sigma_3^2}.
\end{split}
\end{align}
\end{small}

\vspace{2mm} 
\noindent (III) With the above and the form of the FWHM of a univariate Gaussian eq. \eqref{eq:gauss_fwhm}, we have 
\begin{small}
\begin{align}
\begin{split}
\mathbb{V}\left(\frac{\partial}{\partial x_d}Y(x)\right) 
& = \frac{1}{2 \sigma_d^2} 											\\
& = \frac{1}{2 \left((8 \ln 2)^{-\frac{1}{2}} f_{x_d}\right)^2} 	\\
& = \frac{1}{2 (8 \ln 2)^{-1} f_{x_d}^2}							\\ 
& = \frac{1}{2} 8 \ln 2 f_{x_d}^{-2} 								\\
& = 4 \ln 2 f_{x_d}^{-2} 
\end{split}
\end{align}
\end{small}
for $d = 1,2,3$.

$\hfill \Box$

\subsection{Miscellaneous proofs}\label{sec:miscellaneous_proofs}

\subsubsection*{Proof of eq. \eqref{eq:exp_volume}}\label{sec:proof_exp_volume}
Following  \citet[][Section 1.7, pp. 18 - 19]{Adler1981}, eq. \eqref{eq:exp_volume} can be seen as follows: appreciating that the Lebesgue volume of the excursion set can be written using the indicator function as
\begin{equation}
\lambda(E_u) = \int_S 1_{[u,\infty[} \left(X(x)\right) \,d\lambda(x), 
\end{equation}
with respect to the Lebesgue measure, its expected value evaluates to
\begin{align}
\begin{split}
\mathbb{E}(\lambda(E_u)) 
& = \int_S \left(\int_S 1_{[u,\infty[}  (X(x)) \,d\lambda(x) \right) \,d\mathbb{P}(X(x)) \\
& = \int_S 1 \,d \mathbb{P}(X(x) \ge u) \\
& = \int_S \mathbb{P}(X(x) \ge u) \,d\lambda(x). \\
\end{split}
\end{align}
In words, integrating the indicator function for the set of values $x$ for which $X(x) \ge u$ is equivalent to integrating the indicator function of the set $S$, i.e., the constant function $1$ on $S$. This is in turn identical to integrating $\mathbb{P}(X(x) \ge u)$ with respect to Lebesgue measure. Now, because $X$ is stationary, we have
\begin{align}
\begin{split}
\mathbb{E}(\lambda(E_u)) 
& = \int_S \mathbb{P}(X(0) \ge u) \,d\lambda(x) \\
& = \int_S 1_{S}(x) \mathbb{P}(X(0) \ge u) \,d\lambda(x)
\end{split}
\end{align}
and with the linearity of the integral, it follows that
\begin{align}
\begin{split}
\mathbb{E}(\lambda(E_u)) 
& = \int_S 1_{S}(x) \,d\lambda(x) \cdot \mathbb{P}(X(0) \ge u) \\
& = \lambda(S) \cdot \mathbb{P}(X(0) \ge u) \\
& = \lambda(S)(1 - F_X(u)).
\end{split}
\end{align}

$\hfill \Box$

\subsubsection*{Proof of eq. \eqref{eq:k_u_distribution}}\label{sec:proof_k_u_distribution}
Recall that for a real-valued random variable $X$ with probability density function $f_X$ and an invertible and differentiable function $g : \mathbb{R} \to \mathbb{R}$ the probability density function $f_Y$ of the (real-valued) random variable $Y := g(X)$ can be evaluated as
\begin{equation}\label{eq:cov_theorem}
f_Y(y) = \frac{f_X(g^{-1}(y))}{|g'(g^{-1}(y))|},
\end{equation}
where $g'$ and $g^{-1}$ denote the derivative and inverse of $g$, respectively. In the current scenario, the result by \citet{Nosko1969, Nosko1969a, Nosko1970} and reiterated in  \citet[cf.][p. 212]{Friston1994} and \citet[cf.][p.587]{Cao1999} states that the size of a connected component of the excursion set $K_u^{\frac{2}{D}}$ has an exponential distribution with expectation parameter $\kappa$. In light of \eqref{eq:cov_theorem}, we thus have
\begin{equation}
f_{K_u^{\frac{2}{D}}}\left(k^{\frac{2}{D}}\right)
:= \kappa \exp \left(-\kappa  k^{\frac{2}{D}}\right).
\end{equation}
We define
\begin{equation}
g : \mathbb{R} \to \mathbb{R}, x \mapsto g(x) := x^{\frac{D}{2}}
\end{equation}
such that the inverse of $g$ is given by
\begin{equation}
g^{-1} : \mathbb{R} \to \mathbb{R}, y \mapsto g^{-1}(y) = y^{\frac{2}{D}}, 
\end{equation}
because then
\begin{equation}
g^{-1}(g(x)) 
= g^{-1}\left(x^{\frac{D}{2}}\right) 
= \left(x^{\frac{D}{2}}\right)^{\frac{2}{D}} = x.
\end{equation}
Note that
\begin{equation}
g': \mathbb{R} \to \mathbb{R}, x\mapsto g'(x) = \frac{D}{2} x ^{\frac{D}{2} - 1}.
\end{equation}
Substitution in eq. \eqref{eq:cov_theorem} then yields for the probability density function $f_{K_u}(k)$ of $K_u$:

\begin{small}
\begin{align}
\begin{split}
f_{K_u}(k) 
& = \frac{f_{K_u^{\frac{2}{D}}} \left(g^{-1}(k) \right)}{|g'(g^{-1}(k))|} \\
& = \frac{\kappa \exp \left(-\kappa g^{-1}(k)\right)}{\frac{D}{2}(g^{-1}(k))^{\frac{D}{2} -1 }} \\
& = \frac{\kappa \exp \left(-\kappa k^{\frac{2}{D}} \right)}{\frac{D}{2} \left(k^{\frac{2}{D}} \right)^{\frac{D}{2} - 1 }} \\
& = \frac{2\kappa}{D} \exp \left(-\kappa k^{\frac{2}{D}} \right) k^{-\left(\frac{2}{D} \left(\frac{D}{2} - 1 \right) \right)}\\
& = \frac{2\kappa}{D}  \exp \left(-\kappa k^{\frac{2}{D}} \right) k^{-\left(1 - \frac{2}{D} \right)}\\
& = \frac{2\kappa}{D} k^{\frac{2}{D}-1} \exp \left(-\kappa k^{\frac{2}{D}} \right). \\
\end{split}
\end{align}
\end{small}

$\hfill \Box$
 
\subsubsection*{Proof of eq. \eqref{eq:k_u_cdf}}\label{sec:proof_k_u_cdf}
By assuming that the distribution of $K_u$ is absolutely continuous, we can prove eq. \eqref{eq:k_u_cdf} by showing that the functional form of the cumulative density function $F_{K_u}$ is the anti-derivative of the probability density function $f_{K_u}$ defined in eq. \eqref{eq:k_u_distribution}. We have

\begin{small}
\begin{align}
\begin{split}
F_{K_u}'(k) 
& = \frac{d}{dk} \left(1 - \exp \left(-\kappa k^{\frac{2}{D}} \right)\right) \\
& = - \frac{d}{dk} \exp \left(-\kappa k^{\frac{2}{D}} \right) \\
& = - \exp \left(-\kappa k^{\frac{2}{D}} \right) \frac{d}{dk} \left(-\kappa k^{\frac{2}{D}} \right) \\
& = - \exp \left(-\kappa k^{\frac{2}{D}} \right) \left(-\kappa \frac{2}{D} k^{\frac{2}{D} -1 } \right) \\
& =  \frac{2\kappa}{D} k^{\frac{2}{D} -1} \exp \left(-\kappa k^{\frac{2}{D}} \right)  \\
& = f_{K_u}(k)
\end{split}
\end{align}
\end{small}

\noindent and see that indeed, eq. \eqref{eq:k_u_cdf} specifies the anti-derivative of eq. \eqref{eq:k_u_distribution}.

$\hfill \Box$

\subsubsection*{Proof of eq. \eqref{eq:p_q_ku_poiss}}\label{sec:proof_p_q_ku_poiss}
By the definition of marginal and conditional probabilities, we have
\begin{align}
\begin{split}
\mathbb{P}(C_{\ge k, u} = i) 
& = \sum_{j = 1}^{\infty} \mathbb{P}(C_u = j,C_{\ge k,u} = i) \\
& = \sum_{j = 1}^{\infty} \mathbb{P}(C_u = j)\mathbb{P}(C_{\ge k,u} = i|C_u = j).  \\
\end{split}
\end{align}
Substitution of the Poisson and Binomial forms for $\mathbb{P}(C_u = j)$ and $\mathbb{P}(C_{\ge k, u} = i |C_u = j)$ and, for ease of notation, defining $a:=\mathbb{E}(C_u)$ and $b := \mathbb{P}(K_u \ge k)$, then yields
\begin{align}
\begin{split}
\mathbb{P}(C_{\ge k,u} = i) 
& = \sum_{j = 1}^{\infty} 
\frac{a^j \exp(-a)}{j!} 
\frac{j!}{i!(j-i)!}b^i (1 - b)^{j-i}.  \\
\end{split}
\end{align}
Definition of $v := j - i$ (and hence $j = v + i$) then results in
\begin{align}
\begin{split}
\mathbb{P}(C_{\ge k,u} = i) 
& = \sum_{v + i = 1}^{\infty} \frac{a^{v+i} \exp(-a)}{(v+i)!}\frac{(v+i)!}{i!v!}b^i (1 - b)^v  \\
& = \sum_{v = 0}^{\infty}\frac{a^v a^i\exp(-a)}{i!v!}b^i (1 - b)^v  \\
& = \frac{1}{i!}a^i b^i \exp(-a) \sum_{v = 0}^{\infty}\frac{a^v}{v!}(1 - b)^v  \\
& = \frac{1}{i!}(ab)^i \exp(-a) \sum_{v = 0}^{\infty}\frac{(a(1-b))^v}{v!}  \\
& = \frac{1}{i!}(ab)^i \exp(-a) \exp(a(1-b)) \\
& = \frac{1}{i!}(ab)^i \exp(-a + a - ab) \\
& = \frac{1}{i!}(ab)^i \exp(-ab), \\
\end{split}
\end{align}
where the fifth equation follows with  the series definition of the exponential function
\begin{equation}
\exp(x) := \sum_{n = 0}^{\infty} \frac{x^n}{n!}.
\end{equation}
We thus obtain
\begin{align}
\begin{split}
\mathbb{P}(C_{\ge k,u} = i) 
& = \frac{\mathbb{E}(C_u)\mathbb{P}(K_u \ge k)^i \exp(-\mathbb{E}(C_u)\mathbb{P}(K_u \ge k))}{i!} \\
& = \frac{\lambda_{C_{\ge k,u}}^i \exp \left(-\lambda_{C_{\ge k,u}} \right)}{i!} .
\end{split}
\end{align}
$\hfill \Box$

The approximation of $\mathbb{P}(X_m \ge \tilde{t}_i)$ by $\mathbb{P}(C_{\ge0,\tilde{t}_i} \ge 1)$ derives from the approximation 
\begin{equation}
\mathbb{P}(C_{\ge 0,\tilde{t}_i} \ge 1) 
= 1 - \exp\left(-\mathbb{E}(C_{\tilde{t}_i})\right),
\approx \mathbb{E}(C_{\tilde{t}_i})
\end{equation}
because with eqs. \eqref{eq:exp_c_u_approx} and \eqref{eq:maximum_distribution} it then follows that
\begin{equation}
\mathbb{E}(C_{\tilde{t}_i}) \approx \mathbb{E}(\chi(E_{\tilde{t}_i})) \approx \mathbb{P}(X_m \ge \tilde{t}_i).
\end{equation}
We are thus interested in showing that
\begin{equation}
1 - e^{-x} \approx x.
\end{equation}
To this end, recall that
\begin{equation}\label{eq:exp_approx}
e^x := \sum_{i=0}^\infty \frac{x^i}{i!} = \frac{x^0}{0!} + \frac{x^1}{1!} + \sum_{i=2}^\infty \frac{x^i}{i!}.
\end{equation}
Neglecting the series term on the right-hand side of eq.  \eqref{eq:exp_approx} thus yields the approximation
\begin{equation}
e^x \approx 1 + x
\end{equation}
and hence
\begin{equation}
e^x \approx 1 + x
\Leftrightarrow
-x \approx 1 - e^x
\Leftrightarrow
-(-x) \approx 1 - e^{-x}
\Leftrightarrow
x \approx 1 - e^{-x}. 
\end{equation}
$\hfill \Box$

\newpage
\renewcommand{\thesection}{Supplement S2}
\setcounter{definition}{0}
\setcounter{equation}{0}

\section{Hypothesis testing and FWER control}\label{sec:hypothesis_testing_fwer_control}
In this Section we review the formal foundations of multiple testing theory with a focus of controlling the FWER using maximum statistics. To establish intuition and notation, we first consider the single test scenario.

\subsubsection*{The single test scenario}\label{sec:single_test_scenario}
\paragraph{Probabilistic model.} To introduce the notion of a single test, we consider a parametric probabilistic model $P_\theta(Y)$ that describes the probability distribution of a random entity (i.e., a random variable or a random vector) $Y$ and that is governed by a parameter $\theta \in \Theta$. The random entity $Y$ models \textit{data} and is assumed to take on values $y \in \mathbb{R}^n, n \ge 1$. Note that we do not consider the parameter $\theta$ to be a random entity and thus develop the following theory against the background of the classical frequentist scenario. 

\paragraph{Test hypotheses.} In test scenarios, the parameter space $\Theta$  is partitioned into two disjoint subsets, denoted by $\Theta_0$ and $\Theta_1$, such that $\Theta = \Theta_0 \cup \Theta_1$ and $\Theta_0 \cap \Theta_1 = \emptyset$. A \textit{test hypothesis} is a statement about the parameter  governing $P_\theta(Y)$ in relation to these parameter space subsets. Specifically, the statement
\begin{equation}\label{eq:H0}
\theta \in \Theta_0 \Leftrightarrow H = 0 
\end{equation}
is referred to as the \textit{null hypothesis} and the statement
\begin{equation}\label{eq:H1}
\theta \in \Theta_1 \Leftrightarrow H = 1 
\end{equation}
is referred to as the \textit{alternative hypothesis}. Note that we are concerned with the Neyman-Pearson hypothesis testing framework and thus assume that null and alternative hypotheses always exist in an explicitly defined manner. A number of things are noteworthy. First, a statistical hypothesis is a statement about the parameter of a probabilistic model. In the following, we will use the subscript notations $P_{\Theta_0}$ and $P_{\Theta_1}$ to indicate that the parameter $\theta$ of the probabilistic model $P_\theta$ is an element of $\Theta_0$ or $\Theta_1$, respectively. Second, the term null hypothesis is not necessarily the statement that some parameter assumes the value zero, even if this is often the case in practice. Rather, the null hypothesis in a statistical testing problem is the statement about the parameter one is willing to nullify, i.e., reject. Finally, the expressions $H = 0$ and $H = 1$ are not conceived as realizations of a random variable and hence hypothesis-conditional  probability statements are not meaningful. The statements $H = 0$ and $H = 1$ are merely equivalent expressions for $\theta \in \Theta_0$ and $\theta \in \Theta_1$, respectively: $H=0$ refers to the true, but unknown, state of the world that the null hypothesis is true and the alternative hypothesis is false ($\theta \in \Theta_0$), and $H=1$ refers to the true, but unknown, state of the world that the alternative hypothesis is true and the null hypothesis is false ($\theta \in \Theta_1$). In general, hypotheses can be classified as \textit{simple} or \textit{composite}. A \textit{simple hypothesis} refers to a subset of parameter space which contains a single element, for example $\Theta_0 := \{\theta_0\}$. A \textit{composite hypothesis} refers to a subset of parameter space which contains more than one element, for example $\Theta_0 := \mathbb{R}_{\leq0}$. The commonly encountered null hypothesis $\Theta_0 = \{0\}$, also referred to as \textit{nil  hypothesis}, is an example for a simple hypothesis.

\paragraph{Tests.} Given the test hypotheses scenario introduced above, a \textit{test} is defined as a mapping from the data outcome space to the set $\{0,1\}$, formally
\begin{equation}\label{eq:phi}
\phi(Y = \cdot) : \mathbb{R}^n \to \{0,1\}, y \mapsto \phi(Y = y).
\end{equation}
Here, the test value $\phi(Y = y) = 0$ represents the act of not rejecting null hypothesis, while the test value $\phi(Y = y) = 1$ represents the act of rejecting the null hypothesis. Rejecting the null hypothesis is equivalent to accepting the alternative hypothesis, and accepting the null hypothesis is equivalent to rejecting the alternative hypothesis. In the following and in the main text, we suppress the notational dependence of $\phi(Y = \cdot)$ on $y$ and write $\phi(Y)$ instead. Because $Y$ is a random entity, the expression $\phi(Y)$ is also a random entity. All tests $\phi(Y)$ considered in the current study involve the composition of a \textit{test statistic} 
\begin{equation}\label{eq:test_statistic_gamma}
\gamma(Y) : \mathbb{R}^n \to \mathbb{R},
\end{equation}
where $ \mathbb{R}$ models the test statistic's outcome space, and a subsequent \textit{decision rule} 
\begin{equation}
\delta(\gamma(Y)) :  \mathbb{R} \to \{0,1\},
\end{equation}
such that the test can be written as
\begin{equation}\label{eq:phi_concatenated}
\phi(Y) = \delta(\gamma(Y)) :  \mathbb{R}^n \to \{0,1\}.
\end{equation}
Note that, as for the test, we suppress the dependencies of $\gamma(Y)$ and $\delta(\gamma(Y))$ on $y\in \mathbb{R}^n$, such that both $\gamma(Y)$ and $\delta(\gamma(Y))$ should be read as random entities. The subset of the test statistic's outcome space for which the test assumes the value $1$ is referred to as the \textit{rejection region} of the test. Formally, the rejection region is defined as
\begin{equation}
R := \{\gamma(Y) \in  \mathbb{R} \vert \phi(Y) = 1\} \subset  \mathbb{R}.
\end{equation}
The random events $\phi(Y) = 1$ and $\gamma(Y) \in R$ are thus equivalent and associated with the same probability under $P_\theta(Y)$. In a concrete test scenario, it is hence usually the probability distribution of the test statistic that is of principal concern for assessing the test's outcome behaviour. Finally, all test decision rules considered in the context of the current study are based on the test statistic exceeding a \textit{critical value} $u \in \mathbb{R}$. By means of the indicator function, the tests considered here can thus be written 
\begin{equation}\label{eq:phi_critical_value}
\phi(Y = \cdot) : \mathbb{R}^n \to \{0,1\}, y \mapsto \phi(Y = y) 
:= 1_{\{\gamma(Y = y) \ge u \}} 
:= \begin{cases}
1, &\,\gamma(Y = y) \ge u \\ 
0, &\,\gamma(Y = y) < u.
\end{cases}
\end{equation}
Note that \eqref{eq:phi_critical_value} describes the situation of \textit{one-sided} tests. The one-sided one-sample $T$-test is a familiar example of the general test structure described by expression \eqref{eq:phi_critical_value}: using the sample mean and sample standard deviation, a realization of the random entity $Y$ is first transformed into the value of the t-statistic, whose size is then compared to a critical value in order to decide for rejecting the null hypothesis or not.

\paragraph{Tests error probabilities.} When conducting a hypothesis test as just described, two kinds of errors can occur. First, the null hypothesis can be rejected ($\phi(Y)=1$), when it is in fact true ($\theta \in \Theta_0$). This error is referred to as the \textit{Type I error}. Second, the null hypothesis may not be rejected ($\phi(Y)=0$), when it is in fact false ($\theta \in \Theta_1$). The latter error is known as the \textit{Type II error}. The probabilities of Type I and Type II errors under a given probabilistic model are central to the quality of a test: the probability of a Type I error is called the \textit{size} of the test and is commonly denoted by $\alpha \in [0,1]$. It is defined as
\begin{equation}\label{eq:def_alpha}
\alpha := P_{\Theta_0}(\phi(Y) = 1),
\end{equation}
and also routinely referred to as the \textit{Type I error rate} of the test. Its complementary probability, 
\begin{equation}
P_{\Theta_0}(\phi(Y) = 0) = 1 -\alpha,
\end{equation}
is known as the \textit{specificity} of a test. The probability of a Type II error 
\begin{equation}
P_{\Theta_1}(\phi(Y) = 0)
\end{equation}
lacks a common denomination. Its complementary probability
\begin{equation}\label{eq:def_beta}
\beta := P_{\Theta_1}(\phi(Y) = 1) 
\end{equation}
is referred to as the \textit{power} of a test. In words, the power of a test is the probability of accepting the alternative hypothesis (rejecting the null hypothesis), if $\theta \in \Theta_1$, i.e., if the alternative hypothesis is true. Note that basic introductions to test error probabilities often denote the probability of a Type II error by $\beta \in [0,1]$ and thus define power by $1 - \beta$. For our current purposes, we prefer the definition of eq. \eqref{eq:def_beta}, because it keeps the notation concise and is more coherent with common notations of test quality functions. 

\paragraph{Significance level.} It is important to distinguish between the size and the significance level of a test: a test is said to be of \textit{significance level} $\alpha' \in [0,1]$, if its size $\alpha$ is smaller than or equal to $\alpha'$, i.e., if
\begin{equation}
\alpha \le \alpha'.
\end{equation} 
If for a test of significance level $\alpha'$ it holds that $\alpha < \alpha'$, the test is referred to as a \textit{conservative test}. If for a test of significance level $\alpha'$ it holds that $\alpha = \alpha'$, the test is referred to as an \textit{exact test}. Tests with an associated significance level 
$\alpha'$ for which $\alpha > \alpha'$ are sometimes referred to as \textit{liberal tests}. Note, however, that such tests are, strictly speaking, not of significance level $\alpha'$.

\paragraph{The test quality function.} The size and the power of a test are summarized in the test's quality function. For a test $\phi(Y)$, the \textit{test quality function} is defined as
\begin{equation}\label{eq:TQF}
\begin{split}
q : \Theta \to [0,1], 
\theta \mapsto q(\theta):= \mathbb{E}_{P_\theta(Y)}(\phi(Y)).
\end{split}
\end{equation}
In words, the test quality function is a function of the probabilistic model parameter $\theta$ and assigns to each value of this parameter a value in the interval $[0,1]$. This value is given by the expectation of the test $\phi$ under the probabilistic model $P_\theta(Y)$.  The definition of the test quality function is motivated by the value it assumes for $\theta \in \Theta_0$ and $\theta \in \Theta_1$: because the random variable $\phi(Y)$ only takes on values in $\{0,1\}$, the expected value $\mathbb{E}_{P_\theta(Y)}(\phi(Y))$ is identical to the probability of the event $\phi(Y) = 1$ under $P_\theta(Y)$. Thus, for $\theta \in \Theta_0$, the test quality function returns the size of the test (eq. \eqref{eq:def_alpha}) and for $\theta \in \Theta_1$, the test quality function returns the power of the test (eq. \eqref{eq:def_beta}). 

\paragraph{The test power function.} For $\theta \in \Theta_1$, the test quality function is also is referred to as the test's \textit{power function} and is denoted by
\begin{equation}\label{eq:st_power_function}
\beta: \Theta_1 \to [0,1], \theta \mapsto \beta(\theta):= P_{\Theta_1}(\phi(Y) = 1). 
\end{equation}

\paragraph{Test construction.} In both applications and the theoretical development of statistical tests, the probability for a Type I error, i.e., the test size, is usually considered to be more important than the Type II error rate, i.e., the complement of the test's power. In effect, when designing a test, the test's size is usually fixed first, for example by deciding for a significance level such as $\alpha' = 0.05$ and its associated critical value $u_{\alpha'}$ of the test statistic (cf. eq. \eqref{eq:phi_critical_value}). In a second step, different tests or different probabilistic models are then compared in their ability to minimize the probability of the test's Type II error, i.e., maximize the test's power. For example, the celebrated Neyman-Pearson lemma states that for tests of simple hypotheses, the likelihood ratio test achieves the highest power for a given significance level over all conceivable statistical tests \citep{Neyman1933}. Inspired by current discussions about the power of tests in functional neuroimaging, in the current study we primarily target the sample size as a parameter of the probabilistic model to optimize different tests with respect to their Type II error rates given prefixed Type I error rates. 

\subsubsection*{The multiple testing scenario}\label{sec:multiple_testing}
\paragraph{Probabilistic model.} The notion of a multiple hypothesis test can be developed in analogy to the single test scenario. Like the single test scenario, the multiple testing scenario unfolds against the background of a parametric probabilistic model $P_\theta(Y)$ that describes the probability distribution of a random entity $Y$ which models observed data taking on values in $\mathbb{R}^n$. The parameter $\theta$ of the model is assumed to take values in a parameter space $\Theta$. 

\paragraph{Multiple test hypotheses.} In multiple testing scenarios comprising $m \in \mathbb{N}$ tests, the parameter space is partitioned $m$ times into disjoint subsets $\Theta_0^{(i)}$ and $\Theta_1^{(i)}$, such that $\Theta = \Theta_0^{(i)}\cup \Theta_1^{(i)}$ and  $\Theta_0^{(i)} \cap \Theta_1^{(i)} = \emptyset$ for $i \in I := \{1,...,m\}$ and $|I|=m$. In analogy to the single test case, the statements
\begin{equation}
\theta \in \Theta_{0}^{(i)} \Leftrightarrow H^{(i)} = 0 \mbox{ and }
\theta \in \Theta_{1}^{(i)} \Leftrightarrow H^{(i)} = 1 
\end{equation}
about the true, but unknown, value of the parameter $\theta$ are referred to as the $i$th \textit{null} and \textit{alternative hypothesis}, respectively. Collectively, the $m$ null hypotheses and their associated alternative hypotheses are referred to as a \textit{hypotheses family} and the set $I$ is referred to as the \textit{hypotheses index set}. In the following, we will be concerned with the following situations
\begin{itemize}[leftmargin =*]
\item all null hypotheses of the hypotheses family are true and all alternative hypotheses are false,
\item some null hypotheses of the hypotheses family are true and the remaining alternative hypotheses are true,
\item all null hypotheses of the hypotheses family are false and all alternative hypotheses are true.
\end{itemize}
For convenience, we will refer to these scenarios as the \textit{complete null hypothesis}, the \textit{partial alternative hypothesis}, and the \textit{complete alternative hypothesis}, respectively. The following notation is helpful to formally express the complete null hypothesis and complete alternative hypothesis scenarios, respectively:
\begin{equation}\label{eq:H0complete}
\theta \in \Theta_0 := \cap_{i \in I} \Theta_0^{(i)} \Leftrightarrow 
H^{(i)} = 0 \mbox{ for all } i \in I 
\end{equation}
and
\begin{equation}\label{eq:H1complete}
\theta \in \Theta_1 := \cap_{i \in I} \Theta_1^{(i)} \Leftrightarrow 
H^{(i)} = 1 \mbox{ for all } i \in I. 
\end{equation}
Note that despite the identical notation, the difference between the single test scenario null and alternative hypotheses \eqref{eq:H0} and \eqref{eq:H1}, and the multiple testing scenario complete null and complete alternative hypotheses \eqref{eq:H0complete} and \eqref{eq:H1complete} should in general be clear from the context. As above, we will use the subscript notations $P_{\Theta_0}$ and $P_{\Theta_1}$ to indicate that the parameter $\theta$ of the probabilistic model $P_\theta$ is an element of the complete null or alternative hypotheses $\Theta_0$ or $\Theta_1$, respectively. In light of expressions \eqref{eq:H0complete} and \eqref{eq:H1complete}, we denote the partial alternative hypothesis by
\begin{equation}\label{eq:theta_I_1}
\theta \in \cap_{i \in I_1} \Theta_1^{(i)} \mbox{ for } I_1 \subset I \mbox{ with } m_1 := |I_1|,
\end{equation}
and refer to $I_1$ as the \textit{alternative hypotheses index set}. Given the binary nature of the $i$th null and alternative hypothesis, it follows immediately that in the case of \eqref{eq:theta_I_1} it holds that
\begin{equation}\label{eq:theta_I_0}
\theta \in \cap_{i \in I_0} \Theta_0^{(i)} 
\mbox{ for } I_0 := I \setminus I_1 
\mbox{ with }  |I_0| = m - m_1 =: m_0.
\end{equation}
We refer to $I_0$ as the \textit{null hypotheses index set}. The ratio of the cardinality  of the alternative hypotheses index set and the cardinality of the hypotheses index set will be denoted by 
\begin{equation}\label{eq:def_lambda}
\lambda = \frac{m_1}{m}, 
\end{equation}
and will be referred to as the \textit{alternative hypotheses ratio}. Note that $\lambda = 0$ corresponds to the complete null hypothesis, whereas $\lambda = 1$ corresponds to the complete alternative hypothesis. Finally, for $\lambda \in ]0,1[$, we use the subscript notation $P_{\lambda \Theta_1}$ to indicate that the parameter $\theta$ of the probabilistic model $P_\theta$ is an element of a partial alternative hypothesis with alternative hypotheses ratio $\lambda$.

\paragraph{Multiple test.} For the multiple testing scenario, let
\begin{equation}
\phi_i(Y = \cdot ) : \mathbb{R}^n \to \{0,1\}, y \mapsto \phi_i(Y = y) \mbox{ for all } i \in I
\end{equation}
denote a test, such that $\phi_i(Y = y) = 0$ represents the act of accepting the $i$th null hypothesis and rejecting the $i$th alternative hypothesis, while $\phi_i(Y = y) = 1$ represents the act of  rejecting the $i$th null hypotheses and accepting the $i$th alternative hypothesis. Then a multiple test is a mapping
\begin{equation}\label{eq:Phi_multiple_test}
\Phi(Y = \cdot) : \mathbb{R}^n \to \{0,1\}^m, y \mapsto \Phi(Y = y) := \left(\phi_i(Y = y)\right)_{i \in I}. 
\end{equation}
A multiple test can thus be conceived as an $m$-dimensional vector of single tests $\phi_i(Y = \cdot)$, the probability distribution of which is governed by the parametric probabilistic model $P_\theta (Y)$. As in the single test scenario, we will suppress the notational dependence of $\Phi(Y = \cdot)$ on $y$ and write $\Phi(Y)$ instead. Again, because the data $Y$ is modelled as a random entity, the expression $\Phi(Y)$ should be read as a random vector. Similarly, as in the single test scenario we are only concerned with scenarios for which each constituent test $\phi_i(Y)$ of $\Phi(Y)$ is of the form
\begin{equation}\label{eq:phi_i_multiple_test}
\phi_i(Y = \cdot) : \mathbb{R}^n \to \{0,1\}, y \mapsto \phi_i(Y = y) := 1_{\{\gamma_i(Y = y) \ge u_i\}},
\end{equation} 
where 
\begin{equation}\label{eq:gamma_i_multiple_test}
\gamma_i(Y) : \mathbb{R}^n \to \mathbb{R}
\end{equation}
denotes the $i$th test statistic with $i$th rejection region
\begin{equation}\label{eq:R_i_multiple_test}
R_i := \{\gamma_i(Y) \in \mathbb{R}| \phi_i(Y) = 1\} \subset \mathbb{R},
\end{equation}
and $u_i \in \mathbb{R}$ denotes the $i$th critical value. The multiple one-sided one-sample $T$-tests commonly performed for group-level fMRI analyses are a familiar example of the general multiple test structure described by eqs. \eqref{eq:Phi_multiple_test} - \eqref{eq:R_i_multiple_test}: using voxel-specific sample means and sample standard deviations, the data $Y$, usually comprising voxel-wise participant-specific beta parameter estimate contrasts derived from first-level GLM analyses, is projected onto a set of $m$ $T$-statistics. The values of these $m$ $T$-statistics individually evaluated with respect to appropriately defined critical values, and for each of the $m$ voxels, the null hypothesis of zero activation is either rejected or not. 

\paragraph{Multiple test error probabilities.} The multiple testing scenario induces a variety of test error scenarios. While for the single test scenario there exist four possible constellations of true hypotheses and test outcomes ($\theta \in \Theta_j$ and $\phi(Y) = k$ for $j = 0,1$ and $k = 0,1$) there exist $4^m$ such constellations in the multiple testing scenario ($\theta \in \Theta_j^{(i)}$ and $\phi_i(Y) = k$ for $i = 1,...,m, j = 0,1$ and $k = 0,1$). In other words, while a single test $\phi(Y)$ may either result in either a Type I or a Type II error (or a correct result), a multiple test $\Phi(Y)$ may result in the simultaneous occurrence of Type I errors in some of its constituent single tests and Type II error in others of its constituents single tests (and correct results in the remaining single tests). This induces probabilities for the occurrence of a variety of test error scenarios and hence a variety of Type I and Type II error rates. As Type II error rates are complementary probabilities of correct rejections of null hypotheses, different Type II error rates correspond to different notions of power. In the following, we first review the most commonly considered Type I and Type II error rates in multiple testing scenarios. We then consider the family-wise error rate and its control by means of maximum statistics.

\begin{table}[t]
\renewcommand{\arraystretch}{1.3}
\begin{center}
\begin{tabular}{c|c|c|c}
							& $\phi_i(Y) = 0$	& $\phi_i(Y) = 1$	&			\\\hline
$\theta \in \Theta_0^{(i)}$ & $M_{00}$			& $M_{01}$ 			& $m_0$ 	\\\hline
$\theta \in \Theta_1^{(i)}$ & $M_{10}$			& $M_{11}$ 			& $m_1$ 	\\\hline
							& $M_{00} + M_{10}$	& $M_{01} + M_{11}$ & $m$			\\
\end{tabular}\caption{The multiple testing scenario. The numbers $m_0$ and $m_1$ of true null and alternative hypotheses $\theta \in \Theta_0^{(i)}$ and $\theta \in \Theta_1^{(i)}$, are assumed to be fixed and unknown. The outcome of the $i$th test $\phi_i(Y)$, and hence also the aggregate numbers of tests to assume either the value $0$ or $1$, $M_{ij}, i = 0,1, j = 0,1$, as well as their sums, $M_{00} + M_{10}$ and $M_{01} + M_{11}$ are random entities, all of which are governed by the parametric probabilistic model $P_\theta(Y)$ and the functional forms of the test statistics $\gamma_i, i = 1,...,m$.}
\label{tab:multiple_test_errors}
\end{center}
\end{table} 

The test error rates of multiple testing scenarios can be developed quantitatively as follows: as above, let $I_0$ and $I_1$ denote the null and alternative hypotheses index sets, respectively (cf. eqs. \eqref{eq:theta_I_0} and \eqref{eq:theta_I_1}). Note again that the binary single test scenario implies that $I = I_0 \cup I_1$ and $I_0 \cap I_1 = \emptyset$ and that it is assumed that the sets $I_0$ and $I_1$ and their respective cardinalities $m_0$ and $m_1$ are true, but unknown, entities.  Based on the probabilistic binary outcome of each test constituent $\phi_i(Y)$, the following quantities are induced at an aggregate level:
\begin{itemize}[leftmargin = *]
\item the number $M_{00}$ of tests for which $\theta \in \Theta_0^{(i)}$ and $\phi_i(Y) = 0$,
\item the number $M_{01}$ of tests for which $\theta \in \Theta_0^{(i)}$ and $\phi_i(Y) = 1$,
\item the number $M_{10}$ of tests for which $\theta \in \Theta_1^{(i)}$ and $\phi_i(Y) = 0$, and
\item the number $M_{11}$ of tests for which $\theta \in \Theta_1^{(i)}$ and $\phi_i(Y) = 1$.
\end{itemize}
The situation is summarized in \autoref{tab:multiple_test_errors}. Note that the values $m, m_0$ and $m_1$ correspond to true, but unknown, quantities, the four quantities $M_{jk}, j = 0,1, k = 0,1$ correspond to unobservable random variables, and the quantities $M_{00} + M_{10}$ and $M_{01} + M_{11}$, i.e., the total number of accepted and rejected null hypotheses, correspond to observable random variables. Commonly considered Type I error rates in this scenario are 
\begin{itemize}[leftmargin = *]
\item the \textit{family-wise error rate}, defined as the probability for the event $M_{01} \ge 0$, i.e., of one or more Type I errors,
\item the \textit{per-family error rate}, defined as the expectation of the unobservable random variable $M_{01}$, i.e., the expected number of Type I errors,
\item the \textit{per-comparison error rate}, defined as the per-family error rate divided by the number of hypotheses $m$, and
\item the \textit{false-discovery rate}, defined as the expectation of the random variable 
$M_{01}/(M_{01} + M_{11})$ if $M_{01} + M_{11} \neq 0$ and $0$ if $M_{01} + M_{11} = 0$, i.e., the expected proportion of Type I errors among the rejected null hypotheses, or 0, if no hypotheses are rejected. 
\end{itemize}
Notably, in contrast to the Type I error rate in the single test scenario (i.e., the size of a test), the Type I error rates in the multiple testing scenario refer to either probabilities (such as the family-wise error rate) or expectations of the counting random variables $M_{ij}, i =0,1,j=0,1$. In a concrete multiple testing scenario, these probabilities and expectations have to be derived based on the nature of the probabilistic model and the definition of the multiple test.

\paragraph{Multiple test construction.} As in the single test scenario, multiple tests are usually constructed to first and foremost control a chosen Type I error rate at a desired significance level 
$\alpha'$. In a second step, additional test construction measures may then be taken to achieve a desired level of a chosen power type. The random field theory-based fMRI inference framework has traditionally focussed on the family-wise error rate (FWER) as the target for Type I error rate control. In the following, we shall thus further elaborate on the definition of the FWER and establish how the distribution of the maximum statistic can be utilized for its control. 

\paragraph{Maximum statistic-based FWER control.} As introduced above, the FWER of a multiple test is defined as the probability of one or more Type I errors. More formally, let $\Phi(Y) = (\phi_i(Y))_{i \in I}$ denote a multiple test with with hypotheses index set $I$ and null hypotheses index set $I_0 \subseteq I, I_0 \neq \emptyset$. Then the FWER is defined as the probability
\begin{equation}\label{eq:def_fwer}
\alpha_{\tiny \mbox{FWE}} := P_{\cap_{i \in I_0} \Theta_0^{(i)}}
\left(\cup_{i \in I_0} \phi_i(Y) = 1 \right). 
\end{equation}
This expression is to be understood as follows: clearly, the FWER refers to the probability of events $\phi_i(Y) = 1$ under the probabilistic model for the case that at least one null hypothesis holds true, i.e., $I_0 \neq \emptyset$. More specifically, the intersection subscript $\cap_{i \in I_0} \Theta_0^{(i)}$ qualifies that the parameter of the probabilistic model is such, that all null hypotheses with indices in the set $I_0 \subseteq I, I_0 \neq \emptyset$ hold. Complementary, the union statement $\cup_{i \in I_0} \phi_i(Y) = 1$ implies that the event $\phi_{i_1}(Y) = 1$ and/or the event $\phi_{i_2}(Y) = 1$, ..., and/or the event  $\phi_{i_{m_0}}(Y) = 1$ with $i_j \in I_0$ for $j = 1,2,...,m_0$ occurs, i.e., that at least one, but possible more, events $\phi_i(Y) = 1$ with $i \in I_0$ occurs. This is equivalent to the probability of the event $M_{01} \ge 0$ as considered above. In analogy to the significance level in the single test scenario, a multiple test $\Phi(Y)$ is then said to be of \textit{family-wise significance level} $\alpha'_{\tiny \mbox{FWE}}$, if its FWER is equal to or smaller than  $\alpha'_{\tiny \mbox{FWE}}$, i.e., if
\begin{equation}
\alpha_{\tiny \mbox{FWE}} \le \alpha'_{\tiny \mbox{FWE}}.
\end{equation}
Equivalently, such a test is said to \textit{control the FWER at level} $\alpha'_{\tiny \mbox{FWE}}$. If for a test $\Phi(Y)$ it holds that $\alpha_{\tiny \mbox{FWE}} = \alpha'_{\tiny \mbox{FWE}}$, we say that $\Phi(Y)$ offers \textit{exact control of the FWER at level} $\alpha'_{\tiny \mbox{FWE}}$. A general method to establish FWER control for a multiple test of the form \eqref{eq:Phi_multiple_test} - \eqref{eq:R_i_multiple_test} at a level $\alpha_{\tiny FWE}^{\prime}$ is afforded by consideration of the distribution of the \textit{maximum test statistic}
\begin{equation}\label{eq:def_maximum_statistic_I0}
\gamma_{max}^0(Y) := \max_{i \in I_0} \gamma_i(Y).
\end{equation}
The method rests on identifying a common critical value $u_{\alpha'}^{\tiny \mbox{FWE}} \in \mathbb{R}$ for all constituent tests $\phi_i(Y)$ of the form \eqref{eq:phi_i_multiple_test} that satisfies
\begin{equation}\label{eq:fwer_control_critical_value}
P_{\cap_{i \in I_0} \Theta_0^{(i)}} 
(\gamma_{max}^0(Y) \ge u_{\alpha'}^{\tiny \mbox{FWE}}) \le \alpha'_{\tiny \mbox{FWE}}.
\end{equation}
Intuitively, requirement \eqref{eq:fwer_control_critical_value} states that the probability of the maximum of the multiple test's test statistics to assume a value larger than or equal to the critical value $u_{\alpha'}^{\tiny \mbox{FWE}}$ over the set of true null hypotheses is smaller or equal to the the desired FWER control level $\alpha'_{\tiny \mbox{FWE}}$. As shown below, from requirement \eqref{eq:fwer_control_critical_value} it readily follows that
\begin{equation}\label{eq:fwer_control_theorem}
P_{\cap_{i \in I_0} \Theta_0^{(i)}} \left(\cup_{i \in I_0} \phi_i(Y) = 1 \right)
 \le \alpha'_{\tiny \mbox{FWE}}, 
\end{equation}
i.e., that the multiple test controls the FWER at level $\alpha'_{\tiny \mbox{FWE}}$. From an applied perspective, the maximum statistic-based FWER control approach entails that the distribution of the maximum statistic over the set of true null hypotheses needs to be evaluated based on the form of the  probabilistic model and the resulting distributions of the component test statistics $\gamma_i(Y)$. 

\begin{footnotesize}
\vspace{2mm}
\noindent \textbf{Proof of eq. \eqref{eq:fwer_control_theorem}}
\vspace{2mm}

\noindent Let $\Phi(Y)$ denote a multiple test of the form eqs. \eqref{eq:Phi_multiple_test} - \eqref{eq:R_i_multiple_test}, and define $u_i := u_{\alpha'}^{\tiny \mbox{FWE}}$ for all $i \in I_0 $. Further, let the critical value $u_{\alpha'}^{\tiny \mbox{FWE}}$ be such that with the definition of the maximum statistic $\gamma_{max}^0(Y)$ in eq. \eqref{eq:def_maximum_statistic_I0} it holds that
\begin{equation}
P_{\cap_{i \in I_0} \Theta_0^{(i)}} 
\left(\gamma_{max}^0(Y) \ge u_{\alpha'}^{\tiny \mbox{FWE}}\right) 
\le \alpha'_{\tiny \mbox{FWE}}.
\end{equation}
Then, with the definition of the FWER in eq. \eqref{eq:def_fwer}, it follows that
\begin{align}
\begin{split}
P_{\cap_{i \in I_0} \Theta_0^{(i)}} \left( \cup_{i \in I_0} \, \phi_i(Y) = 1 \right)
& = 1 - P_{\cap_{i \in I_0} \Theta_0^{(i)}} \left( \cap_{i \in I_0} \, \phi_i(Y) = 0 \right) \\
& = 1 - P_{\cap_{i \in I_0} \Theta_0^{(i)}} 
\left(\gamma_{i_1}(Y) < u_{\alpha'}^{\tiny \mbox{FWE}},
	  \gamma_{i_2}(Y) < u_{\alpha'}^{\tiny \mbox{FWE}}, ...,
	  \gamma_{i_{m_0}}(Y) < u_{\alpha'}^{\tiny \mbox{FWE}}  \right) \\
& = 1 - P_{\cap_{i \in I_0} \Theta_0^{(i)}}\left(\gamma_{max}^0 (Y) < u_{\alpha'}^{\tiny \mbox{FWE}} \right) \\
& = P_{\cap_{i \in I_0} \Theta_0^{(i)}}\left(\gamma_{max}^0 (Y) \ge u_{\alpha'}^{\tiny \mbox{FWE}} \right) \\
& \le \alpha'_{\tiny \mbox{FWE}},
\end{split}
\end{align}
where on the right-hand side of the second equation $i_j \in I_0$ for $j = 1,2,...,m_0$. In verbose form: the probability of the event that one or more of the component tests $\phi_i(Y), i \in I_0$ of the multiple test $\Phi(Y)$ evaluate to $1$ over the set of true null hypotheses $I_0$ is equal to the complementary probability of the event that all component tests evaluate to $0$ over the set of true null hypotheses $I_0$. Given the form of the multiple test $\Phi(Y)$, this probability in turn corresponds to the probability that all relevant component test statistics assume values smaller than the critical value $u_{\alpha'}^{\tiny \mbox{FWE}}$. The latter event is identical to the event that the maximum statistic $\gamma_{max}^0(Y)$ over the set of true null hypotheses is smaller than $u_{\alpha'}^{\tiny \mbox{FWE}}$. The complementary probability of this event then implies the validity of eq. \eqref{eq:fwer_control_theorem}.

$\hfill \Box$
\vspace{2mm}
\end{footnotesize}

\noindent A note on the usage of the terms ``uncorrected single test'' and ``corrected multiple testing'' inference in the main text may be appropriate here: de-facto, FWER control in multiple testing is not based on some form of correction procedure that turns an ``uncorrected \textit{p}-value'' into a ``corrected \textit{p}-value'', but the two \textit{p}-values of uncorrected and corrected inference instead refer to different statistics. Because the notion of ``correcting for the multiple testing problem'' using ``corrected \textit{p}-values'' is deeply engrained in the fMRI literature, however, we refrain from abandoning this terminology.

\end{document}